\documentclass{aa}
 
\usepackage[varg]{txfonts}

\usepackage[]{hyphenat}
\usepackage[]{microtype}
\usepackage{amssymb}
\usepackage{float}
\usepackage{bm}
\usepackage{epsfig}
\usepackage{soul}
\usepackage{subcaption}
\usepackage{multirow,balance}
\usepackage{amsfonts}
\usepackage{amsmath}
\usepackage{mathtools}
\usepackage{mathrsfs}
\usepackage{aas_macros}
\usepackage{graphicx}
\usepackage{natbib}
\usepackage{color}
\usepackage{listings} 
\usepackage{hyperref}  
\usepackage{url}
\usepackage{cases}
\usepackage{tikz}
\usetikzlibrary{decorations.pathreplacing}

\hypersetup{dvips, colorlinks=true, linkcolor=cyan, citecolor=black, filecolor=black, urlcolor=black}

\newcommand{\ri}{\mathrm{i}}
\newcommand{\re}{\mathrm{e}}
\newcommand{\rd}{\mathrm{d}}
\newcommand{\ra}{\mathrm{a}}
\newcommand{\rb}{\mathrm{b}}
\newcommand{\rc}{\mathrm{c}}
\newcommand{\rr}{\mathrm{r}}
\newcommand{\oF}{\overline{F}}
\newcommand{\oFa}{\overline{F^{\ra}}}
\newcommand{\oFb}{\overline{F^{\rb}}}

\newcommand{\oC}{\overline{\mathcal{C}}}
\newcommand{\oCab}{\overline{\mathcal{C}^{\ra \rb}}}
\newcommand{\oCac}{\overline{\mathcal{C}^{\ra \rc}}}
\newcommand{\oCbc}{\overline{\mathcal{C}^{\rb \rc}}}
\newcommand{\bR}{\bm{\mathcal{E}}}
\newcommand{\oP}{\overline{\Phi}}
\newcommand{\oPa}{\overline{\Phi^{\ra}}}
\newcommand{\oU}{\overline{U}}
\newcommand{\opsi}{\overline{\psi}}

\begin{document}

\setcounter{tocdepth}{3}

\title
{
The secular evolution of discrete  quasi-Keplerian systems.\\ I. Kinetic theory of stellar clusters near black holes
}

\titlerunning{The secular evolution of quasi-Keplerian systems. I. Kinetic theory}

\author{J.-B. Fouvry
\inst{\ref{inst1}}
\and
C. Pichon
\inst{\ref{inst1},\ref{inst2}}
\and
J. Magorrian
\inst{\ref{inst3}}
}

\institute{
Institut d'Astrophysique de Paris and UPMC, CNRS (UMR 7095), 98 bis Boulevard Arago, 75014, Paris, France\\
\email{fouvry@iap.fr}\label{inst1}
\and
Korea Institute of Advanced Studies (KIAS) 85 Hoegiro, Dongdaemun-gu, Seoul, 02455, Republic of Korea\label{inst2}
\and
Rudolf Peierls Centre for Theoretical Physics, University of Oxford, Keble Road, Oxford OX1 3RH, United Kingdom\label{inst3}
}

\date{Received \today /
Accepted --}
\abstract{
We derive the kinetic equation that describes the secular evolution of
a large set of particles orbiting a dominant massive
object, such as stars bound to a supermassive black hole or 
a proto-planetary debris disc encircling a star.
Because the particles move in a quasi-Keplerian potential, their
orbits can be approximated by ellipses whose orientations remain fixed
over many dynamical times.  The kinetic equation is obtained by simply
averaging the BBGKY equations over the fast angle that describes
motion along these ellipses.
This so-called Balescu-Lenard equation describes self-consistently the long-term
evolution of the distribution of quasi-Keplerian orbits around the
central object: it models the
diffusion and drift of their actions, induced through their mutual
resonant interaction.
Hence, it is the master equation that describes the secular effects of
resonant relaxation.
We show how it captures the phenonema of mass segregation and of the
relativistic Schwarzschild barrier recently discovered in $N$-body
simulations.

}

\keywords{ Galaxies: kinematics and dynamics - Galaxies: nuclei - Diffusion - Gravitation
}

\maketitle

\section{Introduction}
\label{sec:introduction}

The stars in a stellar cluster surrounding a dominant supermassive
black hole (BH) move in a quasi-Keplerian potential.
Their orbits are ellipses that maintain their spatial orientation for
many orbital periods.
So, for many purposes, the cluster can be thought of as a system of
massive wires, in which the mass of each star is smeared out along the
path traced by its quasi-Keplerian orbit.
The consequences of this idea were first developed
by~\cite{RauchTremaine1996}, who showed that wire-wire interactions
greatly enhance the relaxation of the stars' angular momenta when compared
to conventional estimates that ignore the coherence of the stars'
orbits over many dynamical times and consider only uncorrelated
two-body encounters.
They named this phenomenon ``resonant relaxation'', because such
enhanced relaxation occurs more generally in any potential in which
the three-dimensional vector of stellar orbital frequencies~$\bm{\Omega}$
satisfies a commensurability condition of the form
$ { \bm{n} \!\cdot\! \bm{\Omega} \!\simeq\! 0 }$ for some vector of
integers~${ \bm{n} \!=\! (n_{1} , n_{2} , n_{3}) }$.

Understanding the effects of these relaxation processes in galactic
nuclei is important when predicting the rates of tidal disruptions of
stars by black holes~\citep[e.g.][]{RauchTremaine1996,RauchIngalls1998}, predicting
merging rates of binary supermassive black holes~\citep[e.g.][]{Yu2002} or of gravitational wave signatures from
star-BH interactions~\citep[e.g.][]{HopmanAlexander2006,Merritt2011}.
Resonant relaxation provides as well a promising framework for explaining
some of the puzzling features of the young stellar populations found
at the centre of our own Galaxy~\citep[e.g.][]{KocsisTremaine2011}.

A first way to study the evolution of such star
clusters is by direct ${N-}$body simulations.
Unfortunately, extracting
physical insights from such simulations is challenging, because the
complex dynamical processes involved are entangled and, more
practically, the computational costs of running the simulations
typically mean that one can run just a few realisations, each with
relatively small~$N$.
For problems that focus on resonant relaxation phenomena, one can often
do better by using ${N-}$wires codes~\citep[e.g.][]{KocsisTremaine2015}
in which individual stars are replaced by orbit-averaged Keplerian
wires.

A complementary way of understanding these systems is by using the
tools of kinetic theory.
For plasmas,~\cite{Balescu1960} and~\cite{Lenard1960} have developed a
rigorous kinetic theory that takes the most important collective
effects into account.
In this theory, the coupled evolution equations for
the system's one-body distribution function and its two-body
correlation function are reduced to a single equation -- the
Balescu-Lenard equation -- that describes the evolution of the
one-body distribution function alone. See~\cite{Chavanis2010jsm,Chavanis2012epjp2,Chavanis2013,FouvryChavanisPichon2016}
  for a review on the early development of kinetic theory for plasmas,
  stellar systems, and other long-range interacting systems.
The original Balescu-Lenard formalism was developed for homogeneous
plasmas.  One way of generalising it to inhomogeneous self-gravitating
systems, such as star clusters, was proposed by~\cite{Gilbert1968}.
\cite{ST1,ST2} have recently applied Gilbert's methods
to the secular evolution of a star cluster around a BH, which is the
problem addressed by the present paper.

An alternative way of generalising the Balescu-Lenard formalism to
inhomogeneous systems has been presented by~\cite{Heyvaerts2010}
and~\cite{Chavanis2012}, who reformulate the non-linear kinetic
equation in terms of the angle-action variables that are appropriate
for spatially inhomogeneous multi-periodic systems.
\cite{fouvry1,fouvry2,fouvry3} have applied this formalism to describe
the secular response of tepid self-gravitating stellar discs.
The resulting inhomogeneous Balescu-Lenard equation accounts for
the self-induced secular orbital diffusion of a self-gravitating system driven by
the internal shot noise due to the finite number~$N$ of particles
involved.
In common with all results based on the Balescu-Lenard formalism, it
is valid to order ${ O(1/N) }$ in a formal expansion of the dynamics
ordered by the small parameter~${1/N}$.
Therefore, it describes the evolution of the system on timescales of about
${ N t_{\rm d} }$, where $t_{\rm d}$ is the dynamical time.
The secular interactions between particles need not be local in space:
they need only correspond to gravitationally amplified long-range
correlations via resonances.

In its original form, however, the Balescu-Lenard formalism assumes that
resonances are localised in action space and not degenerate.
Therefore, it must be re-examined before it can be applied to the
degeneracies inherent to resonant systems.
In this paper, we show how to account for these degeneracies in the
case of a cluster of $N$ particles orbiting a massive, possibly
relativistic, central body.
We first average the equations of motion over the fast angle
associated with the orbital motion of the stars around the BH.
Once such an averaging is carried out, it turns out that the general
formalism of the inhomogenous Balescu-Lenard equation applies
straightforwardly and yields the associated secular collisional
equation.
This equation captures the diffusion and drift of particles' actions
induced through their mutual resonant interaction at the frequency
shifts present in addition to the mean Keplerian dynamics, for instance
induced by the self-gravity of the cluster or relativistic effects.
Hence it is well suited to describe the secular evolution of a large
set of particles orbiting around a massive object, for instance to account
for the long-term evolution of a disc or a sphere of (possibly
relativistic) stars near a galactic centre, or a proto-planetary debris
disc circling a star.
As such, it captures the secular effects of a sequence of polarised
wire-wire interactions (corresponding to scalar or vector resonant
relaxation) on the underlying orbital structure of the cluster.

The paper is organised as follows.
Section~\ref{sec:BBGKY_hierarchy} derives the BBGKY hierarchy of a system with a massive central body
using canonical coordinates to account properly for the black hole's motion.
Section~\ref{sec:AA_coordinates} introduces angle-action coordinates for such quasi-Keplerian systems.
Section~\ref{sec:_Average_equations} averages the corresponding dynamical equations over the fast angles of the Keplerian motion. 
Section~\ref{sec:secular_collisional} presents the degenerate one and multi-component Keplerian Balescu-Lenard equations (Appendix~\ref{sec:secular_collisional_derivation} details both derivations, following the steps of~\cite{Heyvaerts2010}, while also correcting for a minor issue in the multi-component case). 
Section~\ref{sec:applications} discusses applications to the cases of razor-thin axisymmetric and spherical clusters orbiting a massive central object, and compares our results to those of~\cite{ST3} and others, while section~\ref{sec:conclusion} concludes. Appendix~\ref{sec:relativistic_precessions} outlines the relativistic precessions frequencies involved near a massive 
black hole, and Appendix~\ref{sec:FP_to_Langevin} presents the stochastic counterpart of the Keplerian Balescu-Lenard equation.

\section{The BBGKY hierarchy}
\label{sec:BBGKY_hierarchy}
Consider a system of $N$ stars in motion about a central black hole of
mass~$M_\bullet$, in which each star has mass $\mu$. We assume that
the total stellar mass $M_\star\equiv\mu N$ is small enough that the ratio
\begin{equation}
\varepsilon\equiv M_\star/M_\bullet\ll 1.
\end{equation}
Let $\bm{X}_{\bullet}$ be the location of the BH and $\bm{X}_{n}$ be the location of
the $n^{\rm th}$ star referred to an inertial frame.  The Hamiltonian
for the system is then given by
\begin{align}
  H&=\frac{\bm{P}_{\bullet}^{2}}{2 M_{\bullet} } + \sum_{i = 1}^{N} \frac{\bm{P}^2_{i}}{2\mu}  \nonumber
\\
&+ \mu M_{\bullet} \sum_{i = 1}^{N} U(| \bm{X}_{i} \!-\! \bm{X}_{\bullet} |) + \mu^{2} \sum_{i < j}^{N} U(|\bm{X}_{i} - \bm{X}_{j}|)  \nonumber
\\
&+\mu M_{\star} \sum_{i = 1}^{N} \Phi_{\rm rel} (\bm{X}_{i} \!-\! \bm{X}_{\bullet})  \, ,
\label{initial_Hamiltonian}
\end{align}
in which the canonical momenta are given by ${ \bm{P}_{\bullet} \!\equiv\! M_{\bullet} \dot{\bm{X}}_{\bullet} }$ and ${ \bm{P}_{n} \!\equiv\! \mu \dot{\bm{X}}_{n} }$. Here, ${ U(|\bm{X}|) }$ corresponds to the interaction potential, that is ${ U(|\bm{X}|) \!\equiv\! - G/|\bm{X}| }$ in the gravitational context. In equation~\eqref{initial_Hamiltonian}, the first two terms correspond to the kinetic energy of the BH and the stars. The third term corresponds to the Keplerian potential of the BH, while the fourth term is associated with the pairwise interactions among stars. Finally, the third line of equation~\eqref{initial_Hamiltonian} accounts for the relativistic correction forces such as the Schwarzschild and Lense-Thirring precessions occurring in the vicinity of the BH (see Appendix~\ref{sec:relativistic_precessions}), where the normalisation prefactor ${ \mu M_{\star} }$ was added for later convenience. For simplicity, we neglected any additional external perturbations, which could offset the system. This will be the subject of a future work.

Let us now rewrite the Hamiltonian from equation~\eqref{initial_Hamiltonian} as $N$ decoupled Keplerian Hamiltonians plus perturbations. We follow~\cite{Duncan1998} and carry out a canonical transformation to a new set of coordinates, the democratic heliocentric coordinates ${ (\bm{x}_{\bullet} , \bm{x}_{1} , ... , \bm{x}_{N}) }$ defined as
\begin{equation}
\bm{x}_{\bullet} = \frac{1}{M_{\rm tot}} \bigg[ M_{\bullet} \,  \bm{X}_{\bullet} \!+\! \sum_{i=1}^{N} \mu \, \bm{X}_{i} \bigg] \;\;\; ; \;\;\; \bm{x}_{i} = \bm{X}_{i} \!-\! \bm{X}_{\bullet} \, ,
\label{democratic_coordinates}
\end{equation}
where we have introduced the total mass of the system ${ M_{\rm tot} \!=\! M_{\bullet} \!+\! M_{\star} }$. In equation~\eqref{democratic_coordinates}, $\bm{x}_{\bullet}$ corresponds to the position of the system's centre of mass and $\bm{x}_{i}$ to the locations of the stars in the frame centred on the BH. These relations have inversion
\begin{equation}
\bm{X}_{\bullet} = \bm{x}_{\bullet} \!-\! \frac{1}{M_{\rm tot}} \sum_{i = 1}^{N} \mu \, \bm{x}_{i} \;\;\; ; \;\;\; \bm{X}_{i} = \bm{x}_{\bullet} \!+\! \bm{x}_{i} \!-\! \frac{1}{M_{\rm tot}} \sum_{j = 1}^{N} \mu \, \bm{x}_{j} \, .
\label{democratic_inversion}
\end{equation}
As obtained in~\cite{Duncan1998}, the associated canonical momenta ${ (\bm{p}_{\bullet} , \bm{p}_{1} ,..., \bm{p}_{N}) }$ are
\begin{equation}
\bm{p}_{\bullet} = \bm{P}_{\bullet} \!+\! \sum_{i = 1}^{N} \bm{P}_{i} \;\;\; ; \;\;\; \bm{p}_{i} = \bm{P}_{i} \!-\! \frac{\mu}{M_{\rm tot}} \bigg[ \bm{P}_{\bullet} \!+\! \sum_{j = 1}^{N} \bm{P}_{j} \bigg] \, .
\label{democratic_momenta}
\end{equation}
Within these new canonical coordinates, the Hamiltonian from equation~\eqref{initial_Hamiltonian} takes the form
\begin{align}
H & = \sum_{i = 1}^{N} \bigg[ \frac{\bm{p}_{i}^{2}}{2 \mu} \!+\! \mu M_{\bullet} U (|\bm{x}_{i}|) \!+\! \mu M_{\star} \Phi_{\rm rel} (\bm{x}_{i}) \bigg] + \mu^{2} \sum_{i < j}^{N} U (|\bm{x}_{i} \!-\! \bm{x}_{j}|)  \nonumber
\\
& + \frac{\bm{p}_{\bullet}^{2}}{2 M_{\rm tot}} \!+\! \frac{1}{2 M_{\bullet}} \bigg[ \sum_{i = 1}^{N} \bm{p}_{i} \bigg]^{2}  \, ,  
\label{Hamiltonian_democratic}
\end{align}
which consists of $N$ independent Keplerian Hamiltonians (first term of
the first line) plus the two-body couplings among them (second term) plus
additional kinetic terms (second line).
The evolution of the total momentum $\bm{p}_{\bullet}$ is given by ${ \dot{\bm{p}}_{\bullet} \!=\! - \partial H / \partial \bm{x}_{\bullet} \!=\! 0}$. Without loss of generality, we may therefore assume that ${ \bm{p}_{\bullet} \!=\! 0 }$. The evolution of the barycentre position is then given by ${ \dot{\bm{x}}_{\bullet} \!=\! \partial H/\partial \bm{p}_{\bullet} \!=\! \bm{p}_{\bullet}/M_{\rm tot} \!=\! 0 }$, and we therefore set ${ \bm{x}_{\bullet} \!=\! 0 }$.
Introducing the notation ${ \bm{v}_{n} \!\equiv\! \bm{p}_{n} / \mu \!\neq\! \dot{\bm{x}}_{n} }$, the Hamiltonian from equation~\eqref{Hamiltonian_democratic} becomes
\begin{align}
H & = \sum_{i = 1}^{N} \bigg[ \frac{\mu}{2} \bm{v}_{i}^{2} \!+\! \mu M_{\bullet} U (|\bm{x}_{i}|) \!+\! \mu M_{\star} \Phi_{\rm rel} (\bm{x}_{i})  \bigg] + \mu^{2} \sum_{i < j}^{N} U (|\bm{x}_{i} \!-\! \bm{x}_{j}|)  \nonumber
\\
& + \frac{\mu^{2}}{2 M_{\bullet}} \bigg[ \sum_{i = 1}^{N} \bm{v}_{i} \bigg]^{2} \, ,
\label{Hamiltonian_democratic_simpler}
\end{align}
in which one of the kinetic terms in the second line has been
transformed away.

In order to obtain a statistical description of the system, we now introduce its ${N-}$body probability distribution function (PDF) ${ P_{N} (\Gamma_{1} ,..., \Gamma_{N} , t) }$ defined so that  ${ P_{N} (\Gamma_{1} , ... , \Gamma_{N} , t ) \, \rd\Gamma_{1} ... \rd \Gamma_{N} }$ is at time~$t$ the probability of finding particle~1 within the volume element ${ \rd \Gamma_{1} }$ located at the phase space point ${ \Gamma_{1} \!=\! (\bm{x}_{1} , \bm{v}_{1}) }$, particle~2  within ${ \rd \Gamma_{2} }$ of the phase space point ${ \Gamma_{2} \!=\! (\bm{x}_{2} , \bm{v}_{2}) }$, and so on. We normalise $P_{N}$ such that
\begin{equation}
 \!\! \int \!\! \rd\Gamma_{1} ... \rd \Gamma_{N} \, P_{N} (\Gamma_{1} , ... , \Gamma_{N} , t) = 1 \, .
 \label{normalisation_PN}
\end{equation}
It evolves according to Liouville's equation
\begin{equation}
\frac{\partial P_{N}}{\partial t} + \sum_{i = 1}^{N} \bigg[ \dot{\bm{x}}_{i} \!\cdot\! \frac{\partial P_{N}}{\partial \bm{x}_{i}} \!+\! \dot{\bm{v}}_{i} \!\cdot\! \frac{\partial P_{N}}{\partial \bm{v}_{i}} \bigg] = 0 \, ,
\label{Liouville_equation}
\end{equation}
The dynamics of the individual particles are given by Hamilton's equations, ${ \mu \rd \bm{x}_{i} / \rd t \!=\! \partial H / \partial \bm{v}_{i} }$ and ${ \mu \rd \bm{v}_{i} / \rd t \!=\! - \partial H / \partial \bm{x}_{i} }$, where the system's Hamiltonian was obtained in equation~\eqref{Hamiltonian_democratic_simpler}.
From $P_{N}$, we define reduced PDFs,
\begin{equation}
P_{n} (\Gamma_{1} , ... , \Gamma_{n} , t) \equiv
 \!\! \int \!\! \rd \Gamma_{n+1} ... \rd \Gamma_{N} \, P_{N} (\Gamma_{1} , ... , \Gamma_{N} , t) \, ,
\label{definition_Pn}
\end{equation}
by integrating over the phase space locations of particles~${n+1}$ to~$N$. To obtain the evolution equation of any reduced PDF $P_{n}$, we integrate Liouville's equation~\eqref{Liouville_equation} over ${ \rd \Gamma_{n+1} ... \rd \Gamma_{N} }$ and use the fact that $P_{N}$ and $H$ are unchanged under permutations of their arguments. This leads to the general term of the BBGKY hierarchy
\begin{align}
& \, \frac{\partial P_{n}}{\partial t} \!+\! \!\sum_{i = 1}^{n} \!\bigg\{ \bigg[ \bm{v}_{i} \!+\! \frac{\varepsilon}{N} \!\sum_{j = 1}^{n} \bm{v}_{j} \bigg] \!\cdot\! \frac{\partial P_{n}}{\partial \bm{x}_{i}} \!+\! \bigg[\! M_{\bullet} \bm{\mathcal{F}}_{\!i0} \!+\! \mu \!\! \sum_{j = 1 , j \neq i}^{n} \!\!\! \bm{\mathcal{F}}_{\!ij} \!+\! M_{\star} \bm{\mathcal{F}}_{\! i \rr} \!\bigg] \!\cdot\! \frac{\partial P_{n}}{\partial \bm{v}_{i}} \!\bigg\}  \nonumber
\\
& \, + (N \!-\! n) \sum_{i = 1}^{n} \!\! \int \!\! \rd \Gamma_{n+1} \, \bigg[ \frac{\varepsilon}{N} \, \bm{v}_{n+1} \!\cdot\! \frac{\partial P_{n+1}}{\partial \bm{x}_{i}} \!+\! \mu \, \bm{\mathcal{F}}_{\!i , n+1} \!\cdot\! \frac{\partial P_{n+1}}{\partial \bm{v}_{i}} \bigg] = 0 \, .
\label{BBGKY_Pn}
\end{align}
Here, we have written the force exerted by particle $j$ on particle $i$ as ${ \mu \bm{\mathcal{F}}_{ij} \!=\! - \mu \partial U_{ij} / \partial \bm{x}_{i} }$, using the shorthand notation ${ U_{ij} \!=\! U (|\bm{x}_{i} \!-\! \bm{x}_{j}|) }$. The force exerted by the BH on particle $i$ is denoted by ${ M_{\bullet} \bm{\mathcal{F}}_{i 0} \!=\! - M_{\bullet} \partial U_{i0} / \partial \bm{x}_{i} }$ and the force associated with the relativistic corrections as ${ M_{\star} \bm{\mathcal{F}}_{i \rr} \!=\! - M_{\star} \partial \Phi_{\rm rel} / \partial \bm{x}_{i} }$.

It is convenient to replace these PDFs by the reduced distribution functions (DFs)
\begin{equation}
f_{n} (\Gamma_{1} , ... , \Gamma_{n} , t)
\equiv \mu^{n} \frac{N!}{(N \!-\! n)!} P_{n} (\Gamma_{1} , ... , \Gamma_{n} , t),
\label{definition_fn}
\end{equation}
in terms of which equation~\eqref{BBGKY_Pn} can be rewritten as
\begin{align}
& \, \frac{\partial f_{n}}{\partial t} \!+\!\! \sum_{i = 1}^{n} \!\bigg\{ \bigg[ \bm{v}_{i} \!+\! \frac{\varepsilon}{N} \sum_{j = 1}^{n} \bm{v}_{j} \!\bigg] \!\cdot\! \frac{\partial f_{n}}{\partial \bm{x}_{i}} \!+\! \bigg[\! M_{\bullet} \bm{\mathcal{F}}_{\!i0} \!+\! \mu \!\! \sum_{j = 1 , j \neq i}^{n} \!\! \bm{\mathcal{F}}_{\!ij} \!+\! M_{\star} \bm{\mathcal{F}}_{\! i \rr} \!\bigg] \!\cdot\! \frac{\partial f_{n}}{\partial \bm{v}_{i}} \!\bigg\}  \nonumber
\\
& \, + \sum_{i = 1}^{n} \!\! \int \!\! \rd \Gamma_{n+1} \, \bigg[ \frac{1}{M_{\bullet}} \, \bm{v}_{n+1} \!\cdot\! \frac{\partial f_{n+1}}{\partial \bm{x}_{i}} \!+\! \bm{\mathcal{F}}_{\!i , n+1} \!\cdot\! \frac{\partial f_{n+1}}{\partial \bm{v}_{i}} \bigg] = 0 \, .
\label{BBGKY_fn}
\end{align}
To isolate the contributions to $f_{n}$ that arise from correlations among particles, let us introduce the cluster representation of the DFs. We define the ${2-}$body correlation function $g_{2}$ in terms of $f_{1}$ and~$f_{2}$ via
\begin{equation}
f_{2} (\Gamma_{1} , \Gamma_{2}) = f_{1} (\Gamma_{1}) \, f_{1} (\Gamma_{2}) + g_{2} (\Gamma_{1} , \Gamma_{2}) \, .
\label{definition_g2}
\end{equation}
Similarly, the ${3-}$body correlation function $g_{3}$ is defined by
\begin{align}
& \, f_{3} (\Gamma_{1} , \Gamma_{2} , \Gamma_{3}) = f_{1} (\Gamma_{1}) \, f_{1} (\Gamma_{2}) \, f_{1} (\Gamma_{3})   \nonumber
\\
& \, + f_{1} (\Gamma_{1}) \, g_{2} (\Gamma_{2} , \Gamma_{3}) + f_{1} (\Gamma_{2}) \, g_{2} (\Gamma_{1} , \Gamma_{3}) + f_{1} (\Gamma_{3}) \, g_{2} (\Gamma_{1} , \Gamma_{2})    \nonumber
\\
& \, + g_{3} (\Gamma_{1} , \Gamma_{2} , \Gamma_{3}) \, .
\label{definition_g3}
\end{align}
These correlation functions have simple dependence on the number of
particles~$N$. It is straightforward to check that the following normalisations
hold:
\begin{align}
& \, \!\! \int \!\! \rd \Gamma_{1} \, f_{1} (\Gamma_{1}) = \mu N \;\; ; \;\; \!\! \int \!\! \rd \Gamma_{1} \rd \Gamma_{2} \, g_{2} (\Gamma_{1} , \Gamma_{2}) = - \mu^{2} N  \;\; ; \;\;   \nonumber
\\
& \, \!\! \int \!\! \rd \Gamma_{1} \rd \Gamma_{2} \rd \Gamma_{3} \, g_{3} (\Gamma_{1} , \Gamma_{2} , \Gamma_{3}) = 2 \mu^{3} N \, .
\label{normalisations_f1_g2_g3}
\end{align}
As the individual mass scales like ${ \mu \!\sim\! 1/N }$, one immediately has ${ |f_{1}| \!\sim\! 1 }$, ${ |g_{2}| \!\sim\! 1/N }$, and ${ |g_{3}| \!\sim\! 1/N^{2} }$.  Using the decompositions from equations~\eqref{definition_g2} and~\eqref{definition_g3}, after some simple algebra, the first two equations of the BBGKY hierarchy from equation~\eqref{BBGKY_fn} become
\begin{align}
& \, \frac{\partial f_{1}}{\partial t} \!+\! \bigg[ \bm{v}_{1} \!+\! \frac{\varepsilon}{N} \,  \bm{v}_{1} \bigg] \!\cdot\! \frac{\partial f_{1}}{\partial \bm{x}_{1}} \!+\! M_{\bullet} \bm{\mathcal{F}}_{\!10} \!\cdot\! \frac{\partial f_{1}}{\partial \bm{v}_{1}} \!+\! \bigg[ \!\! \int \!\! \rd \Gamma_{2} \, \bm{\mathcal{F}}_{\!12} f_{1} (\Gamma_{2}) \bigg] \!\cdot\! \frac{\partial f_{1}}{\partial \bm{v}_{1}}  \nonumber
\\
& \, \!+\! M_{\star} \bm{\mathcal{F}}_{\! 1 \rr} \!\cdot\! \frac{\partial f_{1}}{\partial \bm{v}_{1}} \!+\! \!\! \int \!\! \rd \Gamma_{2} \, \bm{\mathcal{F}}_{\!12} \!\cdot\! \frac{\partial g_{2} (\Gamma_{1} , \Gamma_{2})}{\partial \bm{v}_{1}}  \nonumber
\\
& \,  \!+\! \frac{1}{M_{\bullet}} \frac{\partial f_{1}}{\partial \bm{x}_{1}} \!\cdot\! \!\! \int \!\! \rd \Gamma_{2} \, \bm{v}_{2} \, f_{1} (\Gamma_{2}) \!+\! \frac{1}{M_{\bullet}} \!\! \int \!\! \rd \Gamma_{2} \, \bm{v}_{2} \!\cdot\! \frac{\partial g_{2} (\Gamma_{1} , \Gamma_{2})}{\partial \bm{x}_{1}} = 0 \, ,
\label{BBGKY_1_fn}
\end{align}
and
\begin{align}
& \, \frac{1}{2} \frac{\partial g_{2}}{\partial t} \!+\! \bigg[ \bm{v}_{1} \!+\! \frac{\varepsilon}{N} \,  (\bm{v}_{1} \!+\! \bm{v}_{2}) \bigg] \!\cdot\! \frac{\partial g_{2}}{\partial \bm{x}_{1}} \!+\! \frac{\varepsilon}{N} \, \bm{v}_{2} \!\cdot\! \frac{\partial f_{1}}{\partial \bm{x}_{1}} \, f_{1} (\Gamma_{2})  \nonumber
\\
& \, \!+\! M_{\bullet} \bm{\mathcal{F}}_{\!10} \!\cdot\! \frac{\partial g_{2}}{\partial \bm{v}_{1}} \!+\! \bigg[ \!\! \int \!\! \rd \Gamma_{3} \, \bm{\mathcal{F}}_{\!13} f_{1} (\Gamma_{3}) \bigg] \!\cdot\! \frac{\partial g_{2}}{\partial \bm{v}_{1}}  \nonumber  
\\
& \, \!+\! M_{\star} \bm{\mathcal{F}}_{\! 1 \rr} \!\cdot\! \frac{\partial g_{2}}{\partial \bm{v}_{1}} + \mu \bm{\mathcal{F}}_{\!12} \!\cdot\! \frac{\partial f_{1}}{\partial \bm{v}_{1}} f_{1} (\Gamma_{2}) \!+\! \bigg[ \!\! \int \!\! \rd \Gamma_{3} \, \bm{\mathcal{F}}_{\!13} g_{2} (\Gamma_{2} , \Gamma_{3}) \bigg] \!\cdot\! \frac{\partial f_{1}}{\partial \bm{v}_{1}}  \nonumber
\\
& \, \!+\! \frac{1}{M_{\bullet}} \frac{\partial f_{1}}{\partial \bm{x}_{1}} \!\cdot\! \!\! \int \!\! \rd \Gamma_{3} \, \bm{v}_{3} g_{2} (\Gamma_{2} , \Gamma_{3}) + \frac{1}{M_{\bullet}} \frac{\partial g_{2}}{\partial \bm{x}_{1}} \!\cdot\! \!\! \int \!\! \rd \Gamma_{3} \, \bm{v}_{3} f_{1} (\Gamma_{3})  \nonumber
\\
& \, \!+\! \mu \bm{\mathcal{F}}_{\!12} \!\cdot\! \frac{\partial g_{2}}{\partial \bm{v}_{1}} \!+\! \!\! \int \!\! \rd \Gamma_{3} \, \bm{\mathcal{F}}_{\!13} \!\cdot\! \frac{\partial g_{3} (\Gamma_{1} , \Gamma_{2} , \Gamma_{3})}{\partial \bm{v}_{1}}  \nonumber
\\
& \, \!+\! \frac{1}{M_{\bullet}} \!\! \int \!\! \rd \Gamma_{3} \, \bm{v}_{3} \!\cdot\! \frac{\partial g_{3} (\Gamma_{1} , \Gamma_{2} , \Gamma_{3})}{\partial \bm{x}_{1}} \!+\! (1 \!\leftrightarrow\! 2) = 0 \, ,
\label{BBGKY_2_fn}
\end{align}
where ${ (1 \!\leftrightarrow\! 2) }$ means that all preceding terms are written out again, but with indices~1 and~2 swapped.

We now use the scalings obtained in equation~\eqref{normalisations_f1_g2_g3} to 
truncate equations~\eqref{BBGKY_1_fn} and~\eqref{BBGKY_2_fn} at order~${1/N}$. Notice that the system includes two small parameters, namely ${1/N}$ associated with the discreteness of the system and ${ \varepsilon \!=\! M_{\star} / M_{\bullet} }$ associated with the amplitude of the non-Keplerian components. As will be emphasised in the upcoming calculations, we will perform kinetic developments, where we only keep terms of the order $\varepsilon$ and ${ \varepsilon/N }$. In equation~\eqref{BBGKY_1_fn}, all the terms are of order ${1/N}$ or larger, and should therefore all be kept. In equation~\eqref{BBGKY_2_fn},  the first four lines are of order ${1/N}$ (except for the correction ${ (\varepsilon/N) (\bm{v}_{1} \!+\! \bm{v}_{2}) \!\cdot\! \partial g_{2} / \partial \bm{x}_{1} }$ which may be neglected), while all the terms from the two last lines are of order ${1/N^{2}}$ and may therefore be neglected.
Notice that the first term on the fifth line of equation~\eqref{BBGKY_2_fn}, which, while being of order ${1/N^{2}}$, can nevertheless get arbitrarily large as particles $1$ and $2$ get closer. This term accounts for strong collisions between two particles, which are not accounted for in the present formalism.
In addition to these truncations, and in order to consider terms of order $1$, let us finally introduce the system's ${1-}$body DF $F$ and its ${2-}$body autocorrelation function $\mathcal{C}$ as
\begin{equation}
F = \frac{f_{1}}{M_{\star}} \;\;\; ; \;\;\; \mathcal{C} = \frac{g_{2}}{\mu M_{\star}} \, .
\label{definition_F_C}
\end{equation}
Moreover, in order to emphasise the various order of magnitude of the forces present in the problem, let us also rescale some of the quantities appearing in equations~\eqref{BBGKY_1_fn} and~\eqref{BBGKY_2_fn}. Let us first rescale the interaction potential using the mass of the central black hole, so as to have the relations
\begin{equation}
\bm{\mathcal{F}}_{\! i j} = - \frac{\partial U_{ij}}{\partial \bm{x}_{i}} \;\; ; \;\; U_{ij} = - \frac{G M_{\bullet}}{|\bm{x}_{i} \!-\! \bm{x}_{j}|} \, .
\label{rescaling_U}
\end{equation}
Similarly, the relativistic potential ${ \Phi_{\rr} \!=\! \Phi_{\rm rel} }$ is also rescaled so that
\begin{equation}
\bm{\mathcal{F}}_{\! i \rr} = - \frac{\partial \Phi_{\rr}}{\partial \bm{x}_{i}} \;\;\; ; \;\;\; \Phi_{\rr} \!\to\! \frac{\Phi_{\rr}}{M_{\bullet}} \;\;\; ; \;\;\; \bm{\mathcal{F}}_{\! i \rr} \!\to\! \frac{\bm{\mathcal{F}}_{\! i \rr}}{M_{\bullet}} \, .
\label{rescale_Phi_a}
\end{equation}
Following these various truncations and renormalisations, equation~\eqref{BBGKY_1_fn} becomes
\begin{align}
& \, \frac{\partial F}{\partial t} \!+\! \bigg[ \bm{v}_{1} \!+\! \frac{\varepsilon}{N} \, \bm{v}_{1} \bigg] \!\cdot\! \frac{\partial F}{\partial \bm{x}_{1}} \!+\! \bm{\mathcal{F}}_{\! 10} \!\cdot\! \frac{\partial F}{\partial \bm{v}_{1}} \!+\! \varepsilon \bigg[ \!\! \int \!\! \rd \Gamma_{2} \, \bm{\mathcal{F}}_{\! 12} F (\Gamma_{2}) \bigg] \!\cdot\! \frac{\partial F}{\partial \bm{v}_{1}}   \nonumber
\\
& \, \!+\! \varepsilon \bm{\mathcal{F}}_{\! 1 \rr} \!\cdot\! \frac{\partial F}{\partial \bm{v}_{1}} \!+\! \frac{\varepsilon}{N} \!\! \int \!\! \rd \Gamma_{2} \, \bm{\mathcal{F}}_{\! 12} \!\cdot\! \frac{\partial \mathcal{C} (\Gamma_{1} , \Gamma_{2})}{\partial \bm{v}_{1}}   \nonumber
\\
& \, \!+\! \varepsilon \frac{\partial F}{\partial \bm{x}_{1}} \!\cdot\! \!\! \int \!\! \rd \Gamma_{2} \, \bm{v}_{2} \, F (\Gamma_{2}) \!+\! \frac{\varepsilon}{N} \!\! \int \!\! \rd \Gamma_{2} \, \bm{v}_{2} \!\cdot\! \frac{\partial \mathcal{C} (\Gamma_{1} , \Gamma_{2})}{\partial \bm{x}_{1}} = 0 \, ,
\label{BBGKY_1_rescaled}
\end{align}
while equation~\eqref{BBGKY_2_fn} becomes
\begin{align}
& \, \frac{1}{2} \frac{\partial \mathcal{C}}{\partial t} \!+\! \bm{v}_{1} \!\cdot\! \frac{\partial \mathcal{C}}{\partial \bm{x}_{1}} \!+\! \bm{\mathcal{F}}_{\! 10} \!\cdot\! \frac{\partial \mathcal{C}}{\partial \bm{v}_{1}}  \!+\! \varepsilon \, \bm{v}_{2} \!\cdot\! \frac{\partial F}{\partial \bm{x}_{1}} \, F (\Gamma_{2})  \nonumber
\\
& \, \!+\! \varepsilon \bigg[ \!\! \int \!\! \rd \Gamma_{3} \, \bm{\mathcal{F}}_{\!13} F (\Gamma_{3}) \bigg] \!\cdot\! \frac{\partial \mathcal{C}}{\partial \bm{v}_{1}} \!+\! \varepsilon \bm{\mathcal{F}}_{\! 1 \rr} \!\cdot\! \frac{\partial \mathcal{C}}{\partial \bm{v}_{1}}   \nonumber
\\
& \, \!+\! \varepsilon \bm{\mathcal{F}}_{\! 12} \!\cdot\! \frac{\partial F}{\partial \bm{v}_{1}} F (\Gamma_{2}) \!+\! \varepsilon \bigg[ \!\! \int \!\!\! \rd \Gamma_{3} \bm{\mathcal{F}}_{\! 13} \mathcal{C} (\Gamma_{2} , \Gamma_{3}) \!\bigg] \!\cdot\! \frac{\partial F}{\partial \bm{v}_{1}}   \nonumber
\\
& \, \!+\! \varepsilon \frac{\partial F}{\partial \bm{x}_{1}} \!\cdot\! \!\! \int \!\! \rd \Gamma_{3} \, \bm{v}_{3} \, \mathcal{C} (\Gamma_{2} , \Gamma_{3}) \!+\! \varepsilon \frac{\partial \mathcal{C}}{\partial \bm{x}_{1}} \!\cdot\! \!\! \int \!\!\rd \Gamma_{3} \, \bm{v}_{3} \, F (\Gamma_{3}) \!+\! (1 \!\leftrightarrow\! 2) \!=\! 0 \, .
\label{BBGKY_2_rescaled}
\end{align}
The next step of the calculation involves rewriting equations~\eqref{BBGKY_1_rescaled} and~\eqref{BBGKY_2_rescaled} within appropriate angle-action coordinates allowing us to capture in a simple manner the dominant mean Keplerian motion due to the central BH. When considering Keplerian potentials, one has to deal with additional dynamical degeneracies between the orbital frequencies, which should be handled with care, as we will now detail.

\section{Degenerate angle-action coordinates}
\label{sec:AA_coordinates}

In equations~\eqref{BBGKY_1_rescaled} and~\eqref{BBGKY_2_rescaled}, one can note the presence of an advection term ${ \bm{v}_{1} \!\cdot\! \partial / \partial \bm{x}_{1} \!+\! \bm{\mathcal{F}}_{\!10} \!\cdot\! \partial / \partial \bm{v}_{1} }$ associated with the Keplerian motion driven by the central black hole. The next step of the derivation is to introduce the appropriate angle-action coordinates~\citep{Goldstein1950,Born1960,BinneyTremaine2008} to simplify this integrable Keplerian motion. We therefore remap the physical coordinates ${ (\bm{x} , \bm{v}) }$ to the Keplerian angle-action ones ${ (\bm{\theta} , \bm{J}) }$. Along the unperturbed Keplerian orbits, the actions $\bm{J}$ are conserved, while the angles $\bm{\theta}$ are ${2\pi-}$periodic, evolving with the frequency $\bm{\Omega}_{\rm Kep}$ defined as
\begin{equation}
\dot{\bm{\theta}} = \bm{\Omega}_{\rm Kep} (\bm{J}) \equiv \frac{\partial H_{\rm Kep} (\bm{J})}{\partial \bm{J}} \, ,
\label{definition_Omega}
\end{equation}
where ${ H_{\rm Kep} }$ is the Hamiltonian associated with the Keplerian motion due to the black hole.
For ${ 3D }$ spherical potentials, the usual angles and actions~\citep{BinneyTremaine2008} are given by
\begin{equation}
(\bm{J} , \bm{\theta}) = (J_{1} , J_{2} , J_{3} , \theta_{1} , \theta_{2} , \theta_{3}) 
= (J_{r} , L , L_{z} , \theta_{1} , \theta_{2} , \theta_{3}) \, ,
\label{usual_3D_actions}
\end{equation}
where $J_{r}$ is the radial action, $L$ the magnitude of the angular momentum, and $L_{z}$ its projection along the ${z-}$axis. The Keplerian Hamiltonian then becomes ${ H_{\rm Kep} \!=\! H_{\rm Kep} (J_{r} \!+\! L) }$.
Another choice of angle-action coordinates in ${3D}$ is given by the Delaunay variables~\citep{ST99,BinneyTremaine2008} defined as
\begin{equation}
(\bm{J} , \bm{\theta}) = (I , L , L_{z} , w , g , h) \, .
\label{Delaunay_3D_actions}
\end{equation}
In equation~\eqref{Delaunay_3D_actions}, ${ (I \!=\! J_{r} \!+\! L , L , L_{z}) }$ are the three actions of the system, while ${ (w , g, h) }$ are the associated angles. Here, $w$ stands for the orbital phase or mean anomaly, $g$ for the angle from the ascending node to the periapse, and $h$ for the longitude of the ascending node. With these variables, one has ${ H_{\rm Kep} \!=\! H_{\rm Kep} (I) }$, so that the angle $w$ advances at the frequency ${ \dot{w} \!=\! \Omega_{\rm Kep} \!=\! \partial H_{\rm Kep} / \partial I }$, while the angles $g$ and $h$ are constant.
The existence of these additional conserved quantities makes the Keplerian potential dynamically degenerate. This can have some crucial consequences on its long-term behaviour, as we will now detail.

To clarify the upcoming discussions we denote as $d$ the dimension of the considered physical space, for instance ${ d \!=\! 2 }$ for a razor-thin disc. In this space, we consider an integrable potential $\psi$ and an associated angle-action mapping ${ (\bm{x} , \bm{v}) \!\mapsto\! (\bm{\theta} , \bm{J}) }$. A potential is said to be degenerate if there exists ${ \bm{n} \!\in\! \mathbb{Z}^{d} }$ such that
\begin{equation}
\forall \bm{J} \; , \; \bm{n} \!\cdot\! \bm{\Omega} (\bm{J}) = 0 \, ,
\label{degeneracy_condition}
\end{equation}
where it is understood that the vector $\bm{n}$ is independent of $\bm{J}$, so that the degeneracy is global. A given potential may have more than one such degeneracy, and we denote as $k$ the degree of degeneracy of a potential, that is the number of linearly independent vectors $\bm{n}$ satisfying equation~\eqref{degeneracy_condition}. For example, for the angle-action coordinates from equation~\eqref{usual_3D_actions}, the frequencies and degeneracy vectors are given by
\begin{equation}
\bm{\Omega}_{\rm 3D} \!=\! (\Omega_{\rm Kep} , \Omega_{\rm Kep} , 0) \; \Rightarrow \; \bm{n}_{1} \!=\! (1 , -1 , 0) \; \text{and} \; \bm{n}_{2} \!=\! (0 , 0 , 1) \, ,
\label{example_degeneracy_3D}
\end{equation}
so that ${ k \!=\! 2 }$. Using the Delaunay angle-action coordinates from equation~\eqref{Delaunay_3D_actions}, one can similarly write
\begin{equation}
\bm{\Omega}_{\rm Del} \!=\! (\Omega_{\rm Kep} , 0 , 0) \; \Rightarrow \; \bm{n}_{1} \!=\! (0 , 1 , 0) \; \text{and} \; \bm{n}_{2} \!=\! (0 , 0 , 1) \, ,
\label{example_degeneracy_Delaunay}
\end{equation}
which also gives ${ k \!=\! 2}$. The degree of degeneracy of the potential is independent of the chosen angle-action coordinates. The Delaunay variables from equation~\eqref{Delaunay_3D_actions} appear as a simpler choice than the usual ones from equation~\eqref{usual_3D_actions}, because of their simpler degeneracy vectors.

For a given degenerate potential, one can always remap the angle-action coordinates to get simpler degeneracies. Indeed, let us assume that in our initial angle-action coordinates ${ (\bm{\theta} , \bm{J}) }$, we have at our disposal $k$ degeneracy vectors $\bm{n}_{1}$, ... , $\bm{n}_{k}$. Thanks to a linear transformation, we may change coordinates ${ (\bm{\theta} , \bm{J}) \!\mapsto\! (\bm{\theta}' , \bm{J}') }$, so that in the new coordinates the $k$ new degeneracy vectors take the simple form ${ \bm{n}_{i}' \!=\! \bm{e}_{i} }$, where $\bm{e}_{i}$ are the natural basis elements of $\mathbb{Z}^{d}$. Following~\cite{Morbidelli2002}, as the vectors $\bm{n}_{i}$ are assumed to be linearly independent, we may complete this family with ${ d \!-\! k }$ vectors ${ \bm{n}_{k+1} , ... , \bm{n}_{d} \!\in\! \mathbb{Z}^{d} }$ to have a basis over $\mathbb{Q}^{d}$. We then define the transformation matrix $\bm{\mathcal{A}}$ of determinant $1$ as
\begin{equation}
\bm{\mathcal{A}} = \big( \bm{n}_{1} , ... , \bm{n}_{d} \big)^{\rm t} \, / \,  | \big( \bm{n}_{1} , ... , \bm{n}_{d} \big) | \, ,
\label{definition_transformation_matrix}
\end{equation}
and the new angle-action coordinates ${ (\bm{\theta}' , \bm{J}') }$ are defined as
\begin{equation}
\bm{\theta}' = \bm{\mathcal{A}} \!\cdot\! \bm{\theta} \;\;\; ; \;\;\; \bm{J} ' = (\bm{\mathcal{A}}^{\rm t})^{-1} \!\cdot\! \bm{J} \, .
\label{definition_new_AA}
\end{equation}
One can check that ${ (\bm{\theta}' , \bm{J}') }$ are indeed new angle-action coordinates, with $\bm{J}'$ conserved and ${\bm{\theta}' \!\in\! [0 , 2 \pi] }$. Within these new coordinates, the $k$ degeneracy vectors are immediately given by ${ \bm{n}_{i}' \!=\! \bm{e}_{i} }$, that is the intrinsic frequencies satisfy ${ \Omega_{i}' \!=\! 0 }$ for ${ 1 \!\leq\! i \!\leq\! k }$. The degeneracies of the potential got simpler.
In the upcoming calculations, we will always consider such simpler angle-action coordinates, and we introduce the notations
\begin{align}
& \, \bm{\theta}^{\rm s} = (\theta_{1} , ... , \theta_{k}) \;\;\; ; \;\;\; \bm{\theta}^{\rm f} = (\theta_{k+1} , ... \theta_{d}) \, ,  \nonumber
\\
& \, \bm{J}^{\rm s} = (J_{1} , ... , J_{k}) \;\;\; ; \;\;\; \bm{J}^{\rm f} = (J_{k+1} , ... , J_{d}) \, ,    \nonumber
\\
& \, \bR = (\bm{J} , \bm{\theta}^{\rm s}) \, , 
\label{notations_angle_action}
\end{align}
where $\bm{\theta}^{\rm s}$ and $\bm{J}^{\rm s}$ respectively stand for the slow angles and actions, while $\bm{\theta}^{\rm f}$ and $\bm{J}^{\rm f}$ stand for the fast angles and actions. Finally, we introduced $\bR$ as the vector of all the conserved quantities (for a Keplerian potential, this corresponds to a Keplerian elliptical wire). For a degenerate potential, the slow angles are the angles for which the associated frequencies are equal to $0$, while these frequencies are non-zero for the fast angles. Let us finally define the degenerate angle-average with respect to the fast angles as
\begin{equation}
\oF (\bm{J} , \bm{\theta}^{\rm s}) \equiv \!\! \int \!\! \frac{\rd \bm{\theta}^{\rm f}}{(2 \pi)^{d - k}} \, F (\bm{J} , \bm{\theta}^{\rm s} , \bm{\theta}^{\rm f}) \, .
\label{definition_degenerate_angle_average}
\end{equation}

We now use these various properties to rewrite equations~\eqref{BBGKY_1_rescaled} and~\eqref{BBGKY_2_rescaled} using the angle-action coordinates appropriate for the Keplerian motion due to the central BH. In these coordinates, the Keplerian advection term becomes
\begin{equation}
\bm{v}_{1} \!\cdot\! \frac{\partial }{\partial \bm{x}_{1}} \!+\! \bm{\mathcal{F}}_{\!10} \!\cdot\! \frac{\partial }{\partial \bm{v}_{1}} = \bm{\Omega}_{\rm Kep} \!\cdot\! \frac{\partial }{\partial \bm{\theta}} \, .
\label{Keplerian_advection_AA}
\end{equation}
A nice property of the average from equation~\eqref{definition_degenerate_angle_average}, is that the collisionless advection term from equation~\eqref{Keplerian_advection_AA} then naturally vanishes, so that one has
\begin{equation}
\overline{\bm{\Omega}_{\rm Kep} \!\cdot\! \frac{\partial F}{\partial \bm{\theta}}}  = \!\! \int \!\! \frac{\rd \theta_{k + 1}}{2 \pi} ... \frac{\rd \theta_{d}}{2 \pi} \sum_{i = k +1}^{d} \Omega_{\rm Kep}^{i} (\bm{J}) \, \frac{\partial F}{\partial \theta_{i}} = 0 \, .
\label{cancel_Keplerian_average}
\end{equation}
Finally, the mapping ${ (\bm{x} , \bm{v}) \!\mapsto\! (\bm{\theta} , \bm{J}) }$ preserves the infinitesimal volumes so that ${ \rd \Gamma \!=\! \rd \bm{x} \rd \bm{v} \!=\! \rd \bm{\theta} \rd \bm{J} }$. In addition, it also preserves Poisson brackets, so that for two functions ${ G_{1} (\bm{x} , \bm{v}) }$, and ${ G_{2} (\bm{x} , \bm{v}) }$, one has
\begin{equation}
\big[ G_{1} , G_{2} \big] = \frac{\partial G_{1}}{\partial \bm{x}} \!\cdot\! \frac{\partial G_{2}}{\partial \bm{v}} \!-\! \frac{\partial G_{1}}{\partial \bm{v}} \!\cdot\! \frac{\partial G_{2}}{\partial \bm{x}}
 = \frac{\partial G_{1}}{\partial \bm{\theta}} \!\cdot\! \frac{\partial G_{2}}{\partial \bm{J}} \!-\! \frac{\partial G_{1}}{\partial \bm{J}} \!\cdot\! \frac{\partial G_{2}}{\partial \bm{\theta}} \, .
\label{definition_Poisson_bracket}
\end{equation}
In order to shorten the notations, let us now introduce the rescaled self-consistent potential $\Phi$ as
\begin{equation}
\Phi (\bm{x}_{1}) = \!\! \int \!\! \rd \Gamma_{2} \, U_{12} \, F (\Gamma_{2}) \;\;\; ; \;\;\; - \frac{\partial \Phi}{\partial \bm{x}_{1}} = \!\! \int \!\! \rd \Gamma_{2} \, \bm{\mathcal{F}}_{\! 12} \, F (\Gamma_{2}) \, .
\label{definition_Phi_self}
\end{equation}
One can now rewrite equation~\eqref{BBGKY_1_rescaled} within these angle-action coordinates and it takes the form
\begin{align}
& \, \frac{\partial F}{\partial t} \!+\! \bm{\Omega}_{\rm Kep}^{1} \!\cdot\! \frac{\partial F}{\partial \bm{\theta}_{1}} \!+\! \varepsilon \big[ F , \Phi \!+\! \Phi_{\rr} \big] \!+\! \frac{\varepsilon}{N} \!\!\! \int \!\!\! \rd \Gamma_{2} \, \big[ \mathcal{C} (\Gamma_{1} , \Gamma_{2}) , U_{12} \big]_{(1)} \nonumber
\\
& \, \!+\! \frac{\varepsilon}{N} \bigg[ F , \frac{\bm{v}_{1}^{2}}{2} \bigg] \!+\! \varepsilon \bigg[ F , \bm{v}_{1} \!\cdot\! \!\! \int \!\! \rd \Gamma_{2} \, \bm{v}_{2} \, F (\Gamma_{2}) \bigg] \nonumber
\\
& \, \!+\! \frac{\varepsilon}{N} \!\! \int \!\! \rd \Gamma_{2} \, \big[ \mathcal{C} (\Gamma_{1} , \Gamma_{2}) , \bm{v}_{1} \!\cdot\! \bm{v}_{2} \big]_{(1)}= 0 \, ,
\label{BBGKY_1_AA}
\end{align}
where we have written ${ \bm{\Omega}_{\rm Kep}^{1} \!=\! \bm{\Omega}_{\rm Kep} (\bm{J}_{1}) }$ and have introduced the notation
\begin{equation}
\big[ G_{1} (\Gamma_{1} , \Gamma_{2}) , G_{2} (\Gamma_{1} , \Gamma_{2}) \big]_{(1)} = \frac{\partial G_{1}}{\partial \bm{\theta}_{1}} \!\cdot\! \frac{\partial G_{2}}{\partial \bm{J}_{1}} \!-\! \frac{\partial G_{1}}{\partial \bm{J}_{1}} \!\cdot\! \frac{\partial G_{2}}{\partial \bm{\theta}_{1}} \, ,
\label{definition_Poisson_1}
\end{equation}
so that it corresponds to the Poisson bracket with respect to the variables~$1$. In equation~\eqref{BBGKY_1_AA}, the terms of the second and third lines are associated with the additional kinetic terms appearing in the Hamiltonian from equation~\eqref{Hamiltonian_democratic_simpler}. As we will emphasise later on, once averaged over the fast Keplerian dynamics, these terms will be negligible at the order considered here. Similarly, one can straightforwardly rewrite equation~\eqref{BBGKY_2_rescaled} as
\begin{align}
& \, \frac{1}{2} \frac{\partial \mathcal{C}}{\partial t} \!+\! \bm{\Omega}_{\rm Kep}^{1} \!\cdot\! \frac{\partial \mathcal{C}}{\partial \bm{\theta}_{1}} \!+\! \varepsilon \big[ \mathcal{C} (\Gamma_{1} , \Gamma_{2}) , \Phi \!+\! \Phi_{\rr} \big]_{(1)} \!+\! \varepsilon \big[ F (\Gamma_{1}) F(\Gamma_{2}) , U_{12} \big]_{(1)}  \nonumber
\\
& \,  \!+\! \varepsilon \!\! \int \!\! \rd \Gamma_{3} \, \mathcal{C} (\Gamma_{2} , \Gamma_{3}) \, \big[ F (\Gamma_{1}) , U_{13} \big]_{(1)}   \nonumber
\\
& \, \!+\! \varepsilon \big[ F (\Gamma_{1}) , \bm{v}_{1} \!\cdot\! \bm{v}_{2} F (\Gamma_{2}) \big]_{(1)} \!+\! \varepsilon \bigg[ F (\Gamma_{1}) , \bm{v}_{1} \!\cdot\! \!\! \int \!\! \rd \Gamma_{3} \, \bm{v}_{3} \, \mathcal{C} (\Gamma_{2} , \Gamma_{3}) \bigg]_{(1)} \nonumber
\\
& \, \!+\! \varepsilon \bigg[ \mathcal{C} (\Gamma_{1} , \Gamma_{2}) , \bm{v}_{1} \!\cdot\! \!\! \int \!\! \rd \Gamma_{3} \, \bm{v}_{3} \, F (\Gamma_{3}) \bigg]_{(1)} \!\!+\! (1 \!\leftrightarrow\! 2) = 0 \, ,
\label{BBGKY_2_AA}
\end{align}
where the terms from the two last lines are associated with the additional kinetic terms from equation~\eqref{Hamiltonian_democratic_simpler}, and will become negligible once averaged over the fast Keplerian dynamics.
The rewriting from equation~\eqref{BBGKY_1_AA} is particularly enlightening, since one can easily identify in its first line the three relevant timescales of the problem. These are: i) the dynamical timescale ${ T_{\rm Kep} \!=\! 1/\Omega_{\rm Kep} }$ associated with the Keplerian advection term ${ \bm{\Omega}_{\rm Kep}^{1} \!\cdot\! \partial F / \partial \bm{\theta}_{1} }$, ii) the secular collisionless timescale of evolution ${ T_{\rm sec} \!=\! \varepsilon^{-1} T_{\rm Kep} }$ associated with the potential contributions ${ \varepsilon [\Phi \!+\! \Phi_{\rr}] }$, and finally iii) the collisional timescale of relaxation ${ T_{\rm relax} \!=\! N T_{\rm sec} }$, associated with the last term in the first line of equation~\eqref{BBGKY_1_AA}.

\section{Fast averaging the evolution equations}
\label{sec:_Average_equations}

Starting from equations~\eqref{BBGKY_1_AA} and~\eqref{BBGKY_2_AA}, let us carry out an average over the degenerate angles as defined in equation~\eqref{definition_degenerate_angle_average}. We recall that the main virtue of such an averaging is to naturally cancel out any contributions associated with the Keplerian advection term, as observed in equation~\eqref{cancel_Keplerian_average}. We start from equation~\eqref{BBGKY_1_AA} and multiply it by ${ \! \int \! \rd \bm{\theta}^{\rm f}/ (2 \pi)^{d-k} }$. In order to estimate the average of the various crossed terms in equation~\eqref{BBGKY_1_AA}, let us assume that the DF of the system can be expanded as
\begin{equation}
F = \oF \!+\! \epsilon f \;\;\; \text{with} \;\;\;
\begin{cases}
\displaystyle f \!\sim\! O(1) \, ,
\\
\displaystyle \overline{f} = 0 \, ,
\end{cases}
\label{DL_DF}
\end{equation}
where ${ \epsilon \!\ll\! 1 }$ is a small parameter of order ${ 1/N }$. This ansatz is the crucial assumption of the present derivation. Indeed, the BH's domination on the dynamics strongly limits the efficiency of violent relaxation or phase mixing to allow for a rapid dissolution of any dependence on~${\bm{\theta}^{\rm f}}$. Hence it is somewhat arbitrarily assumed here that this condition has been achieved, so that, for our purposes, the system starts in a phased-mixed state.

We now discuss in turn how the various terms appearing in equation~\eqref{BBGKY_1_AA} can be averaged with respect to the fast Keplerian angle.
In the first Poisson bracket of equation~\eqref{BBGKY_1_AA}, one should keep in mind that the self-consistent potential $\Phi$, introduced in equation~\eqref{definition_Phi_self}, should be seen as a functional of $F$. As a consequence, this term takes the form
\begin{align}
\varepsilon \overline{\big[ F , \Phi (F) \!+\! \Phi_{\rr} \big]} & \, = \varepsilon \overline{\big[ \oF \!+\! \epsilon f , \Phi (\oF \!+\! \epsilon f) \!+\! \Phi_{\rr} \big]}   \nonumber
\\
& \, = \varepsilon \overline{\big[ \oF , \Phi (\oF) \!+\! \Phi_{\rr} \big]} \!+\! O (\varepsilon \epsilon)   \nonumber
\\
& \, = (2 \pi)^{d - k} \varepsilon \big[ \oF , \oP (\oF) \!+\! \oP_{\rr} \big] \!+\! O (\varepsilon \epsilon) \, ,
\label{rewrite_Poisson_bracket_epsilon}
\end{align}
where the averaged self-consistent potential $\oP$ was introduced as
\begin{equation}
\oP (\bR_{1}) = \!\! \int \!\! \rd \bR_{2} \, \oF (\bR_{2}) \, \oU_{12} (\bR_{1} , \bR_{2} ) \, .
\label{definition_averaged_Phi}
\end{equation}
In equation~\eqref{definition_averaged_Phi}, for clarity, the notation was shortened for the self-consistent potential as ${ \oP \!=\! \oP (\oF) }$. The (doubly) averaged interaction potential $\oU_{12}$ is defined as
\begin{equation}
\oU_{12} (\bR_{1} , \bR_{2}) = \!\! \int \!\! \frac{\rd \bm{\theta}_{1}^{\rm f}}{(2 \pi)^{d - k}} \frac{\rd \bm{\theta}_{2}^{\rm f}}{(2 \pi)^{d - k}} \, U_{12} (\Gamma_{1} , \Gamma_{2}) \, ,
\label{definition_Ubar12}
\end{equation}
while the angle-averaged potential $\oP_{\rr}$ was also introduced as 
\begin{equation}
\oP_{\rr} (\bR) = \frac{1}{(2 \pi)^{d - k}} \!\! \int \!\! \frac{\rd \bm{\theta}^{\rm f}}{(2 \pi)^{d - k}} \, \Phi_{\rr} (\Gamma) \, ,
\label{definition_average_Phi_a}
\end{equation}
where the prefactor ${ 1/ (2 \pi)^{d - k} }$,  was introduced for convenience.
As emphasised in equation~\eqref{rewrite_Poisson_bracket_epsilon}, one should note that at first order in $\varepsilon$ and zeroth order in $\epsilon$, the self-consistent potential has to be computed while only considering the averaged system's DF $\oF$.

To deal with the second Poisson bracket of equation~\eqref{BBGKY_1_AA}, the same double average as introduced in equation~\eqref{definition_Ubar12} should be performed on $\mathcal{C}$. As we did for equation~\eqref{DL_DF}, it is assumed that the ${2-}$body correlation can be developed as
\begin{equation}
\mathcal{C} = \oC \!+\! \epsilon c \;\;\; \text{with} \;\;\;
\begin{cases}
\displaystyle c \!\sim\! O(1) \, ,
\\
\displaystyle \overline{c} = 0 \, .
\end{cases}
\label{DL_C}
\end{equation} 
At first order in $\varepsilon$ and zeroth order in $\epsilon$, the third term from equation~\eqref{BBGKY_1_AA} can immediately be rewritten as
\begin{equation}
\frac{\varepsilon}{N} \!\! \int \!\! \rd \Gamma_{2} \, \overline{\big[ \mathcal{C} (\Gamma_{1} , \Gamma_{2}) , U_{12} \big]_{(1)}} = \frac{ \varepsilon  (2 \pi)^{d - k} }{N} \!\! \int \!\! \rd \bR_{2} \, \big[ \oC (\bR_{1} , \bR_{2}) , \oU_{12} \big]_{(1)} \, .
\label{rewrite_collision_term_epsilon} \nonumber
\end{equation}
Finally, at first order in $\varepsilon$ and zeroth order in $\epsilon$, the terms from the two last lines of equation~\eqref{BBGKY_1_AA} will involve the quantities
\begin{equation}
\int \!\! \rd \bm{\theta}_{1}^{\rm f} \, \bm{v}_{1} = 0 \;\;\; ; \;\;\; \!\! \int \!\! \rd \bm{\theta}_{1}^{\rm f} \, \frac{\bm{v}_{1}^{2}}{2} \propto H_{\rm Kep} (\bm{J}_{1}^{\rm f}) \, .
\label{vanishing_quantity_BBGKY}
\end{equation}
The first identity comes from the fact that Keplerian orbits are closed, so that the mean displacement over one orbit is zero, while the second identity comes from the virial theorem.
As these terms either vanish or do not depend on the slow coordinates $\bm{\theta}^{\rm s}$ and $\bm{J}^{\rm s}$, they will not contribute to the dynamics at the orders considered here once averaged over the fast angle. Therefore, keeping only terms of order $\varepsilon$ and ${ \varepsilon/N }$, one can finally rewrite equation~\eqref{BBGKY_1_AA} as
\begin{equation}
\!\! \frac{\partial \oF}{\partial t} \!+\! \varepsilon (2 \pi)^{d - k} \big[ \oF \!,\! \oP \!+\! \oP_{\rr} \big] 
 \!+\! \frac{\varepsilon (2 \pi)^{d - k}}{N}  \!\!\! \int \!\!\! \rd \bR_{2} \big[ \oC (\bR_{1} , \bR_{2}) ,\! \oU_{12} \big]_{(1)} \!\!=\! 0 .
\label{rewrite_BBGKY_1}
\end{equation}
In equation~\eqref{rewrite_BBGKY_1}, we note that all the functions appearing in the Poisson brackets only depend on ${ \bR_{1} \!=\! (\bm{J}_{1} , \bm{\theta}_{1}^{\rm s}) }$. As a consequence, the Poisson brackets defined in equation~\eqref{definition_Poisson_bracket} take the shortened form
\begin{equation}
\big[ G_{1} (\bR) , G_{2} (\bR) \big] = \frac{\partial G_{1}}{\partial \bm{\theta}^{\rm s}} \!\cdot\! \frac{\partial G_{2}}{\partial \bm{J}^{\rm s}} \!-\! \frac{\partial G_{1}}{\partial \bm{J}^{\rm s}} \!\cdot\! \frac{\partial G_{2}}{\partial \bm{\theta}^{\rm s}} \, ,
\label{shorter_Poisson}
\end{equation}
so that only derivatives with respect to the slow coordinates appear. Let us finally introduce the rescaled time $\tau$ as
\begin{equation}
\tau = (2 \pi)^{d - k} \varepsilon t \, ,
\label{definition_tau}
\end{equation}
so that equation~\eqref{rewrite_BBGKY_1} becomes
\begin{equation}
\frac{\partial \oF}{\partial \tau} \!+\! \big[ \oF , \oP \!+\! \oP_{\rr} \big] \!+\! \frac{1}{N} \!\! \int \!\! \rd \bR_{2} \, \big[ \oC (\bR_{1} , \bR_{2}) , \oU_{12} \big]_{(1)} = 0 \, .
\label{BBGKY_1_final}
\end{equation}

One may use a similar angle-averaging procedure for the second equation of the BBGKY hierarchy. Indeed, multiplying equation~\eqref{BBGKY_2_AA} by ${ \! \int \! \rd \bm{\theta}_{1}^{\rm f} \rd \bm{\theta}_{2}^{\rm f} / (2 \pi)^{2 (d-k)} }$, relying on the developments from equations~\eqref{DL_DF} and~\eqref{DL_C}, and keeping only terms of order $\varepsilon$, equation~\eqref{BBGKY_2_AA} can finally be rewriten as
\begin{align}
& \,\frac{1}{2} \frac{\partial \oC}{\partial \tau} \!+\! \big[ \oC (\bR_{1} , \bR_{2}) , \oP (\bR_{1})\! \!+\! \oP_{\rr} (\bR_{1}) \big]_{(1)} \!\!+\! \frac{\big[ \oF (\bR_{1}) \oF (\bR_{2}) , \oU_{12} \big]_{(1)} }{(2 \pi)^{d - k}}   \nonumber
\\
& \, \!+\! \!\! \int \!\! \rd \bR_{3} \, \oC (\bR_{2} , \bR_{3}) \, \big[ \oF (\bR_{1}) , \oU_{13} \big]_{(1)} \!+\! (1 \!\leftrightarrow\! 2) = 0 \, ,
\label{BBGKY_2_final}
\end{align}
where one can note that all the additional kinetic terms of the two last lines of equation~\eqref{BBGKY_2_AA} vanish at the considered order, when averaged over the fast Keplerian angle.

Equations~\eqref{BBGKY_1_final} and~\eqref{BBGKY_2_final} are the main results of this section.
They describe the coupled evolutions of the system's averaged DF, $\overline F$ and ${2-}$body correlation $\oC$.
A rewriting of the same pair of equations has recently been derived by~\cite{ST1,ST2} using Gilbert's method.
At this stage, one could investigate at least four different dynamical regimes of evolution for the system:
\begin{itemize}
\item[I] Considering equation~\eqref{BBGKY_1_final}, the Keplerian wires could initially be far from a quasi-stationary equilibrium, so that ${ \big[ \oF , \oP \!+\! \oP_{\rr} \big] \!\neq\! 0 }$. One then expects that this out-of-equilibrium system will undergo a phase of violent relaxation~\citep{LyndenBell1967}, allowing it to rapidly reach a quasi-stationary equilibrium. We do not investigate this process here, but still rely on the assumption that the collisionless violent relaxation of the wires' DF can be sufficiently efficient for the system to briefly reach a quasi-stationary stable state, which will then be followed by a much slower secular evolution, either collisionless or collisional.
\\
\item[II] For a given DF of stationary wires, one could also investigate the possible existence of collisionless dynamical instabilities associated with the collisionless part of the evolution equation~\eqref{BBGKY_1_final}, namely ${ \partial \oF / \partial \tau \!+\! \big[ \oF , \oP \!+\! \oP_{\rr} \big] \!=\! 0 }$. Such instabilities are not considered in the present paper, and we will assume, as will be emphasised in the upcoming derivations, that throughout its evolution the system always remains dynamically stable with respect to the collisionless dynamics. See for instance~\cite{Tremaine2005,PolyachenkoShukhman2007,JalaliTremaine2012} for examples of stability investigations in this context.
\\
\item[III] Once it is assumed that the system has reached a quasi-stationary stable state, one can study the secular evolution of this system along quasi-stationary equilibria. Such a long-term evolution can first be induced by the presence of external stochastic perturbations. To capture such a secular collisionless evolution, one should neglect contributions from the collisional term in ${ 1/N }$ in equation~\eqref{BBGKY_1_final}, and look for the long-term effects of stochastic perturbations. The formalism appropriate for such a secular collisionless stochastic forcing is similar to the one presented in~\cite{FouvryPichonPrunet2015} in the context of stellar discs. The specification of such externally forced secular dynamics to the case of dynamically degenerate systems is postponed to a future work.
\\
\item[IV] During its secular evolution along quasi-stationary equilibria, the dynamics of an isolated system can also be driven by finite${-N}$ fluctuations. This amounts to neglecting the effects due to any external stochastic perturbations, and considering the contributions associated with the collisional term in ${1/N}$ in equation~\eqref{BBGKY_1_final}. This requires to solve simultaneously the system of two coupled evolution equations~\eqref{BBGKY_1_final} and~\eqref{BBGKY_2_final}. This approach is presented in section~\ref{sec:secular_collisional}, where the analogs of the (bare, that is without collective effects) Landau equation and (dressed, that is with collective effects) Balescu-Lenard equation are derived in the context of degenerate dynamical systems, such as galactic nuclei. As will be emphasised later on, these diffusion equations, sourced by finite${-N}$ fluctuations capture the known mechanism of resonant relaxation~\citep{RauchTremaine1996}. See~\cite{BarOrAlexander2014} for a similar study of the effect of the finite${-N}$ stochastic internal forcing via the so-called ${\eta-}$formalism.
\end{itemize}

We note that one could also consider the secular evolution of a non-axisymmetric set of eccentric orbits orbiting a black hole as an unperturbed collisionless equilibrium (corresponding to the expected configuration of the galactic centre of M31~\citep{Tremaine1995}).
The derivation of the associated Balescu-Lenard equation for such a configuration would first involve identifying new angle-action variables for the non-axisymmetric configuration  so as to satisfy II, and then extend the formalism accordingly.
This will not be explored any further in this paper.
Regarding item II, we expect that, depending on the relative mass of the considered cluster, there is a regime where the self-induced orbital precession is significant, but the self-gravity of the wires is not strong enough to induce a collisionless instability. In this regime, accounting for the polarisation of the orbits becomes 
important in item III and IV. This motivates the rest of the paper.

\section{The degenerate Balescu-Lenard equation}
\label{sec:secular_collisional}

We now show how to obtain the closed kinetic equations -- the degenerate Balescu-Lenard and Landau equations -- when considering the ${1/N}$ collisional contribution present in the evolution equation~\eqref{BBGKY_1_final}. It will be assumed that the system is isolated so that it experiences no external perturbations. Our aim is to obtain a closed kinetic equation involving $\oF$ only. To do so, we rely on the adiabatic approximation (or Bogoliubov's ansatz) that the system secularly relaxes through a series of collisionless equilibria. In this context, collisionless equilibria are stationary (and stable) steady states of the collisionless advection component of equation~\eqref{BBGKY_1_final}. Therefore, it is assumed that throughout the secular evolution, one has
\begin{equation}
\forall \tau \, , \, \big[ \oF (\tau) , \oP (\tau) \!+\! \oP_{\rr} (\tau) \big] = 0 \, .
\label{adiabatic_approximation}
\end{equation}
As already highlighted, it is expected that such collisionless equilibria are rapidly reached by the system (on a few $T_{\rm sec}$), through an out-of-equilibrium mechanism related to violent relaxation. In addition, the symmetry of the system is expected to be such that the collisionless equilibria are of the form
\begin{equation}
\oF (\bm{J} , \bm{\theta}^{\rm s} , \tau) = \oF (\bm{J} , \tau) \, ,
\label{shape_equilibria}
\end{equation}
so that, during its secular evolution, the system's averaged DF does not have any slow angle dependence. Notice however that, despite the hypothesis from equation~\eqref{shape_equilibria}, the averaged autocorrelation $\oC$ evolving according to equation~\eqref{BBGKY_2_final} still depends on the two slow angles $\bm{\theta}_{1}^{\rm s}$ and $\bm{\theta}_{2}^{\rm s}$. We also assume that the symmetry of the system is such that 
\begin{equation}
\oF = \oF (\bm{J}) \;\; \Rightarrow \;\; \oP = \oP (\bm{J}) \quad \text{and} \quad \oP_{\rr} = \oP_{\rr} (\bm{J}) \, .
\label{assumption_self_consistent_Phi}
\end{equation}
As we will see later on in sections~\ref{sec:case_disc} and~\ref{sec:case_3D}, such symmetry is satisfied for instance for razor-thin axisymmetric discs and {3D} spherical clusters (see also Appendix~\ref{sec:relativistic_precessions} for the expression of the relativistic precession frequencies). Given equations~\eqref{shape_equilibria} and~\eqref{assumption_self_consistent_Phi}, the equilibrium condition from equation~\eqref{adiabatic_approximation} is immediately satisfied. We introduce the precession frequencies ${ \bm{\Omega}^{\rm s} }$ as
\begin{equation}
\bm{\Omega}^{\rm s} (\bm{J}) = \frac{\partial [ \oP \!+\! \oP_{\rr} ]}{\partial \bm{J}^{\rm s}} \, .
\label{definition_precession_frequencies}
\end{equation}
These frequencies correspond to the precession frequencies of the slow angles due to the joint contributions from the system's self-consistent potential and the relativistic corrections. Notice that they do not involve the Keplerian frequencies from equation~\eqref{definition_Omega} anymore and hence are not degenerate a priori. With them, one can for example easily rewrite the collisionless precession advection term from equation~\eqref{BBGKY_2_final} as
\begin{equation}
\big[ \oC (\bR_{1} , \bR_{2}) , \oP (\bR_{1}) \!+\! \oP_{\rr} (\bR_{1}) \big]_{(1)} = \bm{\Omega}_{1}^{\rm s} \!\cdot\! \frac{\partial \oC (\bR_{1} , \bR_{2})}{\partial \bm{\theta}_{1}^{\rm s}} \, .
\label{advection_C_precession}
\end{equation}
where the precession frequencies ${ \bm{\Omega}_{1}^{\rm s} \!=\! \bm{\Omega}^{\rm s} (\bm{J}_{1})}$ associated with the slow angles $\bm{\theta}_{1}^{\rm s}$ come into play.

The two coupled evolution equations~\eqref{BBGKY_1_final} and~\eqref{BBGKY_2_final} are now quasi-identical to the traditional coupled BBGKY equations considered in~\cite{Heyvaerts2010} to derive the inhomogeneous Balescu-Lenard equation for non-degenerate inhomogeneous systems. Various methods have been proposed in the literature to derive the closed kinetic equation satisfied by $\oF$. \cite{Heyvaerts2010} proposed a direct resolution of the BBGKY equations, based on Bogoliubov's ansatz.~\cite{Chavanis2012} considered a rewriting of equations~\eqref{BBGKY_1_final} and~\eqref{BBGKY_2_final} using the Klimontovich equation~\citep{Klimontovich1967}, and relied on a quasi-linear approximation. Finally, in the limit where collective effects are not accounted for,~\cite{FouvryChavanisPichon2016} recently presented a new derivation of the relevant kinetic equation based on functional integrals.

In the present paper, the derivation proposed by~\cite{Heyvaerts2010} will be followed, by directly solving the two first averaged BBGKY equations~\eqref{BBGKY_1_final} and~\eqref{BBGKY_2_final}. The basic idea of this approach is to solve equation~\eqref{BBGKY_2_final}, so as to obtain the system's autocorrelation $\oC$ as a functional of the system's ${1-}$body DF $\oF$. Injecting this expression in equation~\eqref{BBGKY_1_final} yields finally a closed kinetic equation quadratic in $\oF$.
The detailed calculations required to derive the inhomogeneous degenerate Balescu-Lenard equation are presented in Appendix~\ref{sec:secular_collisional_derivation}.

\subsection{The one component Balescu-Lenard equation}

\begin{figure}[!htbp]
\begin{center}
\epsfig{file=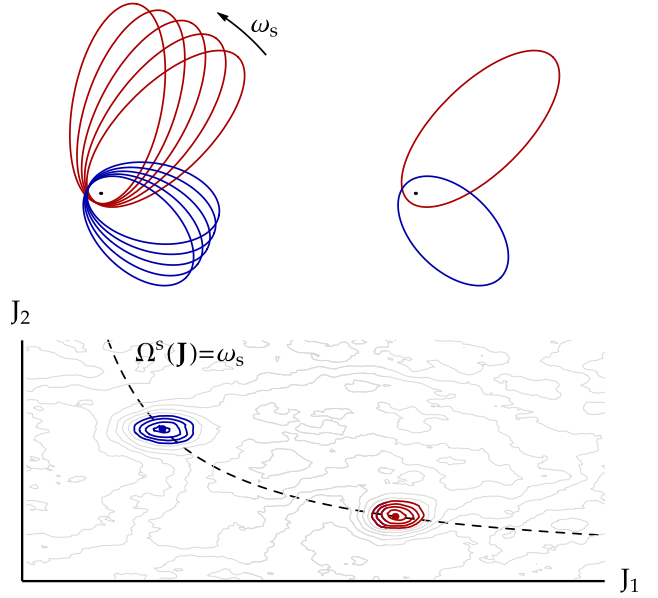,angle=-00,width=0.45\textwidth}
\caption{\small{Illustration of the resonance condition appearing in the degenerate inhomogeneous Balescu-Lenard equation~\eqref{BL_Kepler}. \textbf{Top-left}: a set of two resonant orbits precessing at the same frequency $\omega_{\rm s}$. \textbf{Top-right}: in the rotating frame at frequency $\omega_{\rm s}$ in which the two orbits are in resonance. \textbf{Bottom}: fluctuations of the system's DF in action space caused by finite${-N}$ effects and showing overdensities for the blue and red orbits. The dashed line correspond to the critical resonant line in action space along which the resonance condition ${ \Omega^{\rm s} (\bm{J}) \!=\! \omega_{\rm s} }$ is satisfied. The two set of orbits satisfy a resonance condition for their precession frequencies, and uncorrelated sequences of such interactions lead to a secular diffusion of the system's orbital structure following equation~\eqref{BL_Kepler}. Such resonances are non local in the sense that the resonant orbits need not be close in action space nor in position space. As emphasised in section~\ref{sec:case_disc} for axisymmetric razor-thin discs, symmetry enforces ${ m_{1}^{\rm s} \!=\! m_{2}^{\rm s} }$, so that the two orbits are caught in the same resonance.
}}
\label{figIllustrationBL}
\end{center}
\end{figure}

In its explicitly conservative form, the degenerate inhomogeneous Balescu-Lenard equation reads
\begin{align}
\frac{\partial \oF}{\partial \tau} = & \, \frac{\pi (2 \pi)^{2k-d}}{N} \frac{\partial }{\partial \bm{J}_{1}^{\rm s}} \!\cdot\! \bigg[ \sum_{\bm{m}_{1}^{\rm s} , \bm{m}_{2}^{\rm s}} \bm{m}_{1}^{\rm s} \!\! \int \!\! \rd \bm{J}_{2} \, \frac{\delta_{\rm D} (\bm{m}_{1}^{\rm s} \!\cdot\! \bm{\Omega}_{1}^{\rm s} \!-\! \bm{m}_{2}^{\rm s} \!\cdot\! \bm{\Omega}_{2}^{\rm s})}{| \mathcal{D}_{\bm{m}_{1}^{\rm s} , \bm{m}_{2}^{\rm s}} (\bm{J}_{1} , \bm{J}_{2} , \bm{m}_{1}^{\rm s} \!\cdot\! \bm{\Omega}_{1}^{\rm s}) |^{2}}   \nonumber
\\
& \, \times \bigg( \bm{m}_{1}^{\rm s} \!\cdot\! \frac{\partial }{\partial \bm{J}_{1}^{\rm s}} \!-\! \bm{m}_{2}^{\rm s} \!\cdot\! \frac{\partial }{\partial \bm{J}_{2}^{\rm s}} \bigg) \, \oF (\bm{J}_{1}) \, \oF (\bm{J}_{2}) \bigg] \, .
\label{BL_Kepler}
\end{align}
In equation~\eqref{BL_Kepler}, we recall that $d$ is the dimension of the physical space and $k$ the number of degeneracies of the underlying zeroth-order potential. The r.h.s. of equation~\eqref{BL_Kepler} is the degenerate inhomogeneous Balescu-Lenard collision operator, which describes the secular diffusion induced by dressed finite${-N}$ fluctuations. It describes the distortion of Keplerian orbits as their actions diffuse through their self-interaction. As expected, it vanishes in the large $N$ limit.
Notice the presence of the resonance condition operating on their precession frequencies encapsulated by the Dirac delta ${ \delta_{\rm D} (\bm{m}_{1}^{\rm s} \!\cdot\! \bm{\Omega}_{1}^{\rm s} \!-\! \bm{m}_{2}^{\rm s} \!\cdot\! \bm{\Omega}_{2}^{\rm s}) }$ (using the shortened notation ${ \bm{\Omega}_{i}^{\rm s} \!=\! \bm{\Omega}^{\rm s} (\bm{J}_{i}) }$), where $\bm{m}_{1}^{\rm s}$, $\bm{m}_{2}^{\rm s} \!\in\! \mathbb{Z}^{k}$ are integer vectors. 
In fact, equation~\eqref{BL_Kepler} shows that the diffusion occurs along preferred discrete directions labelled by the resonance vectors $\bm{m}_{1}^{\rm s}$.
The integration over the dummy variable $\bm{J}_{2}$ scans action space for regions where the resonance condition is satisfied, and such resonant (possibly distant) encounters between orbits are the drivers of the collisional evolution. 
The resonance condition is illustrated in figure~\ref{figIllustrationBL}.
Notice also that equation~\eqref{BL_Kepler} involves the antisymmetric operator, ${ \bm{m}_{1}^{\rm s} \!\cdot\! \partial / \partial \bm{J}_{1}^{\rm s} \!-\! \bm{m}_{2}^{\rm s} \!\cdot\! \partial / \partial \bm{J}_{2}^{\rm s} }$, which when applied to $ \oF (\bm{J}_{1}) \, \oF (\bm{J}_{2})$ weighs the relative number of pairwise resonant orbits caught in this resonant configuration.  The quantities ${ 1/\mathcal{D}_{\bm{m}_{1}^{\rm s} , \bm{m}_{2}^{\rm s}} (\bm{J}_{1} , \bm{J}_{2} , \omega) }$ are the so-called dressed susceptibility coefficients: each distribution entering the r.h.s. of equation~\eqref{BL_Kepler} is boosted by this susceptibility.
These dressed coefficients include the effects of the gravitational wake induced by each wire, represented by the last term of equation~\eqref{BBGKY_2_final}; in constrast, bare susceptibility coefficients (introduced later) are obtained without taking this self-gravity into account.
In order to solve Poisson's non-local equation relating the DF's perturbations and the induced potential perturbations, Kalnajs' matrix method~\citep{Kalnajs1976II} can be
used to implement a biorthonormal basis of potentials and densities $\psi^{(p)}$ and $\rho^{(p)}$ such that
\begin{equation}
\psi^{(p)} (\bm{x}) \!=\! \!\! \int \!\!\! \rd \bm{x} ' \, \rho^{(p)} (\bm{x}') \, U (| \bm{x} \!-\! \bm{x}' |) \, ; \, \!\! \int \!\!\! \rd \bm{x} \, \psi^{(p)} (\bm{x}) \, \rho^{(q) *} (\bm{x}) \!=\! - \delta_{p}^{q} \, ,
\label{definition_basis_Kalnajs}
\end{equation}
where $U$ stands for the rescaled interaction potential from equation~\eqref{rescaling_U}. The dressed susceptibility coefficients appearing in equation~\eqref{BL_Kepler} are then given by
\begin{equation}
\frac{1}{\mathcal{D}_{\bm{m}_{1}^{\rm s} , \bm{m}_{2}^{\rm s}} (\bm{J}_{1} , \bm{J}_{2} , \omega)} = \sum_{p , q} \opsi_{\bm{m}_{1}^{\rm s}}^{(p)} (\bm{J}_{1}) \, \big[ \mathbf{I} \!-\! \widehat{\mathbf{M}} (\omega) \big]_{pq}^{-1} \, \opsi_{\bm{m}_{2}^{\rm s}}^{(q) *} (\bm{J}_{2}) \, ,
\label{definition_1/D}
\end{equation}
where $\mathbf{I}$ is the identity matrix, and $\widehat{\mathbf{M}}$ is the system's averaged response matrix defined as
\begin{equation}
\widehat{\mathbf{M}}_{pq} (\omega) = (2 \pi)^{k} \sum_{\bm{m}^{\rm s}} \!\! \int \!\! \rd \bm{J} \, \frac{\bm{m}^{\rm s} \!\cdot\! \partial \oF / \partial \bm{J}^{\rm s}}{\omega \!-\! \bm{m}^{\rm s} \!\cdot\! \bm{\Omega}^{\rm s}} \, \opsi^{(p) *}_{\bm{m}^{\rm s}} (\bm{J}) \, \opsi^{(q)}_{\bm{m}^{\rm s}} (\bm{J}) \, .
\label{Fourier_M}
\end{equation}
In equation~\eqref{Fourier_M}, the averaged basis elements $\opsi^{(p)}$ were defined following equation~\eqref{definition_degenerate_angle_average}. Their Fourier transform with respect to the slow angles was also defined using the convention
\begin{equation}
\opsi^{(p)} \!(\bR) \!=\! \!\! \sum_{\bm{m}^{\rm s}} \opsi^{(p)}_{\bm{m}^{\rm s}} \!(\bm{J}) \, \re^{\ri \bm{m}^{\rm s} \cdot \bm{\theta}^{\rm s}} \; ; \; \opsi^{(p)}_{\bm{m}^{\rm s}} \!(\bm{J}) \!=\! \!\!\! \int \!\!\! \frac{\rd \bm{\theta}^{\rm s}}{(2 \pi)^{k}} \, \re^{- \ri \bm{m}^{\rm s} \cdot \bm{\theta}^{\rm s}} \, \opsi^{(p)} \!(\bR) \, .
\label{definition_Fourier_slow_angles}
\end{equation}
The susceptibility coefficients from equation~\eqref{definition_1/D} quantify the polarisation cloud around each orbit, which triggers sequences of transient wakes~\citep{JulianToomre1966,Toomre1981}. In the secular timeframe, these are assumed to be instantaneous, via the so-called Bogoliubov's ansatz, as shown in Appendix~\ref{sec:solving_autocorrelation}.

One can straightforwardly rewrite the Balescu-Lenard equation~\eqref{BL_Kepler} as an anisotropic non-linear diffusion equation, by introducing the appropriate drift and diffusion coefficients. Equation~\eqref{BL_Kepler} then reads
\begin{equation}
\frac{\partial \oF}{\partial \tau} = \frac{\partial }{\partial \bm{J}_{1}^{\rm s}} \!\cdot\! \bigg[ \sum_{\bm{m}_{1}^{\rm s}} \bm{m}_{1}^{\rm s} \bigg( A_{\bm{m}_{1}^{\rm s}} (\bm{J}_{1}) \oF (\bm{J}_{1}) + D_{\bm{m}_{1}^{\rm s}} (\bm{J}_{1}) \, \bm{m}_{1}^{\rm s} \!\cdot\! \frac{\partial \oF}{\partial \bm{J}_{1}^{\rm s}} \bigg) \bigg] \, ,
\label{rewrite_BL_Kepler}
\end{equation}
where ${ A_{\bm{m}_{1}^{\rm s}} (\bm{J}_{1}) }$ and ${ D_{\bm{m}_{1}^{\rm s}} (\bm{J}_{1}) }$ are respectively the drift and diffusion coefficients associated with a given resonance $\bm{m}_{1}^{\rm s}$.
The secular dependence of these coefficients with the system's averaged DF, $\oF$, is not written out explicitly to simplify the notations but is a central feature of the present formalism. In equation~\eqref{rewrite_BL_Kepler}, the drift coefficients ${ A_{\bm{m}_{1}^{\rm s}} (\bm{J}_{1}) }$ and diffusion coefficients ${ D_{\bm{m}_{1}^{\rm s}} (\bm{J}_{1}) }$ are given by
\begin{align}
& \, A_{\bm{m}_{1}^{\rm s}} (\bm{J}_{1}\!) \!=\! - \frac{\pi (2 \pi)^{2k -d}}{N} \!\sum_{\bm{m}_{2}^{\rm s}} \!\! \int \!\!\! \rd \bm{J}_{2} \, \frac{\delta_{\rm D} (\bm{m}_{1}^{\rm s} \!\cdot\! \bm{\Omega}_{1}^{\rm s} \!-\! \bm{m}_{2}^{\rm s} \!\cdot\! \bm{\Omega}_{2}^{\rm s})}{|\mathcal{D}_{\bm{m}_{1}^{\rm s} , \bm{m}_{2}^{\rm s}} (\bm{J}_{1} ,\! \bm{J}_{2} ,\! \bm{m}_{1}^{\rm s} \!\cdot\! \bm{\Omega}_{1}^{\rm s}\!)|^{2}} \, \bm{m}_{2}^{\rm s} \!\cdot\! \frac{\partial \oF}{\partial \bm{J}_{2}^{\rm s}} \, ,  \nonumber
\\
& \, D_{\bm{m}_{1}^{\rm s}} (\bm{J}_{1}\!) \!=\! \frac{\pi (2 \pi)^{2k - d}}{N} \!\!\sum_{\bm{m}_{2}^{\rm s}} \!\! \int \!\!\!  \rd \bm{J}_{2} \frac{\delta_{\rm D} (\bm{m}_{1}^{\rm s} \!\cdot\! \bm{\Omega}_{1}^{\rm s} \!-\! \bm{m}_{2}^{\rm s} \!\cdot\! \bm{\Omega}_{2}^{\rm s})}{|\mathcal{D}_{\bm{m}_{1}^{\rm s} , \bm{m}_{2}^{\rm s}} \!(\bm{J}_{1} ,\! \bm{J}_{2} ,\! \bm{m}_{1}^{\rm s} \!\cdot\! \bm{\Omega}_{1}^{\rm s} \! )|^{2}} \oF (\bm{J}_{2}).
\label{drift_diff_Kepler}
\end{align}

When collective effects are not accounted for (that is when the last term of equation~\eqref{BBGKY_2_final} is neglected), the degenerate Balescu-Lenard equation~\eqref{BL_Kepler} becomes the degenerate Landau equation (see~\cite{PolyachenkoShukhman1982,Chavanis2013} for the non-degenerate case), which reads
\begin{align}
\frac{\partial \oF}{\partial \tau} & \,= \frac{\pi (2 \pi)^{2k - d}}{N} \frac{\partial }{\partial \bm{J}_{1}^{\rm s}} \!\cdot\! \bigg[ \sum_{\bm{m}_{1}^{\rm s} , \bm{m}_{2}^{\rm s}} \bm{m}_{1}^{\rm s} \!\! \int \!\! \rd \bm{J}_{2} \, \delta_{\rm D} (\bm{m}_{1}^{\rm s} \!\cdot\! \bm{\Omega}_{1}^{\rm s} \!-\! \bm{m}_{2}^{\rm s} \!\cdot\! \bm{\Omega}_{2}^{\rm s})    \nonumber
\\
& \, \times \big| A_{\bm{m}_{1}^{\rm s} , \bm{m}_{2}^{\rm s}} (\bm{J}_{1} , \bm{J}_{2}) \big|^{2} \, \bigg( \bm{m}_{1}^{\rm s} \!\cdot\! \frac{\partial }{\partial \bm{J}_{1}^{\rm s}} \!-\! \bm{m}_{2}^{\rm s} \!\cdot\! \frac{\partial}{\partial \bm{J}_{2}^{\rm s}} \bigg) \, \oF (\bm{J}_{1}) \, \oF (\bm{J}_{2}) \bigg]  .
\label{Landau_Kepler}
\end{align}
Notice that this is just the previous Balescu-Lenard equation~\eqref{BL_Kepler} with the dressed ${ 1/\mathcal{D}_{\bm{m}_{1}^{\rm s} , \bm{m}_{2}^{\rm s}} (\bm{J}_{1} , \bm{J}_{2} , \omega) }$ replaced by the bare susceptibility coefficients ${ A_{\bm{m}_{1}^{\rm s} , \bm{m}_{2}^{\rm s}} (\bm{J}_{1} , \bm{J}_{2}) }$. The latter are related to the (partial) Fourier transform of the interaction potential~\citep{LyndenBell1994,Pichon1994,Chavanis2013} and read
\begin{equation}
A_{\bm{m}_{1}^{\rm s} , \bm{m}_{2}^{\rm s}} (\bm{J}_{1} , \bm{J}_{2}) = \!\! \int \!\! \frac{\rd \bm{\theta}_{1}^{\rm s}}{(2 \pi)^{k}} \frac{\rd \bm{\theta}_{2}^{\rm s}}{(2 \pi)^{k}} \, \oU_{12} (\bR_{1} , \bR_{2}) \, \re^{- \ri (\bm{m}_{1}^{\rm s} \cdot \bm{\theta}_{1}^{\rm s} - \bm{m}_{2}^{\rm s} \cdot \bm{\theta}_{2}^{\rm s})} \, ,
\label{definition_bare_A}
\end{equation}
so that the averaged interaction potential $\oU_{12}$ from equation~\eqref{definition_Ubar12} can be decomposed as
\begin{equation}
\oU_{12} (\bR_{1} , \bR_{2}) = \sum_{\bm{m}_{1}^{\rm s} , \bm{m}_{2}^{\rm s}} A_{\bm{m}_{1}^{\rm s} , \bm{m}_{2}^{\rm s}} (\bm{J}_{1} , \bm{J}_{2})  \, \re^{\ri (\bm{m}_{1}^{\rm s} \cdot \bm{\theta}_{1}^{\rm s} - \bm{m}_{2}^{\rm s} \cdot \bm{\theta}_{2}^{\rm s})} \, .
\label{decomposition_bar_U12}
\end{equation}
One should note that the kinetic equations~\eqref{BL_Kepler} and~\eqref{Landau_Kepler}, while defined on the full action space ${ \bm{J} \!=\! (\bm{J}^{\rm s} , \bm{J}^{\rm f}) }$, do not allow for changes in the fast actions $\bm{J}^{\rm f}$. Indeed, if one defines the marginal DF, $P_{\oF}$, as ${ P_{\oF} = \! \int \! \rd \bm{J}^{\rm s} \oF (\bm{J}) }$, equations~\eqref{BL_Kepler} and~\eqref{Landau_Kepler} immediately give
\begin{equation}
\frac{\partial P_{\oF}}{\partial \tau} = 0 \, ,
\label{marginal_diff}
\end{equation}
so that the collisional secular diffusion occurs only in the directions ${ \bm{J}^{\rm f} \!=\! \text{cst.} }$

\subsection{Multiple components black hole environment}
\label{sec:multiplecomponents}

It is of prime importance to follow the joint long-term evolution of multiple types of stars or black holes orbiting a central supermassive black hole, as it will allow astronomers to capture their relative segregation, when the lighter black holes sink in towards the more massive one. In turn, this could allow us to predict the expected rate of mergers and accretion events.

As already emphasised in~\cite{Heyvaerts2010,Chavanis2012}, the Balescu-Lenard equation can also be written for a system involving multiple components (corresponding to say, a spectrum of stars and low mass black holes or debris of different masses orbiting the central object). The different components will be indexed by the letters ${``\ra"}$ and ${``\rb"}$. The particles of the component ${``\ra"}$ have a mass $\mu_{\ra}$ and follow the DF $F^{\ra}$. As briefly detailed in Appendix~\ref{sec:multicase} (which gives the details of all normalisations), the evolution of each DF is given by
\begin{align}
\frac{\partial \oFa}{\partial \tau} = & \, \pi (2 \pi)^{2k - d} \frac{\partial }{\partial \bm{J}_{1}^{\rm s}} \!\cdot\! \bigg[ \sum_{\bm{m}_{1}^{\rm s} , \bm{m}_{2}^{\rm s}} \! \bm{m}_{1}^{\rm s} \!\! \int \!\! \rd \bm{J}_{2} \frac{\delta_{\rm D} (\bm{m}_{1}^{\rm s} \!\cdot\! \bm{\Omega}_{1}^{\rm s} \!-\! \bm{m}_{2}^{\rm s} \!\cdot\! \bm{\Omega}_{2}^{\rm s})}{|\mathcal{D}_{\bm{m}_{1}^{\rm s} , \bm{m}_{2}^{\rm s}} (\bm{J}_{1} , \bm{J}_{2} , \bm{m}_{1}^{\rm s} \!\cdot\! \bm{\Omega}_{1}^{\rm s})|^{2}}   \nonumber
\\
& \, \times  \sum_{\rb} \bigg\{ \eta_{\rb} \bm{m}_{1}^{\rm s} \!\cdot\! \frac{\partial \oFa}{\partial \bm{J}_{1}^{\rm s}} \oFb (\bm{J}_{2}) \!-\! \eta_{\ra} \oFa (\bm{J}_{1}) \, \bm{m}_{2}^{\rm s} \!\cdot\! \frac{\partial \oFb}{\partial \bm{J}_{2}^{\rm s}} \bigg\} \bigg] \, ,
\label{LB_multi}
\end{align}
where the dimensionless relative mass ${ \eta_{\ra} \!=\! \mu_{\ra} / M_{\star} }$ was introduced, and where ${ M_{\star} \!=\! \sum_{\ra} M_{\star}^{\ra} }$ is the total active mass of the system. In the multi-component case, the dressed susceptibility coefficients are still given by equation~\eqref{definition_1/D}. However, as expected, the response matrix now encompasses all the active components of the system which polarise so that
\begin{equation}
\widehat{\mathbf{M}}_{pq} (\omega) = (2 \pi)^{k} \sum_{\bm{m}^{\rm s}} \!\! \int \!\! \rd \bm{J} \, \frac{\bm{m}^{\rm s} \!\cdot\! \partial (\sum_{\rb} \oFb)/\partial \bm{J}^{\rm s}}{\omega \!-\! \bm{m}^{\rm s} \!\cdot\! \bm{\Omega}^{\rm s}} \, \opsi^{(p) *}_{\bm{m}^{\rm s}} (\bm{J}) \, \opsi^{(q)}_{\bm{m}^{\rm s}} (\bm{J}) \, .  \nonumber
\label{Fourier_M_multi}
\end{equation}
In the limit where only one mass is considered, one has ${ \eta_{\ra} \!=\! 1 / N_{\ra} }$, and the single mass Balescu-Lenard equation~\eqref{BL_Kepler} is recovered. 
Equation~\eqref{LB_multi} describes the evolution of the ${``\ra"}$ population, and differs from equation~\eqref{BL_Kepler} via the weight $\eta_{\ra}$, and the sum over ${ ``\rb" }$ weighted by $\eta_{\rb}$.
As in equation~\eqref{rewrite_BL_Kepler}, one can introduce drift and diffusion coefficients to rewrite equation~\eqref{LB_multi} as
\begin{equation}
\frac{\partial \oFa}{\partial \tau} \!=\! \frac{\partial }{\partial \bm{J}_{1}^{\rm s}} \cdot \bigg[ \! \sum_{\bm{m}_{1}^{\rm s}} \!\bm{m}_{1}^{\rm s} \!\!\sum_{\rb} \!\bigg\{ \eta_{\ra} A_{\bm{m}_{1}^{\rm s}}^{\rb} (\bm{J}_{1}) \, \oFa (\bm{J}_{1}) \!+\! \eta_{\rb} D_{\bm{m}_{1}^{\rm s}}^{\rb} (\bm{J}_{1}) \, \bm{m}_{1}^{\rm s} \!\cdot\! \frac{\partial \oFa}{\partial \bm{J}_{1}^{\rm s}} \bigg\} \bigg]\,,
\label{LB_multi_drift_diff} 
\end{equation}
where the drift and diffusion coefficients ${ A_{\bm{m}_{1}^{\rm s}}^{\rb} (\bm{J}_{1}) }$ and ${ D_{\bm{m}_{1}^{\rm s}}^{\rb} (\bm{J}_{1}) }$ depend on the position in action space $\bm{J}_{1}$, the considered resonance $\bm{m}_{1}^{\rm s}$, and the component ${``\rb"}$ used as the underlying DF to estimate them. The drift coefficients and diffusion coefficients are given by
\begin{align}
& \!\! A_{\bm{m}_{1}^{\rm s}}^{\rb} \!(\bm{J}_{1}\!) \!=\! - \pi (2 \pi)^{2k - d} \!\! \sum_{\bm{m}_{2}^{\rm s}} \!\! \int \!\!\! \rd \bm{J}_{2} \frac{\delta_{\rm D} (\bm{m}_{1}^{\rm s} \!\cdot\! \bm{\Omega}_{1}^{\rm s} \!-\! \bm{m}_{2}^{\rm s} \!\cdot\! \bm{\Omega}_{2}^{\rm s})}{| \mathcal{D}_{\bm{m}_{1}^{\rm s} , \bm{m}_{2}^{\rm s}} \!(\bm{J}_{1} ,\! \bm{J}_{2} ,\! \bm{m}_{1}^{\rm s} \!\cdot\! \bm{\Omega}_{1}^{\rm s} \!) |^{2}} \bm{m}_{2}^{\rm s} \!\cdot\! \frac{\partial \oFb}{\partial \bm{J}_{2}^{\rm s}} \, ,  \nonumber 
\\
& \!\! D_{\bm{m}_{1}^{\rm s}}^{\rb} \!(\bm{J}_{1}\!) \!=\! \pi (2 \pi)^{2k - d} \!\! \sum_{\bm{m}_{2}^{\rm s}} \!\! \int \!\!\! \rd \bm{J}_{2} \frac{\delta_{\rm D} (\bm{m}_{1}^{\rm s} \!\cdot\! \bm{\Omega}_{1}^{\rm s} \!-\! \bm{m}_{2}^{\rm s} \!\cdot\! \bm{\Omega}_{2}^{\rm s})}{| \mathcal{D}_{\bm{m}_{1}^{\rm s} , \bm{m}_{2}^{\rm s}} \!(\bm{J}_{1} ,\! \bm{J}_{2} ,\! \bm{m}_{1}^{\rm s} \!\cdot\! \bm{\Omega}_{1}^{\rm s}\!) |^{2}} \oFb (\bm{J}_{2}) \, .
\label{drift_diff_multi}
\end{align}
Equation~\eqref{LB_multi_drift_diff} can finally be rewritten as
\begin{equation}
\frac{\partial \oFa}{\partial \tau} \!=\! \frac{\partial }{\partial \bm{J}_{1}^{\rm s}} \!\cdot\! \bigg[\! \sum_{\bm{m}_{1}^{\rm s}} \! \bm{m}_{1}^{\rm s} \bigg\{ \eta_{\ra} A_{\bm{m}_{1}^{\rm s}}^{\rm tot} (\bm{J}_{1}) \, \oFa (\bm{J}_{1}) \!+\! D_{\bm{m}_{1}^{\rm s}}^{\rm tot} (\bm{J}_{1}) \, \bm{m}_{1}^{\rm s} \!\cdot\! \frac{\partial \oFa}{\partial \bm{J}_{1}^{\rm s}} \bigg\} \bigg] \, ,
\label{BL_multi_short} 
\end{equation}
where the total drift and diffusion coefficients $A_{\bm{m}_{1}^{\rm s}}^{\rm tot}$ and $D_{\bm{m}_{1}^{\rm s}}^{\rm tot}$ are given by
\begin{equation}
A_{\bm{m}_{1}^{\rm s}}^{\rm tot} (\bm{J}_{1}) = \sum_{\rb} A_{\bm{m}_{1}^{\rm s}}^{\rb} (\bm{J}_{1}) \;\; ; \;\; D_{\bm{m}_{1}^{\rm s}}^{\rm tot} (\bm{J}_{1}) = \sum_{\rb} \eta_{\rb} D_{\bm{m}_{1}^{\rm s}}^{\rb} (\bm{J}_{1}) \, .
\label{A_D_tot_multi} \nonumber 
\end{equation}
In equation~\eqref{BL_multi_short}, the total drift coefficients are multiplied by the dimensionless mass $\eta_{\ra}$ of the considered component. This essentially captures the known process of segregation, when a spectrum of masses is involved, so that components with larger individual masses tend to narrower steady states. Indeed, the multi-component Balescu-Lenard formalism captures the secular effect of multiple resonant (non-local) deflections of lighter particles by the more massive ones: the lighter population will drift towards larger radii, while the massive one will sink in. 
This can be seen for instance by seeking asymptotic stationary solutions to equation~\eqref{BL_multi_short} by nulling the curly brace in its r.h.s.

\subsection{Secular evolution increases Boltzmann entropy}
\label{sec:HTheorem}

Following closely the demonstration presented in~\cite{Heyvaerts2010}, let us define the system's entropy ${ S (\tau) }$ as
\begin{equation}
S (\tau) = - \!\! \int \!\! \rd \bm{J}_{1} \, s (\oF (\bm{J}_{1})) \, , \quad {\rm where} \quad { s(x) \!=\! x\log x  }\,.
\label{definition_entropy}
\end{equation}
 Differentiating equation~\eqref{definition_entropy} once with respect to $\tau$ yields
\begin{equation}
\frac{\rd S}{\rd \tau} = - \!\! \int \!\! \rd \bm{J}_{1} \, s ' (\oF (\bm{J}_{1})) \, \frac{\partial \oF}{\partial t} \, .
\label{derivative_entropy}
\end{equation}
Let us introduce the system's diffusion flux, $ \bm{\mathcal{F}}_{\!\rm tot} (\bm{J}_{1})$, given by
\begin{align}
\bm{\mathcal{F}}_{\!\rm tot} (\bm{J}_{1}) = & \!\! \sum_{\bm{m}_{1}^{\rm s} , \bm{m}_{2}^{\rm s}} \!\! \bm{m}_{1}^{\rm s} \!\! \int \!\! \rd \bm{J}_{2} \, \alpha_{\bm{m}_{1}^{\rm s} , \bm{m}_{2}^{\rm s}} (\bm{J}_{1} , \bm{J}_{2})    \nonumber
\\
& \, \times \bigg[\! \bm{m}_{1}^{\rm s} \!\cdot\! \frac{\partial }{\partial \bm{J}_{1}^{\rm s}} \!-\! \bm{m}_{2}^{\rm s} \!\cdot\! \frac{\partial }{\partial \bm{J}_{2}^{\rm s}} \!\bigg] \, \oF (\bm{J}_{1}) \, \oF (\bm{J}_{2} ) \, ,  \label{definition_flux}
\end{align}
with ${ \alpha_{\bm{m}_{1}^{\rm s} , \bm{m}_{2}^{\rm s}} (\bm{J}_{1} , \bm{J}_{2}) }$ given by
\begin{equation}
\alpha_{\bm{m}_{1}^{\rm s} , \bm{m}_{2}^{\rm s}} (\bm{J}_{1} , \bm{J}_{2}) = \frac{\pi (2 \pi)^{2d -k}}{N} \frac{\delta_{\rm D} (\bm{m}_{1}^{\rm s} \!\cdot\! \bm{\Omega}_{1}^{\rm s} \!-\! \bm{m}_{2}^{\rm s} \!\cdot\! \bm{\Omega}_{2}^{\rm s})}{| \mathcal{D}_{\bm{m}_{1}^{\rm s} , \bm{m}_{2}^{\rm s}} (\bm{J}_{1} , \bm{J}_{2} , \bm{m}_{1}^{\rm s} \!\cdot\! \bm{\Omega}_{1}^{\rm s}) |^{2}}  \geq 0  \, ,
\label{definition_alpha}
\end{equation}
such that equation~\eqref{BL_Kepler} reads
\begin{equation}
\frac{ \partial \oF }{\partial \tau}= \frac{\partial }{ \partial \bm{J}_{1}^{\rm s}} \!\cdot\! \bm{\mathcal{F}}_{\!\rm tot} (\bm{J}_{1}) .
\label{BL_Kepler_div}
\end{equation} 
Using integration by parts in equation~\eqref{derivative_entropy} and ignoring boundary terms leads to
\begin{equation}
\frac{\rd S}{\rd \tau} = \!\! \int \!\! \rd \bm{J}_{1} \, s '' (\oF (\bm{J}_{1})) \, \frac{\partial \oF}{\partial \bm{J}_{1}^{\rm s}} \!\cdot\! \bm{\mathcal{F}}_{\!\rm tot} (\bm{J}_{1}) \, .
\label{by_parts_entropy}
\end{equation}
Given equation~\eqref{definition_flux}, equation~\eqref{by_parts_entropy} can be rewritten as
\begin{equation}
\frac{\rd S}{\rd \tau} \!=\!\! \sum_{\bm{m}_{1}^{\rm s} , \bm{m}_{2}^{\rm s}}  \! \int \!\! \rd \bm{J}_{1} \rd \bm{J}_{2} \alpha_{\bm{m}_{1}^{\rm s} , \bm{m}_{2}^{\rm s}} s''_{1} (\bm{m}_{1}^{\rm s} \!\cdot\! \bm{\oF}_{1}') \bigg[ \oF_{2} (\bm{m}_{1}^{\rm s} \!\cdot\! \bm{\oF}_{1}') \!-\! \oF_{1} (\bm{m}_{2}^{\rm s} \!\cdot\! \bm{\oF}_{2}') \bigg] \, ,
\label{short_flux_II} \nonumber
\end{equation}
with $s''_{i} \!=\! s'' (\oF (\bm{J}_{i}))$, $\oF_{i} \!=\! \oF (\bm{J}_{i})$ and $\bm{\oF}_{i}' \!=\! \partial \oF / \partial \bm{J}_{i}^{\rm s}$.
 This equation can symmetrised via the substitutions ${ \bm{m}_{1}^{\rm s} \!\leftrightarrow\! \bm{m}_{2}^{\rm s} }$ and ${ \bm{J}_{1} \!\leftrightarrow\! \bm{J}_{2} }$, relying on the fact that ${ \alpha_{\bm{m}_{2}^{\rm s} , \bm{m}_{1}^{\rm s}} (\bm{J}_{2} , \bm{J}_{1}) \!=\! \alpha_{\bm{m}_{1}^{\rm s} , \bm{m}_{2}^{\rm s}} (\bm{J}_{1} , \bm{J}_{2}) }$, so that
\begin{align}
\frac{\rd S}{\rd \tau} =&  \, \frac{1}{2} \sum_{\bm{m}_{1}^{\rm s} , \bm{m}_{2}^{\rm s}} \! \int \!\! \rd \bm{J}_{1} \rd \bm{J}_{2} \, \alpha_{\bm{m}_{1}^{\rm s} , \bm{m}_{2}^{\rm s}} (\bm{J}_{1} , \bm{J}_{2})   \times \bigg[\! \oF_{2} s_{1}'' (\bm{m}_{1}^{\rm s} \!\cdot\! \bm{\oF}_{1}')^{2}\nonumber
\\
& \,  \!-\! (\bm{m}_{1}^{\rm s} \!\cdot\! \bm{\oF}_{1}' ) (\bm{m}_{2}^{\rm s} \!\cdot\! \bm{\oF}_{2}') (\oF_{1} s_{1} '' \!+\! \oF_{2} s_{2} '') \!+\! \oF_{1} s_{2}'' (\bm{m}_{2}^{\rm s} \!\cdot\! \bm{\oF}_{2}')^{2} \!\bigg] \, .
\label{short_flux_sym}
\end{align}
As the entropy function satisfies ${ s''(x) \!=\! 1/x }$ (any double primitive of ${1/x}$ would work too), the square braket of equation~\eqref{short_flux_sym} can immediately be factored as
\begin{equation}
\frac{1}{\oF_{1} \oF_{2}} \bigg[ \oF_{2} (\bm{m}_{1}^{\rm s} \!\cdot\! \bm{\oF}_{1}') \!-\! \oF_{1} (\bm{m}_{2}^{\rm s} \!\cdot\! \bm{\oF}_{2}') \bigg]^{2} \geq 0  \, ,
\label{second_line_entropy}
\end{equation}
so that one finally gets ${ \rd S /\rd \tau \!\geq\! 0 }$.
This entropy increase corresponds to heat generation as the orbital structure of the cluster rearranges itself in a more eccentric configuration.
The previous demonstration naturally extends for the multi-component Balescu-Lenard equation~\eqref{LB_multi}. Indeed, defining the system's total entropy $S_{\rm tot}$, summed for all components, as
\begin{equation}
S_{\rm tot} (\tau) = - \!\! \int \!\! \rd \bm{J}_{1} \, \!\! \sum_{\ra} \frac{1}{\eta_{\ra}} \, s (\oFa (\bm{J}_{1})) \, ,
\label{definition_entropy_tot}
\end{equation}
one can again show that for ${ s''(x) \!=\! 1/x }$, one has ${ \rd S_{\rm tot} / \rd \tau \!\geq\! 0 }$, which does not necessarily imply that the entropy of each component increases.

\section{Applications}
\label{sec:applications}

Up to now we have considered the general framework of a system made of a finite number of particles orbiting a central massive object. 
We now examine in turn some more specific configurations of particles orbiting a black hole, and discuss how the results of the previous section can be further extended when considering specific geometries and physical secular processes, to highlight the wealth of possible implications one can draw from this framework. Detailed applications are postponed to follow-up papers.

\subsection{Razor-thin axisymmetric discs}
\label{sec:case_disc}

Let us first specialise the degenerate Balescu-Lenard equation~\eqref{BL_Kepler} to razor-thin axisymmetric discs. For such systems, the dimension of the physical space is given by ${ d \!=\! 2 }$, while the number of dynamical degeneracies of the Keplerian dynamics is ${ k \!=\! 1 }$. Therefore, the resonance condition in equation~\eqref{BL_Kepler} takes the simpler form of a ${1D}$ condition naively reading ${ m_{1}^{\rm s} \Omega_{1}^{\rm s} \!-\! m_{2}^{\rm s} \Omega_{2}^{\rm s} \!=\! 0}$ and the Delaunay angle-action variables from equation~\eqref{Delaunay_3D_actions} become
\begin{equation}
(\bm{J} , \bm{\theta}) \!=\! (J_{1} , J_{2} , \theta_{1} , \theta_{2}) \!=\!  (J^{\rm s} , J^{\rm f} , \theta^{\rm s} , \theta^{\rm f}) \!=\! (L , I , g , w) \, .
\label{Delaunay_disc}
\end{equation}

Symmetries of the interaction potential lead to relationships among the susceptibility coefficients, which simplify the Balescu-Lenard equation. The rescaled interaction potential ${ U_{12} }$ from equation~\eqref{rescaling_U} takes the form
\begin{equation}
U_{12} = - \frac{G M_{\bullet}}{|\bm{x}_{1} \!-\! \bm{x}_{2}|} = - \frac{G M_{\bullet}}{\sqrt{R_{1}^{2} \!+\! R_{2}^{2} \!-\! 2 R_{1} R_{2} \cos (\phi_{1} \!-\! \phi_{2})}} \, ,
\label{symmetry_U}
\end{equation}
in which we introduce the usual polar coordinates ${ (R , \phi) }$. Following equations (3.28a) and (3.28b) of~\cite{BinneyTremaine2008}, the mapping from the physical polar coordinates to the Delaunay angle-action ones can be written as
\begin{equation}
R = a (1 \!-\! e \cos (\eta)) \;\; ; \;\; \phi = g \!+\! f \, ,
\label{mapping_Delaunay_disc}
\end{equation}
where the semi-major axis $a$, eccentricity $e$, true anomaly $f$, and eccentric anomaly $\eta$ are introduced as
\begin{align}
& \, e = \sqrt{1 \!-\! (L / I )^{2}} \;\; ; \;\; a = \frac{I^{2}}{G M_{\bullet}} \;\; ;   \nonumber
\\
& \,f = \tan^{-1} \!\bigg[\! \frac{\sqrt{1 \!-\! e^{2}} \sin(\eta)}{\cos(\eta) \!-\! e} \!\bigg] \;\; ; \;\; w = \eta \!-\! e \sin(\eta) \, .
\label{definition_e_a_eta_f}
\end{align}
Substituting equation~\eqref{mapping_Delaunay_disc} into equation~\eqref{symmetry_U}, we immediately have that 
\begin{equation}
U_{12} \!=\! U (g_{1} \!-\! g_{2} , w_{1} , w_{2} , \bm{J}_{1} , \bm{J}_{2}) \; \Rightarrow \; \oU_{12} \!=\! \oU (g_{1} \!-\! g_{2} , \bm{J}_{1} , \bm{J}_{2}) \, .
\label{dependence_U}
\end{equation}
As a consequence, the bare susceptibility coefficients from equation~\eqref{definition_bare_A} for a razor-thin disc are related to one another through
\begin{equation}
A_{m_{1}^{\rm s} , m_{2}^{\rm s}} (\bm{J}_{1} , \bm{J}_{2}) = \delta_{m_{1}^{\rm s}}^{m_{2}^{\rm s}} \, A_{m_{1}^{\rm s} , m_{1}^{\rm s}} (\bm{J}_{1} , \bm{J}_{2}) \, .
\label{symmetry_A_disc}
\end{equation}
A similar result also holds for the dressed susceptibility coefficients from equation~\eqref{definition_1/D}. Indeed, for any ${2D}$ razor-thin system, one can assume the basis elements from equation~\eqref{definition_basis_Kalnajs} to be generically of the form
\begin{equation}
\psi^{(p)} (R , \phi) = \re^{\ri \ell^{p} \phi} \, \mathcal{U}_{n^{p}}^{\ell^{p}} (R) \, ,
\label{shape_psi_p_disc}
\end{equation}
where $\ell^{p}$ and $n^{p}$ are two integer indices, and $\mathcal{U}_{n}^{\ell}$ are radial functions. Such a decomposition of the basis elements allows us to decouple the azimuthal and radial dependence of the basis elements. Noting that in the mapping from equation~\eqref{mapping_Delaunay_disc} only the azimuthal angle $\phi$ depends on the slow angle $g$, one obtains that the Fourier transformed basis elements satisfy
\begin{equation}
\opsi^{(p)}_{m^{\rm s}} (\bm{J}) = \delta_{\ell^{p}}^{m^{\rm s}} \, \opsi^{(p)}_{m^{\rm s}} (\bm{J}) \, . 
\label{symmetry_psi_p_disc}
\end{equation}
Substituting this into the expression~\eqref{Fourier_M} for the response matrix, we find that
\begin{equation}
\widehat{\mathbf{M}}_{pq} (\omega) = \delta_{\ell^{p}}^{\ell^{q}} \, \widehat{\mathbf{M}}_{pq} (\omega) \, .
\label{symmetry_M_disc}
\end{equation}
Using equations~\eqref{symmetry_psi_p_disc} and~\eqref{symmetry_M_disc}, the dressed susceptibility coefficients satisfy
\begin{equation}
\frac{1}{\mathcal{D}_{m_{1}^{\rm s} , m_{2}^{\rm s}} (\bm{J}_{1} , \bm{J}_{2} , \omega)} = \delta_{m_{1}^{\rm s}}^{m_{2}^{\rm s}} \, \frac{1}{\mathcal{D}_{m_{1}^{\rm s} , m_{1}^{\rm s}} (\bm{J}_{1} , \bm{J}_{2} , \omega)} \, ,
\label{symmetry_1/D_disc}
\end{equation}
just as the bare ones satisfy equation~\eqref{symmetry_A_disc}.

The symmetry corresponding to equation~\eqref{symmetry_1/D_disc} allows us to get rid of the sum over the resonance index $m_{2}^{\rm s}$ in the Balescu-Lenard equation~\eqref{BL_Kepler}, so that it becomes
\begin{align}
\frac{\partial \oF}{\partial \tau} = \frac{\pi}{N} \frac{\partial }{\partial J_{1}^{\rm s}} \bigg[ & \,  \!\! \int \!\! \rd \bm{J}_{2} \, \delta_{\rm D} (\Omega^{\rm s} (\bm{J}_{1}) \!-\! \Omega^{\rm s} (\bm{J}_{2})) \, \frac{1}{|\mathcal{D}_{\rm tot} (\bm{J}_{1} , \bm{J}_{2})|^{2}} \nonumber
\\
& \, \times \bigg[ \frac{\partial }{\partial J_{1}^{\rm s}} \!-\! \frac{\partial }{\partial J_{2}^{\rm s}} \bigg] \, \oF (\bm{J}_{1}) \, \oF (\bm{J}_{2}) \bigg] \, ,
\label{BL_Kepler_disc}
\end{align}
in which we use the relation ${ \delta_{\rm D} (\alpha x) \!=\! \delta_{\rm D} (x) / |\alpha| }$ and introduce the (unique) total dressed susceptibility coefficient
\begin{equation}
\frac{1}{|\mathcal{D}_{\rm tot} (\bm{J}_{1} , \bm{J}_{2})|^{2}} = \sum_{m_{1}^{\rm s}} \frac{|m_{1}^{\rm s}| }{|\mathcal{D}_{m_{1}^{\rm s} , m_{1}^{\rm s}} (\bm{J}_{1} , \bm{J}_{2} , m_{1}^{\rm s} \Omega^{\rm s} (\bm{J}_{1}))|^{2}} \, .
\label{1/D_tot_disc}
\end{equation}
Similarly, if we neglect self-gravity, then the symmetry in equation~\eqref{symmetry_A_disc} applies and equation~\eqref{BL_Kepler_disc} becomes the associated Landau equation, in which the total dressed susceptibility coefficient ${ 1/|\mathcal{D}_{\rm tot} (\bm{J}_{1} , \bm{J}_{2})|^{2} }$ is replaced by the bare one,
\begin{equation}
|A_{\rm tot} (\bm{J}_{1} , \bm{J}_{2}) |^{2} = \sum_{m_{1}^{\rm s}} |m_{1}^{\rm s}| \, | A_{m_{1}^{\rm s} , m_{1}^{\rm s}} (\bm{J}_{1} , \bm{J}_{2}) |^{2} \, .
\label{A_tot_disc}
\end{equation}
This Landau analog of equation~\eqref{BL_Kepler_disc} for razor-thin axisymmetric discs with the bare susceptibility coefficients from equation~\eqref{A_tot_disc} has already been derived in~\cite{ST3} via Gilbert's equation.

The result of these simplifications is that the degenerate Balescu-Lenard equation~\eqref{BL_Kepler_disc} possesses a straightforward resonance condition in which resonant encounters can only occur between two orbits caught in the same resonance, as illustrated in figure~\ref{figIllustrationBL}. To compute the diffusion flux appearing in the r.h.s. of this equation, we employ the generic definition of the composition of a Dirac delta and a function~\citep{Hormander2003}, which in a ${d-}$dimensional setup takes the form
\begin{equation}
\!\! \int_{\mathbb{R}^{d}} \!\! \rd \bm{x} \, f (\bm{x}) \, \delta_{\rm D} (g  (\bm{x})) = \!\! \int_{g^{-1} (0)} \!\!\!\!\!\!\!\! \rd \sigma (\bm{x}) \, \frac{f (\bm{x})}{|\nabla g (\bm{x})|} \, ,
\label{composition_Dirac_Delta}
\end{equation}
where ${ g^{-1} (0) \!=\! \big\{ \bm{x} \, | \, g(\bm{x}) \!=\! 0 \big\} }$ is the hypersurface of dimension ${ (d \!-\! 1) }$ defined by the constraint ${ g(\bm{x}) \!=\! 0 }$, and ${ \rd \sigma (\bm{x}) }$ is the surface measure on ${ g^{-1} (0) }$. In our case, the resonance condition is given by the function
\begin{equation}
g (\bm{J}_{2}) = \Omega^{\rm s} (\bm{J}_{1}) - \Omega^{\rm s} (\bm{J}_{2}) \, .
\label{definition_g_disc}
\end{equation}
For a given value of $\bm{J}_{1}$, and introducing ${ \omega \!=\! \Omega^{\rm s} (\bm{J}_{1}) }$, we define the critical resonant curve ${ \gamma (\omega) }$ as
\begin{equation}
\gamma (\omega) = \big\{ \bm{J}_{2} \, \big| \; \Omega^{\rm s} (\bm{J}_{2}) \!=\! \omega \big\} \, .
\label{definition_gamma_disc}
\end{equation}
This curve corresponds to the set of all orbits which are in resonance with the precessing orbit of action $\bm{J}_{1}$. Once this resonance line has been identified, the diffusion flux from equation~\eqref{BL_Kepler_disc} is straightforward to compute and reads
\begin{equation}
\frac{\partial \oF}{\partial \tau} = \frac{\partial }{\partial J_{1}^{\rm s}} \bigg[ \!\! \int_{\gamma (\Omega^{\rm s} (\bm{J}_{1}))} \!\!\!\!\!\!\!\! \rd \sigma \, \frac{G (\bm{J}_{1} , \bm{J}_{2})}{| \nabla (\Omega^{\rm s} (\bm{J}_{2})) |} \bigg] \, ,
\label{BL_Kepler_disc_resonant}
\end{equation}
where to shorten the notations, we introduced the function ${ G (\bm{J}_{1} , \bm{J}_{2}) }$ as
\begin{equation}
G (\bm{J}_{1} , \bm{J}_{2}) = \frac{\pi}{N} \frac{1}{\mathcal{D}_{\rm tot} (\bm{J}_{1} , \bm{J}_{2})}  \bigg[ \frac{\partial }{\partial J_{1}^{\rm s}} \!-\! \frac{\partial }{\partial J_{2}^{\rm s}} \bigg] \, \oF (\bm{J}_{1}) \, \oF (\bm{J}_{2}) \, ,
\label{definition_G_disc}
\end{equation}
as well as the resonant contribution ${ |\nabla (\Omega^{\rm s} (\bm{J}_{2}))| }$ given by
\begin{equation}
\nabla (\Omega^{\rm s} (\bm{J}_{2})) = \sqrt{\bigg[ \frac{\partial \Omega^{\rm s}}{\partial J_{2}^{\rm s}} \bigg]^{2} \!+\! \bigg[ \frac{\partial \Omega^{\rm s}}{\partial J_{2}^{\rm f}} \bigg]^{2}} \, .
\label{resonant_contribution_disc}
\end{equation}
We note that equation~\eqref{BL_Kepler_disc_resonant} is now a simple one-dimensional integral involving a regular integrand.

In summary, because the quasi-stationary potentials $\oP$ and $\oP_{\rr}$  are known via equations~\eqref{definition_averaged_Phi} and~\eqref{BL_rel_freq}, one can compute the associated precession frequencies $\Omega^{\rm s}$ (and their gradients). This allows for the determination of the critical resonant lines $\gamma$ from equation~\eqref{definition_gamma_disc}. Following equation~\eqref{BL_Kepler_disc_resonant}, it then only remains to integrate along these lines to determine the secular diffusion flux. Such an effective computation for razor-thin discs in the Landau limit is postponed to a follow-up paper, as equation~\eqref{definition_averaged_Phi} involves a singular triple integral over wire-wire interactions.
Similarly, the study of the long-term evolution of quasi-stationary non-axisymmetric razor-thin discs (such as M31) will also be the subject of a future work.

\subsection{Spherical cluster around BH}
\label{sec:case_3D}

We now turn to the application of the degenerate Balescu-Lenard equation~\eqref{BL_Kepler} to spherically symmetric systems.  The general procedure is the same as for the razor-thin disc case, but now the dimension of the physical space is ${ d \!=\! 3 }$, while the number of Keplerian dynamical degeneracies is given by ${ k \!=\! 2 }$. The resonance condition in equation~\eqref{BL_Kepler} becomes two-dimensional and reads ${ \bm{m}_{1}^{\rm s} \!\cdot\! \bm{\Omega}_{1}^{\rm s} \!-\! \bm{m}_{2}^{\rm s} \!\cdot\! \bm{\Omega}_{2}^{\rm s} \!=\! 0 }$. In the ${3D}$ context, the Delaunay variables from equation~\eqref{Delaunay_disc} become
\begin{equation}
(\bm{J}, \bm{\theta}) \!=\! (J_{1}^{\rm s} , J_{2}^{\rm s} , J_{3}^{\rm f} , \theta_{1}^{\rm s} , \theta_{2}^{\rm s} , \theta_{3}^{\rm f}) \!=\! (L , L_{z} , I , g , h , w) \, ,
\label{Delaunay_3D}
\end{equation}
where $g$ stands for the angle from the ascending node to the periapse, $h$ for the longitude of the ascending node and $w$ for the Keplerian orbital phase, that is the mean anomaly.

As was done in equation~\eqref{BL_Kepler_disc} for razor-thin discs, let us now show how the ${3D}$ geometry allows us to further simplify the kinetic equation. Written in spherical coordinates ${ (R , \theta , \phi) }$, the rescaled interaction potential from equation~\eqref{rescaling_U} becomes
\begin{align}
U_{12} & \, \!=\! - \frac{G M_{\bullet}}{|\bm{x}_{1} \!-\! \bm{x}_{2}|}    \nonumber
\\
& \, \!=\! - G M_{\bullet} \bigg[ R_{1}^{2} \!+\! R_{2}^{2} \!-\!  2 R_{1} R_{2} \bigg\{ \sin (\theta_{1}) \sin(\theta_{2}) \cos (\phi_{1} \!-\! \phi_{2})  \nonumber\
\\
& \, \;\;\;\;\;\;\;\;\;\;\;\;\;\;\;\;\;\;\;\;\;\;\;\;\;\;\;\;\;\;\;\;\;\;\;\;\;\; \!+\! \cos (\theta_{1}) \cos (\theta_{2}) \bigg\} \bigg]^{-1/2} \, .
\label{U_3D}
\end{align}
Following equation~(5.20) from~\cite{AstrophysicalBlackHoles}, these ${ (R,\theta,\phi) }$ can be expressed as a function of the Delaunay angle-action variables, so that
\begin{align}
& \, R = a (1 \!-\! e \cos (\eta)) \;\; ; \;\; \phi = h + \tan^{-1} \!\big[\! \cos (i) \tan (g \!+\! f) \big] \, ;   \nonumber
\\
& \, \theta = \cos^{-1} \!\big[\! \sin (i) \sin(g \!+\! f) \big] \, ,
\label{mapping_Delaunay_3D}
\end{align}
where $a$, $e$, $f$ and $\eta$ were introduced in equation~\eqref{definition_e_a_eta_f}, and $i$ is the orbit's inclination, defined through ${ \cos (i) \!=\! L_{z} / L }$. Therefore the interaction potential of equation~\eqref{U_3D} and its angle-averaged version (equation~\eqref{definition_Ubar12}) have the symmetries
\begin{align}
& \, U_{12} = U (g_{1} , g_{2} , h_{1} \!-\! h_{2} , w_{1} , w_{2} , \bm{J}_{1} , \bm{J}_{2}) \, ,\nonumber
\\
& \, \oU_{12} = \oU (g_{1} , g_{2} , h_{1} \!-\! h_{2} , \bm{J}_{1} , \bm{J}_{2}) \, .
\label{symmetry}
\end{align}
From this, it immediately follows that the bare susceptibility coefficients (equation~\eqref{definition_bare_A}) are related to one another via
\begin{equation}
A_{\bm{m}_{1}^{\rm s} , \bm{m}_{2}^{\rm s}} (\bm{J}_{1} , \bm{J}_{2}) = \delta_{m_{1,h}^{\rm s}}^{m_{2,h}^{\rm s}} \, A_{\bm{m}_{1}^{\rm s} , \bm{m}_{2}^{\rm s}} (\bm{J}_{1} , \bm{J}_{2}) \, .
\label{symmetry_A_3D}
\end{equation}
Here, we have written the resonance vectors as ${ \bm{m}_{1}^{\rm s} \!=\! (m_{1 ,g}^{\rm s} , m_{1, h}^{\rm s}) }$, so that the coefficient $m_{1,h}^{\rm s}$ is the one associated with the slow angle $h$. A similar result holds for the dressed susceptibility coefficients defined in equation~\eqref{definition_1/D}. Indeed, for any ${3D}$ system, the basis elements in equation~\eqref{definition_basis_Kalnajs} can be written as
\begin{equation}
\psi^{(p)} (R , \theta , \phi) = Y_{\ell^{p}}^{m^{p}} \!(\theta , \phi) \, \mathcal{U}_{n^{p}}^{\ell^{p}} (R) \, ,
\label{shape_basis_3D}
\end{equation}
where $\ell^{p}$, $m^{p}$ and $n^{p}$ are three integer indices, $Y_{\ell}^{m}$ are the usual spherical harmonics, and $\mathcal{U}_{n}^{\ell}$ are radial functions. We note in the mappings from equation~\eqref{mapping_Delaunay_3D} that only the azimuthal angle $\phi$ depends on the slow angle $h$. Because the spherical harmonics are of the form ${ Y_{\ell}^{m} (\theta , \phi) \!\propto\! P_{\ell}^{m} (\cos \theta) \, \re^{\ri m \phi}}$, where $P_{\ell}^{m}$ are the associated Legendre polynomials, one immediately finds that the Fourier transformed basis elements satisfy
\begin{equation}
\opsi^{(p)}_{\bm{m}^{\rm s}} (\bm{J}) = \delta_{m^{p}}^{m_{h}^{\rm s}} \, \opsi^{(p)}_{\bm{m}^{\rm s}} (\bm{J}) \, .
\label{symmetry_psi_p_3D}
\end{equation}
As a consequence, the expression~\eqref{Fourier_M} of the response matrix immediately gives
\begin{equation}
\widehat{\mathbf{M}}_{pq} (\omega) = \delta_{m^{p}}^{m^{q}} \, \widehat{\mathbf{M}}_{pq} (\omega) \, .
\label{symmetry_M_3D}
\end{equation}
Equations~\eqref{symmetry_psi_p_3D} and~\eqref{symmetry_M_3D} allow us to rewrite the dressed susceptibility coefficients as
\begin{equation}
\frac{1}{\mathcal{D}_{\bm{m}_{1}^{\rm s} , \bm{m}_{2}^{\rm s}} (\bm{J}_{1} , \bm{J}_{2} , \omega)} = \delta_{m_{1 , h}^{\rm s}}^{m_{2 , h}^{\rm s}} \, \frac{1}{\mathcal{D}_{\bm{m}_{1}^{\rm s} , \bm{m}_{2}^{\rm s}} (\bm{J}_{1} , \bm{J}_{2} , \omega)} \, ,
\label{symmetry_1/D_3D}
\end{equation}
showing that they are related to one another in the same way as the bare susceptibility coefficients of equation~\eqref{symmetry_A_3D}.

As a consequence, when considering a spherically symmetric system, one can simplify the resonance condition, and the Balescu-Lenard equation~\eqref{BL_Kepler} becomes 
\begin{align}
\frac{\partial \oF}{\partial \tau} = & \, \frac{2 \pi^{2}}{N} \frac{\partial }{\partial \bm{J}_{1}^{\rm s}} \!\cdot\! \bigg[ \!\! \sum_{\bm{m}_{1}^{\rm s} , m_{2 , g}^{\rm s}} \!\! \bm{m}_{1}^{\rm s} \!\! \int \!\! \rd \bm{J}_{2} \, \frac{\delta_{\rm D} (\bm{m}_{1}^{\rm s} \!\cdot\! \bm{\Omega}_{1}^{\rm s} \!-\! (m_{2 , g}^{\rm s} , m_{1 , h}^{\rm s}) \!\cdot\! \bm{\Omega}_{2}^{\rm s})}{|\mathcal{D}_{\bm{m}_{1}^{\rm s} , (m_{2,g}^{\rm s} , m_{1,h}^{\rm s})} (\bm{J}_{1} , \bm{J}_{2} , \bm{m}_{1}^{\rm s} \!\cdot\! \bm{\Omega}_{1}^{\rm s} )|^{2}}   \nonumber
\\
& \, \times \bigg[ \bm{m}_{1}^{\rm s} \!\cdot\! \frac{\partial }{\partial \bm{J}_{1}^{\rm s}} \!-\! (m_{2 , g}^{\rm s} , m_{1 , h}^{\rm s}) \!\cdot\! \frac{\partial }{\partial \bm{J}_{2}^{\rm s}} \bigg] \, \oF (\bm{J}_{1}) \, \oF (\bm{J}_{2}) \bigg] \, .
\label{BL_Kepler_3D}
\end{align}
To neglect collective effects in equation~\eqref{BL_Kepler_3D}, one only has to make the substitution ${ 1/|\mathcal{D}|^{2} \!\to\! |A|^{2} }$. Here, it is important to note that the 1.5PN relativistic precession frequencies obtained in Appendix~\ref{sec:relativistic_precessions} do depend on the action $L_{z}$, so that at this stage further simplifications of equation~\eqref{BL_Kepler_3D} are not possible.  The computation of the diffusion flux in equation~\eqref{BL_Kepler_3D}  proceeds as in equation~\eqref{BL_Kepler_disc_resonant} by identifying the critical surfaces of resonance. We do not detail these calculations here.

\subsection{Relativistic barrier crossing in the vicinity of BHs}
\label{sec:relativisticcrossing}

We now show how the degenerate Balescu-Lenard equation~\eqref{BL_Kepler} naturally accounts for the presence of the ``Schwarzschild barrier'' encountered by stars diffusing towards the central BH. 
This Schwarzschild barrier was discovered by~\cite{Merritt2011} in their simulations of spherically symmetric star clusters.
Here, we show how it arises in the simpler case of a razor-thin axisymmetric disc of stars around the BH, but the same fundamental idea applies to the ${3D}$ case.
The secular collisional evolution of such a disc is governed by equation~\eqref{BL_Kepler_disc}.
The resonance condition in that equation is ${ \Omega_{1}^{\rm s} (\bm{J}_{1}) \!-\! \Omega_{2}^{\rm s} (\bm{J}_{2}) \!=\! 0 }$, in which the precession frequency ${ \Omega^{\rm s} (\bm{J}) }$ of each of the two wires, defined in equation~\eqref{definition_precession_frequencies}, is composed of two parts. The first is the contribution from the system's
self-consistent Newtonian potential,
\begin{equation} 
\Omega^{\rm s}_{\rm self} (L_{1} , I_{1}) = \frac{\partial }{\partial L_{1}} \bigg[ \oP (L_{1} , I_{1}) \bigg]  = \frac{\partial }{\partial L_{1}} \bigg[ \!\! \int \!\! \rd \bR_{2} \, \oF (\bR_{2}) \, \oU_{12} \bigg] \, .
\label{BL_self_freq_main}
\end{equation}
The second is the additional contribution from relativistic effects. We derive it in Appendix~\ref{sec:relativistic_precessions}. In the case of a razor-thin disc, it reads
\begin{equation} 
\Omega^{\rm s}_{\rm rel} (L , I) = \frac{1}{2 \pi} \frac{M_{\bullet}}{M_{\star}} \frac{(G M_{\bullet})^{4}}{c^{2}} \bigg[ - \frac{3}{I^{3} L^{2}} \!+\! \frac{G M_{\bullet}}{c} \frac{6 s}{I^{3} L^{3}} \bigg] .
\label{BL_rel_freq_main}
\end{equation}

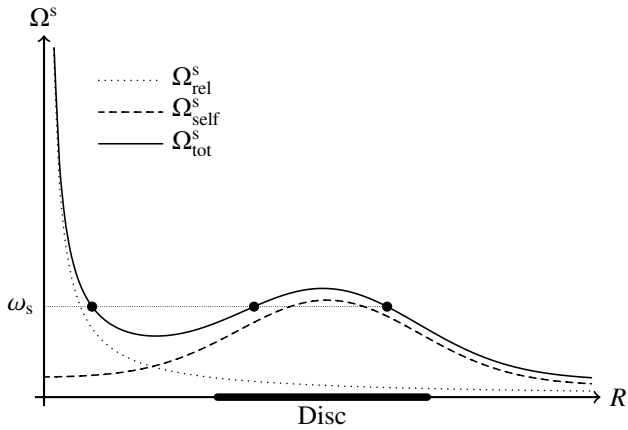
\begin{figure}
\begin{center}
\begin{tikzpicture}[scale=1.2]
\pgfmathsetmacro{\xmin}{0} ; \pgfmathsetmacro{\ymin}{0} ; 
\pgfmathsetmacro{\xmax}{6} ; \pgfmathsetmacro{\ymax}{4.};
\pgfmathsetmacro{\nbp}{100} ; 
\draw [->] [thick] ({\xmin - 0.1},\ymin) -- ({ \xmax + 0.1 }, \ymin) ; \draw ({ \xmax + 0.1} , \ymin) node[font = \normalsize , right] {$R$} ;
\draw [->] [thick] ({\xmin} , {\ymin - 0.1}) -- (\xmin , \ymax) ; \draw (\xmin , \ymax) node[font = \normalsize , above] {$\Omega^{\text{\scriptsize s}}$} ;
\pgfmathsetmacro{\xminrel}{0.104} ; \pgfmathsetmacro{\xmaxrel}{\xmax} ; 
\draw [dotted , semithick , line cap=round , dash pattern=on 0pt off 4 \pgflinewidth , smooth , samples = \nbp] plot [domain = \xminrel : \xmaxrel ] (\x , {0.4/\x}) ; 
\pgfmathsetmacro{\xminself}{\xmin} ; \pgfmathsetmacro{\xmaxself}{\xmax} ; 
\draw [dashdotted , semithick , line cap=round , dash pattern=on 2pt off 2pt on \the\pgflinewidth off 0pt , smooth , samples = \nbp] plot [domain = \xminself : \xmaxself ] (\x , { (0.3+1.5 * exp(- ((\x - 3.2) * (\x - 3.2))/(2 * (1.0) * (1.0)))) * (0.5 - rad(atan((\x-5)/6))/(3.1415926535)) }) ;
\pgfmathsetmacro{\xmintot}{0.11} ; \pgfmathsetmacro{\xmaxtot}{\xmax} ; 
 \draw [ semithick , smooth , line cap = round, samples = \nbp] plot [domain = \xmintot : \xmaxtot ] (\x , { (0.4/\x) + (0.3+1.5 * exp(- ((\x - 3.2) * (\x - 3.2))/(2 * (1.0) * (1.0)))) * (0.5 - rad(atan((\x-5)/6))/(3.1415926535)) }) ;
\pgfmathsetmacro{\omegares}{1.} ;
\pgfmathsetmacro{\xa}{0.526886} ; \pgfmathsetmacro{\xaa}{2.30186} ; \pgfmathsetmacro{\xaaa}{3.76036} ;
\pgfmathsetmacro{\psize}{0.05} ;
\draw [fill] ({\xa , \omegares}) circle [radius = \psize] ;
\draw [fill] ({\xaa , \omegares}) circle [radius = \psize] ; 
\draw [fill] ({\xaaa , \omegares}) circle [radius = \psize] ;
\draw [dotted , very thin , line cap=round , dash pattern=on 0pt off 4 \pgflinewidth , smooth ] (\xmin , \omegares) -- (\xaaa , \omegares) ; \draw (\xmin , \omegares) node[font=\normalsize , left] {$\omega_{\text{\scriptsize s}}$} ; 
\pgfmathsetmacro{\xminleg}{0.6} ; \pgfmathsetmacro{\xsleg}{0.7} ; \pgfmathsetmacro{\xmaxleg}{\xminleg + \xsleg} ; 
\pgfmathsetmacro{\yleg}{3.5} ; \pgfmathsetmacro{\ysleg}{0.35};
\pgfmathsetmacro{\ylegrel}{\yleg};
\pgfmathsetmacro{\ylegself}{\yleg - \ysleg};
\pgfmathsetmacro{\ylegtot}{\yleg - 2 * \ysleg};
\draw [dotted , semithick , line cap=round , dash pattern=on 0pt off 4 \pgflinewidth]  (\xminleg , \ylegrel) -- (\xmaxleg , \ylegrel) ; \draw (\xmaxleg , \ylegrel) node[font=\normalsize , right] {$\Omega_{\text{\scriptsize rel}}^{\text{\scriptsize s}}$} ; 
\draw [dashdotted , semithick , line cap=round , dash pattern=on 2pt off 2pt on \the\pgflinewidth off 0pt]  (\xminleg , \ylegself) -- (\xmaxleg , \ylegself) ; \draw (\xmaxleg , \ylegself) node[font=\normalsize , right] {$\Omega_{\text{\scriptsize self}}^{\text{\scriptsize s}}$} ; 
\draw [semithick , smooth , line cap = round]  (\xminleg , \ylegtot) -- (\xmaxleg , \ylegtot) ; \draw (\xmaxleg , \ylegtot) node[font=\normalsize , right] {$\Omega_{\text{\scriptsize tot}}^{\text{\scriptsize s}}$} ; 
\pgfmathsetmacro{\xmindisc}{1.9} ; \pgfmathsetmacro{\xmaxdisc}{4.2} ;
\draw [line width = 0.1cm, line cap=round] (\xmindisc,\ymin) -- (\xmaxdisc , \ymin) ; 
\draw ({(\xmindisc+\xmaxdisc)/(2)} , \ymin) node[font=\normalsize , below] {$\text{Disc}$} ;
\end{tikzpicture}
\caption{\small{Illustration of the typical dependence of the precession frequencies $\Omega_{\rm self}^{\rm s}$ and $\Omega_{\rm rel}^{\rm s}$ (equations~\eqref{BL_self_freq_main} and~\eqref{BL_rel_freq_main}) as a function of the distance to the central BH. The relativistic precession frequencies, $\Omega_{\rm rel}^{\rm s}$ diverge as the star gets closer to the central BH, while the self-consistent ones $\Omega_{\rm self}^{\rm s}$ are typically the largest for stars in the neighbourhood of the considered disc. The black dots give all the locations, whose precession frequency is equal to $\omega_{\rm s}$ (illustrated by the dotted horizontal line). These positions are in resonance and will therefore have a non-vanishing contribution in the Balescu-Lenard equation~\eqref{BL_Kepler_disc}. Because equation~\eqref{BL_Kepler_disc} involves the product of the system's DF in the two resonating locations, the resonant coupling between the two outer points (which belong to the region where the disc dominates) will be much stronger than the couplings involving the inner point (which does not belong to core of the disc). As stars move inward, their precession frequencies increase up to a point where this prevents any resonant coupling with the disc's region. This effectively stops the secular diffusion, and induces a diffusion barrier.
}}
\label{fig_Resonance_Barrier}
\end{center}
\end{figure}

We now study how these precession frequencies depend on distance to the central BH. Following the timescale comparisons of~\cite{KocsisTremaine2011}, one expects the relativistic precession frequency $\Omega_{\rm rel}^{\rm s}$ to dominate close to BH (and, in fact, to diverge as the star gets closer to capture), while the self-consistent one, $\Omega_{\rm self}^{\rm s}$, will be the largest for orbits in the vicinity of the considered disc. Such a behaviour is qualitatively illustrated in figure~\ref{fig_Resonance_Barrier}, where we represent the typical dependence of the precession frequencies as a function of the distance to the central BH.
Figure~\ref{fig_Resonance_Barrier} shows that, for a given precession frequency $\omega_{\rm s}$, one can identify the actions $\bm{J}$ within the disc for which the resonance condition ${ \Omega^{\rm s}_{\rm tot} (\bm{J}) \!-\! \omega_{\rm s} \!=\! 0 }$ is satisfied.

Equation~\eqref{BL_Kepler_disc} involves the quadratic
factor ${ \oF (\bm{J}_{1}) \, \oF (\bm{J}_{2}) }$, which is the
product of the system's density at the two locations that are in
resonance. As shown in figure~\ref{fig_Resonance_Barrier}, because the
disc is only located in the outer regions of the BH, the resonant
coupling between two locations within the disc will be much stronger
than one involving a resonant location inside the inner edge of the
disc, very close to the BH. Therefore, in
figure~\ref{fig_Resonance_Barrier}, the coupling between the two outer
black dots will be much larger than the couplings involving the inner
dot. The situation becomes even worse if one wants to couple a region
even closer to the BH, for which the precession frequency is too large
to resonate with any part of the disc. In this situation, no efficient
resonant couplings are possible and the secular diffusion is
drastically suppressed.
In short, the divergence of the relativistic precession frequencies in
the neighbourhood of the BH means that stars whose orbits diffuse
inwards closer to the BH experience a rise in their precession
frequency, which prevents them from resonating anymore with the disc,
strongly suppressing further inward diffusion. This is the so-called
Schwarzschild barrier.

This explanation of the Schwarzschild barrier using the notion of resonant coupling is directly related to the explanation proposed in~\cite{BarOrAlexander2014}, which relies on the concept of adiabatic invariance. 
In their picture, a test star can undergo resonant relaxation only if the timescale associated with its relativistic precession is longer than the coherence time of the perturbations induced by the field stars and felt by the test star. Because the typical coherence time of the perturbations scales as the inverse of the typical precession frequency of the field stars (which lie within the cluster), the requirement for an efficient diffusion from the adiabatic invariance point of view is equivalent to the requirement from the point of view of the Balescu-Lenard resonance condition.

\subsection{Solving the Balescu-Lenard equation by Monte Carlo sampling}
\label{sec:relativisticcrossing_Langevin}

This suppression of diffusion in the neigbourhood of the BH can also
be illustrated by considering the orbit-averaged motion of individual wires.
Equation~\eqref{BL_Kepler_disc} takes the form of a diffusion equation
in action space, where one follows self-consistently the evolution of the
system's DF. One could also be interested in describing the stochastic
evolution of individual stellar wires, whose ensemble average is
described by this diffusion equation.  To do so, let us
rewrite equation~\eqref{BL_Kepler_disc} as
\begin{equation}
\frac{\partial \oF}{\partial \tau} = \frac{\partial }{\partial L} \bigg[ A (\bm{J} , \tau) \, \oF (\bm{J} , \tau) \!+\! D (\bm{J} , \tau) \, \frac{\partial \oF}{\partial L} \bigg] \, .
\label{rewrite_BL_Kepler_disc}
\end{equation}
Then, as detailed in
Appendix~\ref{sec:FP_to_Langevin}, one can write the corresponding
Langevin equation that captures the dynamics of individual test wires. Here, we consider the case of a razor-thin axisymmetric disc and denote as ${ \bm{\mathcal{J}} (\tau) \!=\! (\mathcal{L} (\tau) , \mathcal{I} (\tau)) }$ the position at time $\tau$ of a test wire in action space ${ \bm{J} \!=\! (L , I) }$. Following equation~\eqref{Langevin_equation}, the dynamics of the test wire takes the form
\begin{equation}
\frac{\rd \mathcal{L}}{\rd \tau} = h (\bm{\mathcal{J}} , \tau) \!+\! g (\bm{\mathcal{J}} , \tau) \, \Gamma (\tau) \;\;\; ; \;\;\; \frac{\rd \mathcal{I}}{\rd \tau} = 0,
\label{Langevin_disc}
\end{equation}
in which the ${1D}$ Langevin coefficients that describe the diffusion
of the wire in the ${L-}$direction are given by
\begin{equation}
h = - A \!+\! \frac{\partial D}{\partial L} \!-\! \sqrt{D} \, \frac{\partial \sqrt{D}}{\partial L} \;\; ; \;\; g  = \sqrt{D} \, ,
\label{h_g_disc}
\end{equation}
and the Langevin stochastic force ${ \Gamma (\tau) }$ satisfies
equation~\eqref{Langevin_forces}.  As in
equation~\eqref{marginal_diff}, the individual fast action ${ J^{\rm f} \!=\! I }$
is preserved during the wire's evolution.

Equation~\eqref{Langevin_disc} describes the diffusion of an
individual test wire when embedded in the self-induced noisy environment that is
described by the drift and diffusion coefficients from
equation~\eqref{BL_Kepler_disc}.  As such it could be used iteratively jointly with equation~\eqref{1/D_tot_disc} -- which depends on the sampled position of all orbits in action space via equation~\eqref{Fourier_M} -- to effectively integrate equation~\eqref{BL_Kepler_disc} over cosmic time. This would simply involve discretising equation~\eqref{Langevin_disc}  in time 
as ${ \mathcal{L}_{i+1} \!=\! \mathcal{L}_{i} \!+\! ({\rd \mathcal{L}}/{\rd \tau})_{i} \Delta \tau }$, while sampling initial $ \mathcal{L}_{0}$s to match the original distribution.
Strikingly, equation~\eqref{Langevin_disc} shares some similarity with the individual Hamilton's equations associated with the Hamiltonian from equation~\eqref{Hamiltonian_democratic_simpler}, but the significant gain of the present work 
is to allow for individual timesteps, ${ \Delta \tau }$, which are orders of magnitudes larger than the original one required to solve for the trajectories of individual stars.
It also deals seamlessly with post-Newtonian orbit integration over the fast and slow angles.

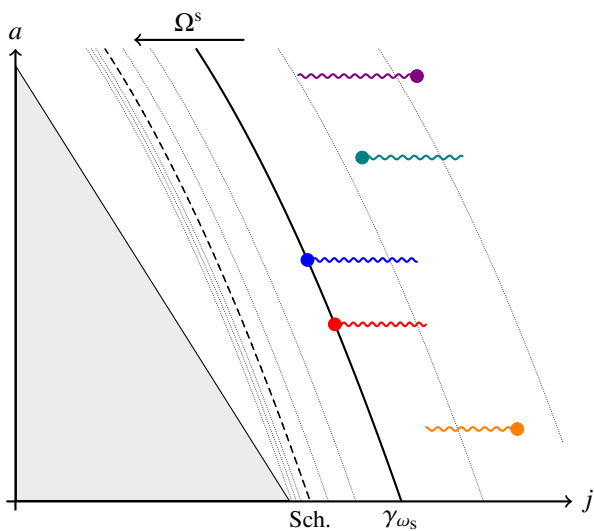
\begin{figure}
\begin{center}
\begin{tikzpicture}[scale = 1.2]
\pgfmathsetmacro{\xmin}{0} ; \pgfmathsetmacro{\ymin}{0} ; 
\pgfmathsetmacro{\xmax}{6} ; \pgfmathsetmacro{\ymax}{5.};
\pgfmathsetmacro{\nbp}{100} ; 
\draw [->] [thick] ({\xmin - 0.1},\ymin) -- ({ \xmax + 0.1 }, \ymin) ; \draw ({ \xmax + 0.1} , \ymin) node[font = \normalsize , right] {$j$} ;
\draw [->] [thick] ({\xmin} , {\ymin - 0.1}) -- (\xmin , \ymax) ; \draw (\xmin , \ymax) node[font = \normalsize , above] {$a$} ;
\pgfmathsetmacro{\ymaxbarr}{4.8} ; \pgfmathsetmacro{\xmaxbarrier}{3.} ; 
\draw [fill = gray!15] (\xmin+0.01,\ymin+0.01) -- (\xmin+0.01,\ymaxbarr) -- (\xmaxbarrier , \ymin+0.01) -- (\xmin + 0.01 , \ymin+0.01) ; 
\pgfmathsetmacro{\decal}{0.6} ; \pgfmathsetmacro{\xmaxdecal}{4.23046} ; \pgfmathsetmacro{\xmindecal}{1.98199} ; 
\pgfmathsetmacro{\decala}{1.1} ; \pgfmathsetmacro{\xmaxdecala}{3.73046} ; \pgfmathsetmacro{\xmindecala}{1.48199} ; 
\pgfmathsetmacro{\decalaa}{1.4} ; \pgfmathsetmacro{\xmaxdecalaa}{3.43046} ; \pgfmathsetmacro{\xmindecalaa}{1.18199} ; 
\pgfmathsetmacro{\decalaaa}{1.6} ; \pgfmathsetmacro{\xmaxdecalaaa}{3.23046} ; \pgfmathsetmacro{\xmindecalaaa}{0.981989} ; 
\pgfmathsetmacro{\decalaaaa}{1.7} ; \pgfmathsetmacro{\xmaxdecalaaaa}{3.13046} ; \pgfmathsetmacro{\xmindecalaaaa}{0.881989} ; 
\pgfmathsetmacro{\decalaaaaa}{1.75} ; \pgfmathsetmacro{\xmaxdecalaaaaa}{3.08046} ; \pgfmathsetmacro{\xmindecalaaaaa}{0.831989} ; 
\pgfmathsetmacro{\decalaaaaaa}{1.8} ; \pgfmathsetmacro{\xmaxdecalaaaaaa}{3.03046} ; \pgfmathsetmacro{\xmindecalaaaaaa}{0.781989} ; 
\pgfmathsetmacro{\decalaaaaaaa}{-0.3} ; \pgfmathsetmacro{\xmaxdecalaaaaaaa}{5.13046} ; \pgfmathsetmacro{\xmindecalaaaaaaa}{2.88199} ; 
\pgfmathsetmacro{\decalaaaaaaaa}{-1.4} ; \pgfmathsetmacro{\xmaxdecalaaaaaaaa}{\xmax} ; \pgfmathsetmacro{\xmindecalaaaaaaaa}{3.98199} ; 
\draw [thick , line cap = round , smooth] plot[domain = \xmindecal : \xmaxdecal] (\x , {- 0.3*pow(\x+\decal,2) + \ymax+2} ) ;
\draw [dotted , very thin , line cap=round , dash pattern=on 0pt off 4 \pgflinewidth , smooth] plot[domain = \xmindecala : \xmaxdecala] (\x , {- 0.3*pow(\x+\decala,2) + \ymax+2}) ;
\draw [dotted , very thin , line cap=round , dash pattern=on 0pt off 4 \pgflinewidth , smooth] plot[domain = \xmindecalaa : \xmaxdecalaa] (\x , {- 0.3*pow(\x+\decalaa,2) + \ymax+2}) ; 
\draw [dashdotted , semithick , line cap=round , dash pattern=on 2pt off 2pt on \the\pgflinewidth off 0pt , smooth] plot[domain = \xmindecalaaa : \xmaxdecalaaa] (\x , {- 0.3*pow(\x+\decalaaa,2) + \ymax+2}) ; 
\draw [dotted , very thin , line cap=round , dash pattern=on 0pt off 4 \pgflinewidth , smooth] plot[domain = \xmindecalaaaa : \xmaxdecalaaaa] (\x , {- 0.3*pow(\x+\decalaaaa,2) + \ymax+2}) ; 
\draw [dotted , very thin , line cap=round , dash pattern=on 0pt off 4 \pgflinewidth , smooth] plot[domain = \xmindecalaaaaa : \xmaxdecalaaaaa] (\x , {- 0.3*pow(\x+\decalaaaaa,2) + \ymax+2}) ; 
\draw [dotted , very thin , line cap=round , dash pattern=on 0pt off 4 \pgflinewidth , smooth] plot[domain = \xmindecalaaaaaa : \xmaxdecalaaaaaa] (\x , {- 0.3*pow(\x+\decalaaaaaa,2) + \ymax+2}) ; 
\draw [dotted , very thin , line cap=round , dash pattern=on 0pt off 4 \pgflinewidth , smooth] plot[domain = \xmindecalaaaaaaa : \xmaxdecalaaaaaaa] (\x , {- 0.3*pow(\x+\decalaaaaaaa,2) + \ymax+2}) ; 
\draw [dotted , very thin , line cap=round , dash pattern=on 0pt off 4 \pgflinewidth , smooth] plot[domain = \xmindecalaaaaaaaa : \xmaxdecalaaaaaaaa] (\x , {- 0.3*pow(\x+\decalaaaaaaaa,2) + \ymax+2}) ; 
\pgfmathsetmacro{\xmina}{3.5} ; \pgfmathsetmacro{\xmaxa}{4.5} ; \pgfmathsetmacro{\xa}{3.5} ; \pgfmathsetmacro{\ya}{{- 0.3*pow(\xa+\decal,2) + \ymax+2}} ;
\draw [red , thick , line cap = round , smooth , samples = 100 ] plot[domain = \xmina : \xmaxa ] (\x, { \ya + 0.02 * cos(\x * 3500) }) ; 
\draw [red , fill] ({\xa , \ya}) circle [radius = 0.07] ;
\pgfmathsetmacro{\xminaa}{3.2} ; \pgfmathsetmacro{\xmaxaa}{4.4} ; \pgfmathsetmacro{\xaa}{3.2} ; \pgfmathsetmacro{\yaa}{{- 0.3*pow(\xaa+\decal,2) + \ymax+2}} ; 
\draw [blue , thick , line cap = round , smooth , samples = 100 ] plot[domain = \xminaa : \xmaxaa ] (\x, { \yaa + 0.02 * cos(\x * 3400) }) ; 
\draw [blue , fill] ({\xaa , \yaa}) circle [radius = 0.07] ;
\pgfmathsetmacro{\xminaaa}{3.1} ; \pgfmathsetmacro{\xmaxaaa}{4.4} ; \pgfmathsetmacro{\xaaa}{4.4} ; \pgfmathsetmacro{\yaaa}{4.7} ; 
\draw [violet , thick , line cap = round , smooth , samples = 100 ] plot[domain = \xminaaa : \xmaxaaa ] (\x, { \yaaa + 0.02 * cos(\x * 3000) }) ; 
\draw [violet , fill] ({\xaaa , \yaaa}) circle [radius = 0.07] ;
\pgfmathsetmacro{\xminaaaa}{3.8} ; \pgfmathsetmacro{\xmaxaaaa}{4.9} ; \pgfmathsetmacro{\xaaaa}{3.8} ; \pgfmathsetmacro{\yaaaa}{3.8} ; 
\draw [teal , thick , line cap = round , smooth , samples = 100 ] plot[domain = \xminaaaa : \xmaxaaaa ] (\x, { \yaaaa + 0.02 * cos(\x * 2800) }) ; 
\draw [teal , fill] ({\xaaaa , \yaaaa}) circle [radius = 0.07] ;
\pgfmathsetmacro{\xminaaaaa}{4.5} ; \pgfmathsetmacro{\xmaxaaaaa}{5.5} ; \pgfmathsetmacro{\xaaaaa}{5.5} ; \pgfmathsetmacro{\yaaaaa}{0.8} ; 
\draw [orange , thick , line cap = round , smooth , samples = 100 ] plot[domain = \xminaaaaa : \xmaxaaaaa ] (\x, { \yaaaaa + 0.02 * cos(\x * 2800) }) ; 
\draw [orange , fill] ({\xaaaaa , \yaaaaa}) circle [radius = 0.07] ;
\pgfmathsetmacro{\xposgamma}{\xmaxdecal} ; \pgfmathsetmacro{\yposgamma}{0} ;
\draw (\xposgamma , \yposgamma) node[font = \normalsize , below] {$\gamma_{\scriptstyle \omega_{\text{\scriptsize s}}}$} ;
\pgfmathsetmacro{\xposbarrier}{\xmaxdecalaaa} ; \pgfmathsetmacro{\yposbarrier}{0} ; 
\draw (\xposbarrier , \yposbarrier) node[font = \normalsize , below] {$\text{\small Sch.}$} ; 
\pgfmathsetmacro{\xminomega}{1.3} ; \pgfmathsetmacro{\xmaxomega}{2.5} ; \pgfmathsetmacro{\yomega}{\ymax+0.1} ;
\draw [<-] [thick , smooth , line cap = round] (\xminomega , \yomega) -- (\xmaxomega , \yomega) ;
\draw ({(\xminomega+\xmaxomega) * 0.5} , \yomega) node[font = \normalsize , above] {$\Omega^{\text{\scriptsize s}}$} ; 
\end{tikzpicture}
\caption{\small{Illustration of the individual dynamics of stars in the $(j , a) \!=\! (L/I , I^{2} /G M_{\bullet})$ space, as given by the Langevin equation~\eqref{Langevin_disc}. The grey region corresponds to the capture region, within which stars inevitably sink into the BH. As $I$ is an invariant of the diffusion (see equation~\eqref{Langevin_disc}), stars' diffusion is one-dimensional, conserves $a$, and occurs only in the ${j-}$direction. The background dotted curves illustrate the contour lines of the precession frequency given by the function ${ (j , a) \!\mapsto\! \Omega^{\rm s} (j , a) }$. As illustrated in figure~\ref{fig_Resonance_Barrier}, the precession frequencies get larger as particles approach the central BH, due to the contributions from the relativistic precession frequencies. The blue and red orbits precess at the same frequency $\omega_{\rm s}$, so that they belong to the same critical resonant line $\gamma_{\omega_{\rm s}}$, which allows them to resonate one with another. As the precession frequencies diverge in the vicinity of the BH, such resonant couplings are significantly less likely as stars get closer to the BH, which effectively creates a diffusion barrier in action space, the so-called Schwarzschild barrier.
}}
\label{fig_Illustration_Barrier}
\end{center}
\end{figure}

A qualitative description of the dynamics of  individual orbits from equation~\eqref{Langevin_disc} is illustrated in figure~\ref{fig_Illustration_Barrier}.
Following the representations from~\cite{BarOrAlexander2016}, figure~\ref{fig_Illustration_Barrier} represents the diffusion of stars in the ${ (j , a) \!=\! (L/I , I^{2} / (G M_{\bullet})) }$ space. As observed in equation~\eqref{Langevin_disc}, the fast action $I$ of the stars is conserved during the diffusion, so that stars diffuse only in the ${j-}$direction along ${ a \!=\! \text{cst.} }$ lines. When diffusing, individual particles may resonate with stars which precess at the same frequency, such as the blue and red particles in figure~\ref{fig_Illustration_Barrier}. However, as already illustrated in figure~\ref{fig_Resonance_Barrier}, the precession frequencies diverge as stars get closer to the BH. This increase in the precession frequencies will then forbid any resonant coupling between a star in this internal region and stars belonging the disc itself, where precession frequencies are much smaller. Resonances becoming impossible, the diffusion is stopped and stars cannot keep diffusing closer to the central BH. This suppression of the diffusion is the Schwarzschild barrier. A quantitative illustration of this damping of resonant couplings is postponed to a later paper, where we will effectively compute the precession frequencies ${ \Omega^{\rm s}_{\rm tot} \!=\! \Omega^{\rm s}_{\rm rel} \!+\! \Omega^{\rm s}_{\rm self} }$ in action space for a physically motivated razor-thin axisymmetric disc.

\subsection{Evolution of BH mass and spin}

The calculations above successfully explain the existence of the
so-called Schwarzschild barrier, which strongly suppresses the supply
of tightly bound matter to the black hole. We note that the analysis
of~\cite{Merritt2011} suggests that, in practice, this suppression is
probably tempered by simple
two-body relaxation (not accounted for in the orbit-averaged
approach followed in this paper), which provides an additional
mechanism for transporting stars even closer to the BH,
once resonant relaxation becomes inefficient.
This mechanism was recently demonstrated in detail in~\cite{BarOrAlexander2016}, which showed that adiabatic invariance (in other words the damping of resonant relaxation) limits the effects of resonant relaxation to a region well away of the loss lines, so that the dynamics of accretion of stars by the BH is only very moderately affected by the presence of resonances.
Nevertheless, one can calculate the rate at which stars are
transported across any boundary in phase space within which resonant relaxation
dominates, which is important for quantifying the growth rate of the central black hole.
Consider then a fixed boundary $\mathcal{S}$ in action space.
From the divergence theorem, the flux of mass, ${ \rd M / \rd
  \tau }$ through that boundary,  $\mathcal{S}$, due to secular
diffusion is proportional to
\begin{equation}
\frac{\rd M}{\rd \tau} \propto \! \sum_{\bm{m}^{\rm s}} \!\! \int_{\mathcal{S}} \!\!\! \rd S (\bm{m}^{\rm s} \!\cdot\! \bm{n}) \left[ A_{\bm{m}^{\rm s}} (\bm{J}) \, \oF (\bm{J}) \!+\! D_{\bm{m}^{\rm s}} (\bm{J}) \, \bm{m}^{\rm s} \!\cdot\! \frac{\partial \oF}{\partial \bm{J}^{\rm s}} \!\right] ,
\label{flux_through_S}
\end{equation}
where $\bm{n}$ is the exterior pointing normal vector. In equation~\eqref{flux_through_S}, one can note that the contribution from a given resonance $\bm{m}^{\rm s}$ takes the form of a preferential diffusion in the direction of $\bm{m}^{\rm s}$. This diffusion is therefore anisotropic because it is maximum for ${ \bm{n} \!\propto\! \bm{m}^{\rm s} }$ and equal to $0$ for ${ \bm{n} \!\cdot\! \bm{m}^{\rm s} \!=\! 0 }$. 
We note that if a set of stars of various masses or black holes orbit the
galactic centre, the net flux of each component can also be computed
via equation~\eqref{LB_multi_drift_diff} as
\begin{equation}
\frac{\rd M_{\ra}}{\rd \tau} \!\propto\! 
 \! \sum_{\bm{m}^{\rm s}} \!\! \int_{\mathcal{S}} \!\!
 \rd S (\bm{m}^{\rm s} \!\cdot \bm{n}) \bigg[   \!\!\sum_{\rb} \!\bigg\{ \eta_{\ra} A_{\bm{m}^{\rm s}}^{\rb} (\bm{J}) \, \oFa (\bm{J}) \!+\! \eta_{\rb} D_{\bm{m}^{\rm s}}^{\rb} (\bm{J}) \, \bm{m}^{\rm s} \!\cdot\! \frac{\partial \oFa}{\partial \bm{J}^{\rm s}} \bigg\} \bigg]\,.
\label{multi_flux_through_S}  \nonumber
\end{equation}
This is likely to be of particular interest for predicting the
distribution of heavy compact remnants, which, from
equipartition arguments, are expected to sink more rapidly towards
the centre.
Similarly, the flux of angular momentum, ${ \rd \bm{L} / \rd \tau }$, can be computed, and is proportional to
 \begin{equation}
 \frac{\rd \bm{L}}{\rd \tau} \propto \! \sum_{\bm{m}^{\rm s}} \!\! \int_{\mathcal{S}} \!\!\! \rd S (\bm{m}^{\rm s} \!\cdot\! \bm{n})\,\bm{L}\, \left[ A_{\bm{m}^{\rm s}} (\bm{J}) \, \oF (\bm{J}) \!+\! D_{\bm{m}^{\rm s}} (\bm{J}) \, \bm{m}^{\rm s} \!\cdot\! \frac{\partial \oF}{\partial \bm{J}^{\rm s}} \!\right].
 \label{AM_flux_through_S}
 \end{equation}
and could contribute to either spinning up or down the central black hole, once the self-consistent evolution of the black hole's spin and loss of angular momentum via gravitational wave emission are taken care of.
We note that if the disc is sufficiently self-gravitating, the diffusion
in action space is likely to be dominated by a specific
resonance~\citep[as was shown in][]{fouvry2}.

\section{Discussion and conclusion}
\label{sec:conclusion}

Supermassive black holes absorb stars and debris whose orbits reach the loss-cone, the region of phase space corresponding to orbits on which they are either taken directly into the black hole or close enough to interact strongly with it.
Such accretion affects the secular evolution of the SMBH's mass and spin, which is of interest in understanding black hole demographics and AGN feedback~\citep{Volonteri2016}.
It also affects the matter that remains. For instance, the continuous loss of stars can resupply and reshape the central stellar distribution~\citep[e.g.][]{genzel2000}. 
These dynamical processes have observable signatures, such as binary capture and hyper-velocity star ejection~\citep{Hyper-velocity}, the tidal heating and disruption of stars~\citep{Rees1976}, gravitational waves produced by inspiraling compact remnants~\citep{LIGOdetect}. All these signatures provide possible indirect evidence of the existence of the black hole and offer the opportunity of probing the theory of relativity in the strong field limit~\citep{blanchet2014}. Understanding the dynamics of stars in the vicinity of supermassive black holes is in fact one of the prime goal of the new generation of interferometers such as Gravity~\citep{gravity}.

In this paper, we have specialised the recently developed kinetic theory of
self-gravitating systems of $N$
particles~\citep{Heyvaerts2010,Chavanis2012} to quasi-Keplerian
systems dominated by a massive central object, deriving the
equation that governs the secular evolution of such systems to leading
order in ${ 1/N }$.
The self-consistent dressed equations
(equation~\eqref{BL_Kepler} and its multi-component and stochastic
counterparts, equations~\eqref{LB_multi} and~\eqref{Langevin_equation}
respectively) account for the dynamical
degeneracies in quasi-Keplerian systems.
Because purely Keplerian orbits do not precess, the dynamical
evolution of such degenerate systems may differ significantly
from that of fully self-gravitating systems, such as discs and spheroids.
In particular, to a good approximation stars behave as if they were
smeared out into orbit-averaged Keplerian wires and the evolution of
the system modelled by following the dressed interactions among such wires.
The coupling among these wires generates sequences of uncorrelated transient polarised density waves, which make the underlying stars' orbits diffuse in phase space.

The quasi-Keplerian Balescu-Lenard equation~\eqref{BL_Kepler} is quadratic in the phase-averaged distribution function and 
describes 
i) the self-gravity of the orbiting particles,
ii) the discreteness of the cluster,
iii) the resonances between such orbits,
iv) a full spectrum of masses, via equation~\eqref{LB_multi},
and v) possible post-Newtonian corrections, including relativistic precession induced by the rotation of the central black hole, if present. 
These last effects are encoded in the frequency shifts occurring in the resonance condition from the diffusion and drift coefficients.
It is therefore the quasi-linear self-consistent master equation quantifying the effect of resonant relaxation.
As such it provides a very rich framework to describe the evolution of galactic centres for cosmic times, or the
secular evolution of debris discs -- which is an interesting venue in
the context of planet formation~\citep[e.g.][]{Tremaine1998}.

A key step in the derivation of this equation is the phase averaging of the first two equations of the BBGKY hierarchy over the fast angles associated with the orbital motion of the bodies on their Keplerian orbits. 
In order to derive equations~\eqref{BL_Kepler} and~\eqref{LB_multi}, we assumed that the (spherical or coplanar) cluster was dynamically relaxed at every stage of its secular evolution.
As the equations are averaged over the Keplerian fast angles, the corresponding actions are adiabatically preserved. Because of this phase average, the Keplerian Balescu-Lenard equation cannot capture mean motion resonances.
Hence a limitation of the present formalism is that it is restricted
to systems with a high degree of symmetry.

More generally, the averaging over fast angles means that traditional
non-resonant two-body relaxation is not accounted for in the
Balescu-Lenard equations we derive here. This is usually
appropriate though, because the derivation of these equations ignores terms
of order ${ O(1/N^{2}) }$, which means that they are valid only on
timescales ${ \!\lesssim\! N t_{\rd} }$, where $t_{\rd}$ is the
dynamical timescale. Such timescales are typically expected to be much shorter
than the non-resonant two-body relaxation time. When investigating specifically the vicinity of supermassive black holes, we found that the quasi-Keplerian Balescu-Lenard equation captures naturally the presence of a Schwarzschild barrier,  explains why it is not fully impermeable, 
and why it allows us to estimate for instance the mass and angular momentum fluxes  of each component through its boundary.
In its multi-component formulation, the Balescu-Lenard equation  also captures mass segregation and radial migration as entropy increases.

\subsection{Comparison to other work}

A number of other recent papers have tackled the dynamics of
quasi-Keplerian stellar systems. The closest to the present paper is
the recent sequence of papers by Sridhar \& Touma~\citep{ST1,ST2}, who
have already obtained equations equivalent to our equations~\eqref{BBGKY_1_final} and~\eqref{BBGKY_2_final}
following a different route starting from the approach of~\cite{Gilbert1968}, which itself extended the work of~\cite{Balescu1960,Lenard1960} from plasma physics. The ``passive
response'' approximation they make in their analysis of razor-thin
axisymmetric discs~\citep{ST3} corresponds to the Landau limit in
which one uses the bare susceptibility
coefficients from equation~\eqref{A_tot_disc} in the Balescu-Lenard
equation~\eqref{BL_Kepler_disc}.

Another way of modelling such dynamics is by using some form of Monte Carlo
approach in which the noise due to the discrete number of stars is treated as an
externally imposed perturbation~\citep[e.g.][]{Madigan2011,BarOrAlexander2014}.
This basic idea is very powerful, particularly if one wants to investigate additional
perturbations that are genuinely external to the cluster.
For example, the ${\eta-}$formalism introduced recently in~\cite{BarOrAlexander2014}
and implemented in detail in~\cite{BarOrAlexander2016} is one such scheme. Imposing plausible constraints on the power spectrum of the discreteness noise, these papers recovered the location of the Schwarzschild barrier (explained in terms of adiabatic invariance), and investigated the role of ${2-}$body relaxation for the loss-cone problem. They showed that on longer timescales, ${2-}$body non-resonant relaxation completely erases the Schwarzschild barrier, and also argued that resonant relaxation is effective only in a restricted region of action space away from the loss-lines, so that its overall effect on plunge rates is small.

The Balescu-Lenard equation has a couple of important conceptual advantages over the ${\eta-}$formalism (and similar Monte Carlo schemes).  
First, the ${\eta-}$formalism requires assumptions about the statistical characteristics of the externally imposed discreteness noise felt by each wire. The Balescu-Lenard equation requires no such external input, because the system's discreteness is described self-consistently.
Second, in the ${\eta-}$formalism the self-gravity of the response to the noise is difficult to account for. In the Balescu-Lenard equation, this full response is naturally present in the dressed susceptibility coefficients (equation~\eqref{definition_1/D}). Such collective effects can be crucial in systems close to marginal stability, where the associated polarisation can get very large (see e.g.~\cite{fouvry2} for an illustration in the case of razor-thin stellar discs).
We note that, just as in Monte Carlo schemes, the Balescu-Lenard approach also offers a natural way of including external potential fluctuations (see point III of Section~\ref{sec:_Average_equations}).

At the heart of the ${\eta-}$formalism lies a distinction between ``field'' and ``test'' stars: the dynamics of the test stars are followed as they undergo the stochastic perturbations generated by the field stars. Such a split is also used in the restricted ${N-}$body calculations recently presented in~\cite{Hamers2014}: the motion of each field star is followed along their precessing Keplerian orbits (with a precession induced by both relativistic effects and the system's self-consistent potential), but interactions among field stars are ignored. The test stars are then followed by direct integration of their motion in the time-varying potential due to the field stars: it does not rely on the averaging approximation. Such an approach is especially useful in order to get a better grasp of the typical stochastic perturbations generated by the cluster of field stars. Like the ${\eta-}$formalism, it ignores the interactions among field stars (and indeed among test stars) and there is no back-influence of the test stars on the field ones.

\subsection{Future work}

The Langevin formulation of the Balescu-Lenard equation
(Section~\ref{sec:relativisticcrossing_Langevin} and
Appendix~\ref{sec:FP_to_Langevin}) combines the flexibility of
Monte Carlo methods with a self-consistent treatment of the dynamics.
A subsequent improvement is offered by the possibility
of adding the secondary effects of two-body relaxation and gravitational-wave losses to the
resonant relaxation dynamics, on which the present paper was focused.
Eventually, one could evolve jointly the BH and its environment.
This would involve considering that the frequencies in
equation~\eqref{BL_Kepler} are time
dependent, via the variation of the BH's mass and spin as outlined in
Section~\ref{sec:relativisticcrossing}.
In the context of razor-thin axisymmetric discs,
one would compute the drift and diffusion coefficients given by
equation~\eqref{BL_Kepler_disc}, following the steps of~\cite{fouvry2}
transposed to the Keplerian framework. A few difficulties have to be overcome to perform such a computation.
The first is the computation of the wire-wire interaction potential (see e.g.~\cite{ToumaTremaine2009,ToumaSridhar2012}) and its harmonic transform over the slow angles.
Then, in order to account for the system's self-gravity, one has to compute the system's averaged response matrix from equation~\eqref{Fourier_M}, which asks for the integration of a resonant function over action space, a daunting numerical task. Finally, on secular timescales, one has to deal with the self-consistency of the diffusion, so that the system's drift and diffusion coefficients should be updated along the diffusion.
As was shown in the present paper, the net effect of equation~\eqref{BL_Kepler_disc}
will be to induce diffusion along preferred ridges, whose properties
are set by the distribution of stars within the cluster and their
self-gravity.
Depending on their starting point in action space, some orbits will be
driven near the region where the black hole dominates diffusion.
This will allow us to quantify for instance the relative importance of black
hole spin on barrier crossing, and the efficiency at which a
supermassive black hole is fed by its surrounding stellar cluster (as
discussed in section~\ref{sec:relativisticcrossing}).

\begin{acknowledgements}
JBF and CP thank the CNRS Inphyniti programme for funding and the KIAS for hospitality while this project was initiated.
We thank Pierre-Henri Chavanis and Simon Prunet for comments.
Special thanks to Walter Dehnen for his suggestion to consider democratic heliocentric coordinates.
CP thanks Scott Tremaine for stimulating discussions on how to tackle this problem. 
This work has made use of the Horizon cluster, hosted by the Institut d'Astrophysique de Paris.
Special thanks to Stephane Rouberol for customising the cluster for our purposes.
This work is partially supported by the Spin(e) grants ANR-13-BS05-0005 of the French Agence Nationale de la Recherche
(${\text{\url{http://cosmicorigin.org}}}$).
\end{acknowledgements}

\bibliographystyle{aa}
\bibliography{references}

\appendix

\section{Relativistic precessions}
\label{sec:relativistic_precessions}

Let us briefly detail the relativistic precessions encompassed in particular by the averaged potential correction $\oP_{\rr}$ present in equations~\eqref{BBGKY_1_final} and~\eqref{BBGKY_2_final}. In order to obtain explicit expressions for these corrections, the ${3D}$ Delaunay variables from equation~\eqref{Delaunay_3D_actions} will be used. In addition, we assume here that the spin of the BH is aligned with the ${z-}$direction and introduce the BH's spin parameter ${ 0 \!\leq\! s \!\leq\! 1 }$. To recover the expression of these precession frequencies, let us follow~\cite{AstrophysicalBlackHoles}. Equation~(5.103) therein gives us that during one Keplerian orbit of duration ${ T_{\rm Kep} \!=\! 2 \pi / \Omega_{\rm Kep} \!=\! 2 \pi I^{3} / (G M_{\bullet})^{2} }$, the 1PN Schwarzschild precession effect leads to a modification of the slow angle $g$ given by
\begin{equation}
\Delta g_{\rm rel}^{\rm 1PN} = g (T_{\rm Kep}) \!-\! g (0) = \frac{6 \pi G M_{\bullet}}{c^{2} a (1 \!-\! e^{2})} \, . 
\label{Delta_g_1PN}
\end{equation}
This is straightforwardly associated with a precession frequency ${ \dot{g}_{\rm rel}^{\rm 1PN} \!=\! \Delta g_{\rm rel}^{\rm 1PN} / T_{\rm Kep} }$ given by
\begin{equation}
\dot{g}_{\rm rel}^{\rm 1PN} = \frac{3 (G M_{\bullet})^{4}}{c^{2} I^{3} L^{2}} = \frac{\partial H_{\rm rel}^{\rm 1PN}}{\partial L} \, ,
\label{1PN_precession}
\end{equation}
where the semi-major axis $a$ and the eccentricity $e$ satisfy ${ a \!=\! I^{2} / (G M_{\bullet}) }$ and ${ 1 \!-\! e^{2} \!=\! (L / I)^{2} }$. We also introduced the Hamiltonian $H_{\rm rel}^{\rm 1PN}$ as
\begin{equation}
H_{\rm rel}^{\rm 1PN} (I , L) = - \frac{3 (G M_{\bullet})^{4}}{c^{2}} \frac{1}{I^{3} L} \, .
\label{definition_H_rel_1}
\end{equation}
Similarly, equation (5.118) of~\cite{AstrophysicalBlackHoles} gives that the 1.5PN Lense-Thirring precession during one Keplerian orbit leads to a precession of the slow angle $g$ given by
\begin{equation}
\Delta g_{\rm rel}^{\rm 1.5PN} = g (T_{\rm Kep}) \!-\! g (0) = - \frac{12 \pi s}{c^{3}} \bigg[ \frac{G M_{\bullet}}{(1 \!-\! e^{2}) a} \bigg]^{3/2} \! \cos (i) \, ,
\label{Delta_g_1.5PN}
\end{equation}
where it is assumed that the spin of the BH was aligned with the ${z-}$direction.
This is immediately associated with a precession frequency ${ \dot{g}_{\rm rel}^{\rm 1.5PN} }$ given by
\begin{equation}
\dot{g}_{\rm rel}^{\rm 1.5PN} = - \frac{6s}{c^{3}} \frac{(G M_{\bullet})^{5} L_{z}}{I^{3} L^{4}} = \frac{\partial H_{\rm rel}^{\rm 1.5PN}}{\partial L} \, ,
\label{1.5PN_precession}
\end{equation}
relying on the relation ${ L_{z} \!=\! L \cos (i) }$. Hence the Hamiltonian $H_{\rm rel}^{\rm 1.5PN}$ which accounts for the rotation of the BH reads
\begin{equation}
H_{\rm rel}^{\rm 1.5PN} (I , L , L_{z}) = \frac{2 s (G M_{\bullet})^{5}}{c^{3}} \frac{L_{z}}{I^{3} L^{3}} \, .
\label{definition_H_rel_1.5}
\end{equation}
Such a Hamiltonian also induces relativistic precessions with respect to the second slow angle $h$ associated with $L_{z}$. We do not detail here how these precessions are indeed correctly described by the Hamiltonian $H_{\rm rel}^{\rm 1.5PN}$.
Paying a careful attention to the normalisation prefactors used in equations~\eqref{initial_Hamiltonian},~\eqref{rescale_Phi_a}, and~\eqref{definition_average_Phi_a}, one finally gets the expression of the averaged 1PN and 1.5PN relativistic corrections $\oP_{\rr}$ appearing in equations~\eqref{BBGKY_1_final} and~\eqref{BBGKY_2_final}. These read
\begin{equation}
\oP_{\rr} (I , L , L_{z}) = \frac{1}{(2 \pi)^{d -k}} \frac{M_{\bullet}}{M_{\star}} \bigg[ H_{\rm rel}^{\rm 1PN} (I , L ) \!+\! H_{\rm rel}^{\rm 1.5PN} (I , L , L_{z}) \bigg] \, .
\label{sum_up_relativistic}
\end{equation}
From this potential correction, following equation~\eqref{definition_precession_frequencies}, one can immediately compute the associated precession frequencies $\bm{\Omega}^{\rm s}_{\rm rel}$ with respect to the slow angles $\bm{\theta}^{\rm s}$. They read
\begin{equation}
\bm{\Omega}^{\rm s}_{\rm rel} \!=\! \frac{\partial  \oP_{\rr} }{\partial \bm{J}^{\rm s}} \!=\! \frac{{M_{\bullet}}}{\!(2 \pi)^{d -k}\!}   \frac{(G M_{\bullet})^{4}}{{M_{\star}} c^{2}}\frac{\partial}{\partial \bm{J}^{\rm s}}   \bigg[ \!-\! \frac{3}{I^{3} L} \!+\! \frac{2 G M_{\bullet}\!}{c} \frac{s L_{z}}{I^{3} L^{3}}  \bigg] \, .
\label{BL_rel_freq}
\end{equation}
Note finally that gravitational waves emissions, along with the associated dissipations, are not considered here.

\section{The degenerate collisional equation}
\label{sec:secular_collisional_derivation}
For completeness,
let us revisit the derivation of the Balescu-Lenard equation presented in~\cite{Heyvaerts2010} in this new quasi-Keplerian regime.
The starting point of this derivation is the two coupled averaged equations~\eqref{BBGKY_1_final} and~\eqref{BBGKY_2_final}, which involve the system's averaged ${1-}$body DF $\oF$, and its averaged ${2-}$body autocorrelation $\oC$. The heart of the present derivation is the following: first one must solve the evolution equation~\eqref{BBGKY_2_final}, so as to obtain ${ \oC \!=\! \oC \big[ \oF \big] }$ (Section~\ref{sec:solving_autocorrelation}). Injecting this relation in equation~\eqref{BBGKY_1_final}, one obtains a closed kinetic equation involving $\oF$ only.
Its simplification will be carried out in section~\ref{sec:rewriting_collision}.
Section~\ref{sec:multicase} will present the specifics of the corresponding multi-component derivation.
\subsection{Solving for the autocorrelation}
\label{sec:solving_autocorrelation}

With the assumption of stationarity from equation~\eqref{adiabatic_approximation} and the Bogoliubov's ansatz from equation~\eqref{shape_equilibria}, one can rewrite equation~\eqref{BBGKY_1_final} as
\begin{equation}
\frac{\partial \oF}{\partial \tau} = \mathscr{C} \big[ \oF \big] \, ,
\label{rewrite_BBGKY_1_solving}
\end{equation}
where the collision operator ${\mathscr{C}}$ is introduced as
\begin{align}
\mathscr{C} \big[ \oF \big] & \, = - \frac{1}{N} \!\! \int \!\! \rd \bR_{2} \, \big[ \oC(\bR_{1} , \bR_{2}) , \oU_{12} \big]_{(1)}   \nonumber
\\
& \, = \frac{1}{N} \!\! \int \!\! \rd \bR_{2} \, \frac{\partial }{\partial \bm{J}_{1}^{\rm s}} \!\cdot\! \bigg[ \!\! \int \!\! \frac{\rd \bm{\theta}_{1}^{\rm s}}{(2 \pi)^{k}} \, \frac{\partial \oU_{12}}{\partial \bm{\theta}_{1}^{\rm s}} \, \oC (\bR_{1} , \bR_{2}) \bigg] \, ,
\label{definition_collision_operator}
\end{align}
relying on the fact that during the secular diffusion, $\oF$ is of the form ${ \oF \!=\! \oF (\bm{J}_{1}) }$, allowing us to perform an angle-average with respect to $\bm{\theta}_{1}^{\rm s}$.
Similarly, relying on the definition of the precession frequencies from equation~\eqref{definition_precession_frequencies}, equation~\eqref{BBGKY_2_final} can be rewritten as
\begin{align}
& \, \frac{\partial \oC}{\partial \tau} \!+\! \bm{\Omega}_{1}^{\rm s} \!\cdot\! \frac{\partial \oC}{\partial \bm{\theta}_{1}^{\rm s}} \!+\! \bm{\Omega}_{2}^{\rm s} \!\cdot\! \frac{\partial \oC}{\partial \bm{\theta}_{2}^{\rm s}} - \!\! \int \!\! \rd \bR_{3} \, \oC (\bR_{2} , \bR_{3}) \, \frac{\partial \oF}{\partial \bm{J}_{1}^{\rm s}} \!\cdot\! \frac{\partial \oU_{13}}{\partial \bm{\theta}_{1}^{\rm s}}   \nonumber
\\
& \, \!-\! \!\! \int \!\! \rd \bR_{3} \, \oC (\bR_{1} , \bR_{3}) \, \frac{\partial \oF}{\partial \bm{J}_{2}^{\rm s}} \!\cdot\! \frac{\partial \oU_{23}}{\partial \bm{\theta}_{2}^{\rm s}} = S_{2} (\bR_{1} , \bR_{2} , \tau) \, ,
\label{rewrite_BBGKY_2_solving}
\end{align}
where in the r.h.s. the source term ${ S_{2} (\bR_{1} , \bR_{2} , \tau) }$ obeys
\begin{equation}
S_{2} (\bR_{1} , \bR_{2} , \tau) = \frac{1}{(2 \pi)^{d - k}} \frac{\partial \oU_{12}}{\partial \bm{\theta}_{1}^{\rm s}} \!\cdot\! \bigg[ \frac{\partial }{\partial \bm{J}_{1}^{\rm s}} \!-\! \frac{\partial }{\partial \bm{J}_{2}^{\rm s}} \bigg] \, \oF (\bm{J}_{1}) \, \oF (\bm{J}_{2}) \, .
\label{definition_source}
\end{equation}
Notice that equation~\eqref{rewrite_BBGKY_2_solving} is linear in $\oC$, symmetric in 1 and 2, and can therefore be solved by working out the Green's function, ${ \mathcal{G}^{(2)} (\bR_{1} , \bR_{2} , \bR_{1}' , \bR_{2}' , \tau') }$, associated with the linear differential operator of the l.h.s. of equation~\eqref{rewrite_BBGKY_2_solving}. The solution for ${ \oC (\bR_{1} , \bR_{2} , \tau) }$ may therefore be written as
\begin{align}
\oC (\bR_{1} , \bR_{2} , \tau) = \!\! \int_{0}^{+ \infty} \!\!\!\!\!\! \rd \tau' \!\! \int \!\! \rd \bR_{1}' \rd \bR_{2}' \, & \, \mathcal{G}^{(2)} (\bR_{1} , \bR_{2} , \bR_{1}' , \bR_{2}' , \tau')   \nonumber
\\
& \, \times S_{2} (\bR_{1}' , \bR_{2}' , \tau \!-\! \tau') \, .
\label{generic_solution_C}
\end{align}
Injecting equation~\eqref{generic_solution_C} into equation~\eqref{rewrite_BBGKY_2_solving}, one gets the propagation equation satisfied by $\mathcal{G}^{(2)}$. It reads
\begin{align}
 \, \frac{\partial \mathcal{G}^{(2)}}{\partial \tau'} & \!+\! \bm{\Omega}_{1}^{\rm s} \!\cdot\! \frac{\partial \mathcal{G}^{(2)}}{\partial \bm{\theta}_{1}^{\rm s}} \!+\! \bm{\Omega}_{2}^{\rm s} \!\cdot\! \frac{\partial \mathcal{G}^{(2)}}{\partial \bm{\theta}_{2}^{\rm s}}   \nonumber  
\\
& \, \!-\! \!\! \int \!\! \rd \bm{\mathcal{R}}_{3} \, \mathcal{G}^{(2)} (\bm{\mathcal{R}}_{3} , \bm{\mathcal{R}}_{2} , \bm{\mathcal{R}}_{1}' , \bm{\mathcal{R}}_{2}' , \tau') \, \frac{\partial \oF}{\partial \bm{J}_{1}^{\rm s}} \!\cdot\! \frac{\partial \oU_{13}}{\partial \bm{\theta}_{1}^{\rm s}}   \nonumber
\\
& \, \!-\! \!\! \int \!\! \rd \bm{\mathcal{R}}_{3} \, \mathcal{G}^{(2)} (\bm{\mathcal{R}}_{1} , \bm{\mathcal{R}}_{3} , \bm{\mathcal{R}}_{1}' , \bm{\mathcal{R}}_{2}' , \tau') \, \frac{\partial\oF}{\partial \bm{J}_{2}^{\rm s}} \!\cdot\! \frac{\partial \oU_{23}}{\partial \bm{\theta}_{1}^{\rm s}} = 0 \, ,
\label{propagation_equation_G2}
\end{align}
where we assumed that the source term ${ S_{2} (t) }$ was effectively turned on only for ${ t \!\geq\! 0 }$, so that ${ S_{2} (t \!<\! 0) \!=\! 0 }$. In addition, the Green's function $\mathcal{G}^{(2)}$ has to satisfy the initial condition ${ \mathcal{G}^{(2)} (\bR_{1} , \bR_{2} , \bR_{1}' , \bR_{2}' , 0) \!=\! \delta_{\rm D} (\bR_{1} \!-\! \bR_{1}') \, \delta_{\rm D} (\bR_{2} \!-\! \bR_{2}') }$.
When considering equation~\eqref{propagation_equation_G2}, it is worth noting that this propagation equation acts separately on the variables ${ (\bR_{1} , \bR_{1}') }$ and ${ (\bR_{2} , \bR_{2}') }$ (and the initial condition of $\mathcal{G}^{(2)}$ is also separable). We may then solve equation~\eqref{propagation_equation_G2} by factoring the ${2-}$body Green's function as the product of two ${1-}$body Green's function so that
\begin{equation}
\mathcal{G}^{(2)} (\bR_{1} , \bR_{2} , \bR_{1}' , \bR_{2}' , \tau') \!=\! \mathcal{G}^{(1)} (\bR_{1} , \bR_{1}' , \tau') \, \mathcal{G}^{(1)} (\bR_{2} , \bR_{2}' , \tau') \, ,
\label{factorisation_G2}
\end{equation}
where the ${1-}$body Green's function $\mathcal{G}^{(1)}$ satisfies the linearised ${1-}$body Vlasov equation, namely
\begin{align}
 \, \frac{\partial \mathcal{G}^{(1)} (\bR_{1} , \bR_{1}' , \tau')}{\partial \tau '}& \!+\! \bm{\Omega}_{1}^{\rm s} \!\cdot\! \frac{\partial \mathcal{G}^{(1)} (\bR_{1} , \bR_{1}' , \tau')}{\partial \bm{\theta}_{1}^{\rm s}}   \nonumber
\\
& \hskip -1cm \, \!-\! \!\! \int \!\! \rd \bR_{2} \, \mathcal{G}^{(1)} (\bR_{2} , \bR_{1}' , \tau ') \, \frac{\partial \oF}{\partial \bm{J}_{1}^{\rm s}} \!\cdot\! \frac{\partial \oU_{12}}{\partial \bm{\theta}_{1}^{\rm s}} = 0 \, ,
\label{propagation_equation_G1}
\end{align}
with the initial condition ${ \mathcal{G}^{(1)} (\bR_{1} , \bR_{1}' , 0 ) \!=\! \delta_{\rm D} (\bR_{1} \!-\! \bR_{1}') }$.
\cite{Heyvaerts2010} interestingly notes that, if one were to account for contributions associated with strong collisions, such as in the fifth line of equation~\eqref{BBGKY_2_fn}, the property of separability from equation~\eqref{factorisation_G2} would not hold anymore.
Because of causality, equation~\eqref{propagation_equation_G1} only has to be solved for ${ \tau' \!\geq\! 0 }$. To do so, we rely once again on Bogoliubov's ansatz, and assume that the system's ${1-}$body DF $\oF$ evolves on a slow secular collisional timescale $T_{\rm relax}$, while the fluctuations evolve much faster on a secular collisionless timescale $T_{\rm sec}$. As a consequence, in equation~\eqref{propagation_equation_G1}, which describes the evolution of fluctuations, we may assume $\oF$ to be frozen. Therefore, the correlations at a given time $\tau$ can be seen as functionals of $\oF$ evaluated at the very same time. To solve equation~\eqref{propagation_equation_G1}, we introduce the Laplace transfom following the convention
\begin{equation}
\widetilde{f} (\omega) = \!\! \int_{0}^{+ \infty} \!\!\!\!\!\! \rd t \, f (t) \, \re^{\ri \omega t} \;\;\; ; \;\;\; f (t) = \frac{1}{2 \pi} \!\! \int_{\mathcal{B}} \! \rd \omega \, \widetilde{f} (\omega) \, \re^{- \ri \omega t} \, . 
\label{definition_Laplace}
\end{equation}
In equation~\eqref{definition_Laplace}, for the inverse Laplace transform, the Bromwich contour $\mathcal{B}$ in the complex ${\omega-}$plane should pass above all the poles of the integrand, that is ${ \text{Im} [\omega] }$ should be large enough. The Laplace transform of equation~\eqref{propagation_equation_G1} gives
\begin{align}
& \, - \ri \omega \widetilde{\mathcal{G}}^{(1)} (\bR_{1} , \bR_{1}' , \omega) + \bm{\Omega}_{1}^{\rm s} \!\cdot\! \frac{\partial \widetilde{\mathcal{G}}^{(1)} (\bR_{1} , \bR_{1}' , \omega)}{\partial \bm{\theta}_{1}^{\rm s}}  \nonumber
\\
& \, - \!\! \int \!\! \rd \bR_{2} \, \widetilde{\mathcal{G}}^{(1)} (\bR_{2} , \bR_{1}' , \omega) \, \frac{\partial \oF}{\partial \bm{J}_{1}^{\rm s}} \!\cdot\! \frac{\partial \oU_{12}}{\partial \bm{\theta}_{1}^{\rm s}} = \delta_{\rm D} (\bR_{1} \!-\! \bR_{1}') \, ,
\label{Laplace_G1}
\end{align}
where the source term on the r.h.s. comes from the initial condition.
We now perform a Fourier transform with respect to the slow angles $\bm{\theta}_{1}^{\rm s}$ of equation~\eqref{Laplace_G1}, following the convention from equation~\eqref{definition_Fourier_slow_angles}. We multiply equation~\eqref{Laplace_G1} by ${ 1/(2\pi)^{k} \! \int \! \rd \bm{\theta}_{1}^{\rm s} \re^{- \ri \bm{m}_{1}^{\rm s} \cdot \bm{\theta}_{1}^{\rm s}} }$ and get 
\begin{align}
& \, - \ri \omega \widetilde{\mathcal{G}}^{(1)}_{\bm{m}_{1}^{\rm s}} (\bm{J}_{1} , \bR_{1}' , \omega) \!+\! \ri \bm{m}_{1}^{\rm s} \!\cdot\! \bm{\Omega}_{1}^{\rm s} \, \widetilde{\mathcal{G}}^{(1)}_{\bm{m}_{1}^{\rm s}} (\bm{J}_{1} , \bR_{1}' , \omega)   \nonumber
\\
& \, - \ri \bm{m}_{1}^{\rm s} \!\cdot\! \frac{\partial \oF}{\partial \bm{J}_{1}^{\rm s}} (2 \pi)^{k} \sum_{\bm{m}_{2}^{\rm s}} \!\! \int \!\! \rd \bm{J}_{2} \, \widetilde{\mathcal{G}}^{(1)}_{\bm{m}_{2}^{\rm s}} (\bm{J}_{2} , \bR_{1}' , \omega) \, A_{\bm{m}_{1}^{\rm s} , \bm{m}_{2}^{\rm s}} (\bm{J}_{1} , \bm{J}_{2})   \nonumber
\\
& \, \quad \quad = \frac{\re^{- \ri \bm{m}_{1}^{\rm s} \cdot \bm{\theta}_{1}^{\rm s \prime} }}{(2 \pi)^{k}} \, \delta_{\rm D} (\bm{J}_{1} \!-\! \bm{J}_{1}') \, .
\label{Laplace_Fourier_G1}
\end{align}
Equation~\eqref{Laplace_Fourier_G1} introduced the bare susceptibility coefficients $A_{\bm{m}_{1}^{\rm s} , \bm{m}_{2}^{\rm s}}$, associated with the Fourier transform of the interaction potential and defined in equation~\eqref{definition_bare_A}. Equation~\eqref{Laplace_Fourier_G1} can easily be rewritten as
\begin{align}
& \, \widetilde{\mathcal{G}}^{(1)}_{\bm{m}_{1}^{\rm s}} (\bm{J}_{1} , \bR_{1}' , \omega)   \nonumber
\\
& \,  + (2 \pi)^{k} \frac{\bm{m}_{1}^{\rm s} \!\cdot\! \partial \oF / \partial \bm{J}_{1}^{\rm s}}{\omega \!-\! \bm{m}_{1}^{\rm s} \!\cdot\! \bm{\Omega}_{1}^{\rm s}} \sum_{\bm{m}_{2}^{\rm s}} \!\! \int \!\! \rd \bm{J}_{2} \, \widetilde{\mathcal{G}}^{(1)}_{\bm{m}_{2}^{\rm s}} (\bm{J}_{2} , \bR_{1}' , \omega) \, A_{\bm{m}_{1}^{\rm s} , \bm{m}_{2}^{\rm s}} (\bm{J}_{1} , \bm{J}_{2})   \nonumber
\\
& \, \quad \quad = \frac{\ri}{(2 \pi)^{k}} \frac{\re^{- \ri \bm{m}_{1}^{\rm s} \cdot \bm{\theta}_{1}^{\rm s \prime}}}{\omega \!-\! \bm{m}_{1}^{\rm s} \!\cdot\! \bm{\Omega}_{1}^{\rm s}} \, \delta_{\rm D} (\bm{J}_{1} \!-\! \bm{J}_{1} ' ) \, .
\label{Laplace_Fourier_G1_rewrite}
\end{align}
At this stage, it is important to note that equation~\eqref{Laplace_Fourier_G1_rewrite} takes the form of a Fredholm equation for the Green's function $\mathcal{G}^{(1)}_{\bm{m}_{1}^{\rm s}}$, as it appears twice in the l.h.s., in particular under the form of an integral term. The trick to solve such an equation is to rely on Kalnaj's matrix method~\citep{Kalnajs1976II}, and introduce a basis of potential and densities ${ (\psi^{(p)} , \rho^{(p)}) }$ presented in equation~\eqref{definition_basis_Kalnajs}, thanks to which potential perturbations may be decomposed. Let us first decompose the rescaled interaction potential $U$ from equation~\eqref{rescaling_U} on these basis elements. We consider the function ${ \bm{x}_{1} \!\mapsto\! U (|\bm{x}_{1} \!-\! \bm{x}_{2}|) }$ and project it on the basis ${ \psi^{(p)} (\bm{x}_{1}) }$. This takes the form ${ U (|\bm{x}_{1} \!-\! \bm{x}_{2}|) \!=\! \sum_{p} \! u_{p} (\bm{x}_{2}) \, \psi^{(p)} (\bm{x}_{1}) }$, where the coefficients ${ u_{p} (\bm{x}_{2}) }$ are given by
\begin{equation}
u_{p} (\bm{x}_{2}) = - \!\! \int \!\! \rd \bm{x}_{1} \, U ( | \bm{x}_{1} \!-\! \bm{x}_{2} | ) \, \rho^{(p) *} (\bm{x}_{1}) = - \psi^{(p) *} (\bm{x}_{2}) \, .
\label{decomposition_U}
\end{equation}
As they were defined in equation~\eqref{definition_bare_A} as the Fourier transform in angles of the averaged interaction potential $\oU$, the bare susceptibility coefficients ${ A_{\bm{m}_{1}^{\rm s} , \bm{m}_{2}^{\rm s}} (\bm{J}_{1} , \bm{J}_{2}) }$ can immediately be written as
\begin{equation}
A_{\bm{m}_{1}^{\rm s} , \bm{m}_{2}^{\rm s}} (\bm{J}_{1} , \bm{J}_{2}) = -  \sum_{p} \opsi_{\bm{m}_{1}^{\rm s}}^{(p)} (\bm{J}_{1}) \, \opsi^{(p) *}_{\bm{m}_{2}^{\rm s}} (\bm{J}_{2}) \, ,
\label{simpler_A}
\end{equation}
where the averaged Fourier-transformed basis elements were introduced in equation~\eqref{definition_Fourier_slow_angles}. In order to invert equation~\eqref{Laplace_Fourier_G1_rewrite}, we perform on $\widetilde{\mathcal{G}}_{\bm{m}_{1}^{\rm s}}^{(1)}$ the same operations than those acting on $\widetilde{\mathcal{G}}_{\bm{m}_{2}^{\rm s}}^{(1)}$. This amounts to multiplying equation~\eqref{Laplace_Fourier_G1_rewrite} by ${ (2 \pi)^{k} \sum_{\bm{m}_{1}^{\rm s}} \! \int \! \rd \bm{J}_{1} \psi_{\bm{m}_{1}^{\rm s}}^{(q) *} (\bm{J}_{1}) }$, so that it becomes
\begin{align}
& \bigg[ (2 \pi)^{k} \sum_{\bm{m}_{1}^{\rm s}} \!\! \int \!\! \rd \bm{J}_{1} \, \widetilde{\mathcal{G}}^{(1)}_{\bm{m}_{1}^{\rm s}} (\bm{J}_{1} , \bR_{1}' , \omega) \,  \opsi_{\bm{m}_{1}^{\rm s}}^{(q) *} (\bm{J}_{1}) \bigg]   \nonumber
\\
& - \sum_{p} \bigg\{ \bigg[ (2 \pi)^{k} \sum_{\bm{m}_{1}^{\rm s}} \!\! \int \!\! \rd \bm{J}_{1} \, \frac{\bm{m}_{1}^{\rm s} \!\cdot\! \partial \oF / \partial \bm{J}_{1}^{\rm s}}{\omega \!-\! \bm{m}_{1}^{\rm s} \!\cdot\! \bm{\Omega}_{1}^{\rm s}} \opsi^{(p)}_{\bm{m}_{1}^{\rm s}} (\bm{J}_{1}) \, \opsi^{(q) *}_{\bm{m}_{1}^{\rm s}} (\bm{J}_{1}) \bigg]   \nonumber
\\
& \times \bigg[ (2 \pi)^{k} \sum_{\bm{m}_{2}^{\rm s}} \!\! \int \!\! \rd \bm{J}_{2} \, \widetilde{\mathcal{G}}^{(1)}_{\bm{m}_{2}^{\rm s}} (\bm{J}_{2} , \bR_{1}' , \omega) \, \opsi_{\bm{m}_{2}^{\rm s}}^{(p) *} (\bm{J}_{2})  \bigg] \bigg\}   \nonumber
\\
& \, \;\;\;\;\;\;\;\;\;\;\; = \sum_{\bm{m}_{1}^{\rm s}} \frac{\ri \re^{- \ri \bm{m}_{1}^{\rm s} \cdot \bm{\theta}_{1}^{\rm s \prime}}}{\omega \!-\! \bm{m}_{1}^{\rm s} \!\cdot\! \bm{\Omega}_{1}^{\rm s \prime}} \, \opsi_{\bm{m}_{1}^{\rm s}}^{(q) *} (\bm{J}_{1}') \, .
\label{calculation_inversion_G1}
\end{align}
In order to further simplify equation~\eqref{calculation_inversion_G1}, let us introduce the notations
\begin{align}
& \, K_{p} (\bR_{1}' , \omega) = (2 \pi)^{k} \sum_{\bm{m}^{\rm s}} \!\! \int \!\! \rd \bm{J} \, \widetilde{\mathcal{G}}^{(1)}_{\bm{m}^{\rm s}} (\bm{J} , \bR_{1}' , \omega) \, \opsi_{\bm{m}^{\rm s}}^{(p) *} (\bm{J}) \, , \nonumber
\\
& \, L_{p} (\bR_{1}' , \omega) = \sum_{\bm{m}^{\rm s}} \frac{\ri \re^{- \ri \bm{m}^{\rm s} \cdot \bm{\theta}_{1}^{\rm s \prime}}}{\omega \!-\! \bm{m}^{\rm s} \!\cdot\! \bm{\Omega}_{1}^{\rm s \prime}} \, \opsi_{\bm{m}^{\rm s}}^{(p) *} (\bm{J}_{1}') \, .
\label{short_notation_K_L}
\end{align}
Using the response matrix $\widehat{\mathbf{M}}$ introduced in equation~\eqref{Fourier_M}, equation~\eqref{calculation_inversion_G1} then becomes
\begin{equation}
K_{q} (\bR_{1}' , \omega) - \sum_{p} \widehat{\mathbf{M}}_{qp} (\omega) \, K_{p} (\bR_{1}' , \omega) = L_{q} (\bR_{1}' , \omega) \, .
\label{rewrite_inversion_G1}
\end{equation}
Assuming that the system always remains dynamically stable, so that ${ \big[ \mathbf{I} \!-\! \widehat{\mathbf{M}} (\omega) \big] }$ can be inverted (where $\mathbf{I}$ stands for the identity matrix), equation~\eqref{rewrite_inversion_G1} leads to
\begin{equation}
K_{q} (\bR_{1}' , \omega) = \sum_{p} \big[ \mathbf{I} \!-\! \widehat{\mathbf{M}} (\omega) \big]^{-1}_{qp} \, L_{p} (\bR_{1}' , \omega) \, .
\label{inversion_G1_K}
\end{equation}
Injecting this relation into equation~\eqref{Laplace_Fourier_G1_rewrite}, ${ \widetilde{\mathcal{G}}_{\bm{m}_{1}^{\rm s}}^{(1)} }$ can be written as
\begin{align}
& \, \widetilde{\mathcal{G}}^{(1)}_{\bm{m}_{1}^{\rm s}} (\bm{J}_{1} , \bR_{1}' , \omega) = \frac{1}{(2 \pi)^{k}} \frac{\ri \re^{- \ri \bm{m}_{1}^{\rm s} \cdot \bm{\theta}_{1}^{\rm s \prime}}}{\omega \!-\! \bm{m}_{1}^{\rm s} \!\cdot\! \bm{\Omega}_{1}^{\rm s}} \delta_{\rm D} (\bm{J}_{1} \!-\! \bm{J}_{1}')   \nonumber
\\
& \, + \frac{\bm{m}_{1}^{\rm s} \!\cdot\! \partial \oF / \partial \bm{J}_{1}^{\rm s}}{\omega \!-\! \bm{m}_{1}^{\rm s} \!\cdot\! \bm{\Omega}_{1}^{\rm s}} \sum_{\bm{m}_{1}^{\rm s \prime}} \frac{1}{\mathcal{D}_{\bm{m}_{1}^{\rm s} , \bm{m}_{1}^{\rm s \prime}} (\bm{J}_{1} , \bm{J}_{1}' , \omega)} \frac{\ri \re^{- \ri \bm{m}_{1}^{\rm s \prime} \cdot \bm{\theta}_{1}^{\rm s \prime}}}{\omega \!-\! \bm{m}_{1}^{\rm s \prime} \!\cdot\! \bm{\Omega}_{1}^{\rm s \prime}} \, ,
\label{inversion_G1}
\end{align}
where the dressed susceptibility coefficients ${ 1 / \mathcal{D}_{\bm{m}_{1}^{\rm s} , \bm{m}_{1}^{\rm s \prime}} }$ were introduced in equation~\eqref{definition_1/D}. Relying on the inverse Fourier transform in angles from equation~\eqref{definition_Fourier_slow_angles}, we finally obtain the expression of the ${1-}$body Green's function ${ \widetilde{\mathcal{G}}^{(1)} }$ as
\begin{align}
\widetilde{\mathcal{G}}^{(1)} (\bR_{1} , \bR_{1} ' , \omega) & \, = \sum_{\bm{m}_{1}^{\rm s} , \bm{m}_{1}^{\rm s \prime}} \frac{\ri \re^{\ri (\bm{m}_{1}^{\rm s} \cdot \bm{\theta}_{1}^{\rm s} - \bm{m}_{1}^{\rm s \prime} \cdot \bm{\theta}_{1}^{\rm s \prime})}}{\omega \!-\! \bm{m}_{1}^{\rm s} \!\cdot\! \bm{\Omega}_{1}^{\rm s}} \bigg[ \frac{\delta_{\bm{m}_{1}^{\rm s}}^{\bm{m}_{1}^{\rm s \prime}}}{(2 \pi)^{k}} \delta_{\rm D} (\bm{J}_{1} \!-\! \bm{J}_{1}')   \nonumber
\\
& \, \!+\! \frac{\bm{m}_{1}^{\rm s} \!\cdot\! \partial \oF / \partial \bm{J}_{1}^{\rm s}}{(\omega \!-\! \bm{m}_{1}^{\rm s \prime} \!\cdot\! \bm{\Omega}_{1}^{\rm s \prime}) \mathcal{D}_{\bm{m}_{1}^{\rm s} , \bm{m}_{1}^{\rm s \prime}} (\bm{J}_{1} , \bm{J}_{1} ' , \omega)} \bigg]   \nonumber
\\
& \, = \sum_{\bm{m}_{1}^{\rm s} , \bm{m}_{1}^{\rm s \prime}} \widetilde{\mathcal{G}}^{(1)}_{\bm{m}_{1}^{\rm s} , \bm{m}_{1}^{\rm s \prime}} (\bm{J}_{1} , \bm{J}_{1} ' , \omega) \, \re^{\ri (\bm{m}_{1}^{\rm s} \cdot \bm{\theta}_{1}^{\rm s} - \bm{m}_{1}^{\rm s \prime} \cdot \bm{\theta}_{1}^{\rm s \prime})} \, .
\label{final_inversion_G1}
\end{align}

\subsection{Simplifying the collision operator}
\label{sec:rewriting_collision}

Given the explicit calculation of the ${1-}$body Green's function in equation~\eqref{final_inversion_G1}, one may proceed to the evaluation of the collision operator from equation~\eqref{definition_collision_operator}. Relying again on Bogoliubov's ansatz in equation~\eqref{generic_solution_C}, we may perform the replacement ${ S_{2} (\bR_{1}' , \bR_{2}', \tau \!-\! \tau') \!\to\! S_{2} (\bR_{1} , \bR_{1}' , \tau) }$. Given the factorisation of the Green's function from equation~\eqref{factorisation_G2}, and the inverse Laplace transform from equation~\eqref{definition_Laplace}, the collision operator then takes the form
\begin{align}
\mathscr{C} \big[ \oF \big] = & \, \!\! \int_{0}^{+ \infty} \!\!\!\!\!\! \rd \tau' \, \!\! \int \!\! \rd \bR_{2} \rd \bR_{1}' \rd \bR_{2}' \,  \frac{\rd \bm{\theta}_{1}^{\rm s}}{(2 \pi)^{k}} \!\! \int_{\mathcal{B}} \! \frac{\rd \omega}{2 \pi} \!\! \int_{\mathcal{B}'} \! \frac{\rd \omega'}{2 \pi} \, \re^{- \ri (\omega + \omega') \tau'}  \nonumber
\\
& \times  \, \frac{(2 \pi)^{k - d}}{N} \frac{\partial }{\partial \bm{J}_{1}^{\rm s}} \!\cdot\! \bigg\{ \frac{\partial \oU_{12}}{\partial \bm{\theta}_{1}^{\rm s}} \, \widetilde{\mathcal{G}}^{(1)} (\bR_{1} , \bR_{1}' , \omega) \, \widetilde{\mathcal{G}}^{(1)} (\bR_{2} , \bR_{2}' , \omega ')   \nonumber
\\
& \times \, \frac{\partial \oU_{1'2'}}{\partial \bm{\theta}_{1}^{\rm s \prime}} \!\cdot\! \bigg[ \frac{\partial }{\partial \bm{J}_{1}^{\rm s \prime}} \!-\! \frac{\partial }{\partial \bm{J}_{2}^{\rm s \prime}} \bigg] \, \oF(\bm{J}_{1}') \, \oF (\bm{J}_{2} ') \bigg\} \, ,
\label{collision_operator_rewrite}
\end{align} 
where the Laplace transformed ${1-}$body Green's functions were introduced in equation~\eqref{final_inversion_G1}. Let us then rewrite equation~\eqref{collision_operator_rewrite} simply as a function of the system's ${1-}$body DF only. Integrating equation~\eqref{collision_operator_rewrite} with respect to $\bm{\theta}_{1}^{\rm s}$, $\bm{\theta}_{2}^{\rm s}$, $\bm{\theta}_{1}^{\rm s \prime}$, and $\bm{\theta}_{2}^{\rm s \prime}$, one gets
\begin{align}
& \, \mathscr{C} \big[ \oF \big] = \!\! \int_{0}^{+ \infty} \!\!\!\!\!\! \rd \tau ' \, \!\! \int \!\! \rd \bm{J}_{2} \rd \bm{J}_{1} ' \rd \bm{J}_{2} ' \, \!\! \int_{\mathcal{B}} \! \frac{\rd \omega}{2 \pi} \!\! \int_{\mathcal{B}'} \! \frac{\rd \omega'}{2 \pi} \, \re^{- \ri (\omega + \omega')\tau'} \frac{(2 \pi)^{4k - d}}{N}   \nonumber
\\
& \times \frac{\partial }{\partial \bm{J}_{1}^{\rm s}} \!\cdot\! \bigg\{ \sum_{\bm{m}_{1}^{\rm s} , \bm{m}_{2}^{\rm s}} \sum_{\bm{m}_{1}^{\rm s \prime} , \bm{m}_{2}^{\rm s \prime}}  \widetilde{\mathcal{G}}^{(1)}_{\bm{m}_{1}^{\rm s} , \bm{m}_{1}^{\rm s \prime}} (\omega) \, \widetilde{\mathcal{G}}^{(1)}_{\bm{m}_{2}^{\rm s}  , \bm{m}_{2}^{\rm s \prime}} (\omega') \, \bm{m}_{1}^{\rm s} \, A_{- \bm{m}_{1}^{\rm s} , \bm{m}_{2}^{\rm s}}   \nonumber
\\
& \times \bigg[ A_{\bm{m}_{1}^{\rm s \prime} , - \bm{m}_{2}^{\rm s \prime}} \, \bm{m}_{1}^{\rm s \prime} \!\cdot\! \frac{\partial \oF}{\partial \bm{J}_{1}^{\rm s \prime}} \oF (\bm{J}_{2}') \!+\! A_{\bm{m}_{2}^{\rm s \prime} , - \bm{m}_{1}^{\rm s \prime}} \, \bm{m}_{2}^{\rm s \prime} \!\cdot\! \frac{\partial \oF}{\partial \bm{J}_{2}^{\rm s \prime}} \oF (\bm{J}_{1}') \bigg] \bigg\} \, ,
\label{collision_operator_1}
\end{align}
with the notations ${ \widetilde{\mathcal{G}}^{(1)}_{\bm{m}_{1}^{\rm s} , \bm{m}_{1}^{\rm s \prime}} (\omega) \!=\! \widetilde{\mathcal{G}}^{(1)}_{\bm{m}_{1}^{\rm s} , \bm{m}_{1}^{\rm s \prime}} (\bm{J}_{1} , \bm{J}_{1} ' , \omega) }$ and ${ A_{\bm{m}_{1}^{\rm s} , \bm{m}_{2}^{\rm s}} \!=\! A_{\bm{m}_{1}^{\rm s} , \bm{m}_{2}^{\rm s}} (\bm{J}_{1} , \bm{J}_{2}) }$. Using the explicit expression of the Fourier coefficients of the ${1-}$body Green's function from equation~\eqref{final_inversion_G1}, equation~\eqref{collision_operator_1} becomes
\begin{align}
& \, \mathscr{C} \big[ \oF \big] = - \!\! \int_{0}^{+ \infty} \!\!\!\!\!\! \rd \tau'  \, \!\! \int \rd \bm{J}_{2} \rd \bm{J}_{1}' \rd \bm{J}_{2}' \, \!\! \int_{\mathcal{B}} \! \frac{\rd \omega}{2 \pi} \!\! \int_{\mathcal{B}'} \! \frac{\rd \omega'}{2 \pi} \, \re^{- \ri (\omega + \omega')\tau'} \frac{(2 \pi)^{2k - d}}{N}   \nonumber
\\
& \, \times \frac{\partial }{\partial \bm{J}_{1}^{\rm s}} \!\cdot\! \bigg\{ \sum_{\bm{m}_{1}^{\rm s} , \bm{m}_{2}^{\rm s}} \sum_{\bm{m}_{1}^{\rm s \prime} , \bm{m}_{2}^{\rm s \prime}} \frac{1}{\omega \!-\! \omega_{1}} \frac{1}{\omega' \!-\! \omega_{2}} \bm{m}_{1}^{\rm s} \, A_{- \bm{m}_{1}^{\rm s} , \bm{m}_{2}^{\rm s}}  \nonumber
\\
& \, \times \bigg[ \delta_{\bm{m}_{1}^{\rm s}}^{\bm{m}_{1}^{\rm s \prime}} \delta_{\rm D} (\bm{J}_{1} \!-\! \bm{J}_{1}') \!+\! (2 \pi)^{k} \frac{\bm{m}_{1}^{\rm s} \!\cdot\! \partial \oF / \partial \bm{J}_{1}^{\rm s}}{(\omega \!-\! \omega_{1}') \, \mathcal{D}_{\bm{m}_{1}^{\rm s} , \bm{m}_{1}^{\rm s \prime}} (\omega)} \bigg]  \nonumber
\\
& \, \times \bigg[ \delta_{\bm{m}_{2}^{\rm s}}^{\bm{m}_{2}^{\rm s \prime}} \delta_{\rm D} (\bm{J}_{2} \!-\! \bm{J}_{2}') \!+\! (2 \pi)^{k} \frac{\bm{m}_{2}^{\rm s} \!\cdot\! \partial \oF / \partial \bm{J}_{2}^{\rm s}}{(\omega' \!-\! \omega_{2}') \, \mathcal{D}_{\bm{m}_{2}^{\rm s} , \bm{m}_{2}^{\rm s \prime}} (\omega')} \bigg]   \nonumber
\\
& \, \times \bigg[ A_{\bm{m}_{1}^{\rm s \prime} , - \bm{m}_{2}^{\rm s \prime}} \bm{m}_{1}^{\rm s \prime} \!\cdot\! \frac{\partial \oF}{\partial \bm{J}_{1}^{\rm s \prime}} \oF(\bm{J}_{2}') + A_{\bm{m}_{2}^{\rm s \prime} , - \bm{m}_{1}^{\rm s \prime}} \bm{m}_{2}^{\rm s \prime} \!\cdot\! \frac{\partial \oF}{\partial \bm{J}_{2}^{\rm s \prime}} \oF (\bm{J}_{1}') \bigg] \bigg\} \, ,
\label{collision_operator_2}
\end{align}
with the shortened notations ${ 1/ \mathcal{D}_{\bm{m}_{1}^{\rm s} , \bm{m}_{1}^{\rm s \prime}} (\omega) \!=\! 1/ \mathcal{D}_{\bm{m}_{1}^{\rm s} , \bm{m}_{1}^{\rm s \prime}} (\bm{J}_{1} , \bm{J}_{1} ' , \omega) }$, as well as ${ \omega_{1\!/\!2} \!=\! \bm{m}_{1\!/\!2}^{\rm s} \!\cdot\! \bm{\Omega}_{1\!/\!2} }$ and ${ \omega_{1\!/\!2}' \!=\! \bm{m}_{1\!/\!2}^{\rm s \prime} \!\cdot\! \bm{\Omega}_{1\!/\!2}^{\rm s \prime} }$.
The rest of this section is devoted to simplifying equation~\eqref{collision_operator_2}, which still involves a triple integration over action space, two integrals over frequency space and one time integration.
In the following, we will first integrate over  two actions, then over time, and over the two remaining frequencies, the trickiest step. 

 Let us first deal with the integration and sum with respect to $\bm{J}_{2}$ and $\bm{m}_{2}^{\rm s}$. It requires to evaluate
\begin{align}
 \sum_{\bm{m}_{2}^{\rm s}} \!\! \int \!\! \rd \bm{J}_{2} \, & \frac{A_{- \bm{m}_{1}^{\rm s} , \bm{m}_{2}^{\rm s}}}{\omega' \!-\! \omega_{2}} \bigg[ \delta_{\bm{m}_{2}^{\rm s}}^{\bm{m}_{2}^{\rm s \prime}} \delta_{\rm D} (\bm{J}_{2} \!-\! \bm{J}_{2}' ) \!+\! (2 \pi)^{k} \frac{\bm{m}_{2}^{\rm s} \!\cdot\! \partial \oF / \partial \bm{J}_{2}^{\rm s}}{(\omega' \!-\! \omega_{2}') \mathcal{D}_{\bm{m}_{2}^{\rm s} , \bm{m}_{2}^{\rm s \prime}} (\omega')} \bigg]   \nonumber
\\
& = - \frac{1}{\omega' \!-\! \omega_{2}' } \frac{1}{\mathcal{D}_{- \bm{m}_{1}^{\rm s} , \bm{m}_{2}^{\rm s \prime}} (\omega ')} \, .
\label{integration_J2_collision}
\end{align}
To obtain equation~\eqref{integration_J2_collision}, we relied on the intrinsic definition of the dressed susceptibility coefficients~\citep[see equation~(A.8) in][]{Chavanis2012} reading
\begin{align}
& \, \frac{1}{\mathcal{D}_{\bm{m}_{1}^{\rm s} , \bm{m}_{2}^{\rm s}} (\bm{J}_{1} , \bm{J}_{2} , \omega)} = - A_{\bm{m}_{1}^{\rm s} , \bm{m}_{2}^{\rm s}} (\bm{J}_{1} , \bm{J}_{2})   \nonumber
\\
& \, - (2 \pi)^{k} \!\! \sum_{\bm{m}_{3}^{\rm s}} \!\! \int \!\! \rd \bm{J}_{3} \, \frac{\bm{m}_{3}^{\rm s} \!\cdot\! \partial \oF / \partial \bm{J}_{3}^{\rm s}}{\omega \!-\! \bm{m}_{3}^{\rm s} \!\cdot\! \bm{\Omega}_{3}^{\rm s}} \frac{A_{\bm{m}_{1}^{\rm s} , \bm{m}_{3}^{\rm s}} (\bm{J}_{1} , \bm{J}_{3})}{\mathcal{D}_{\bm{m}_{3}^{\rm s} , \bm{m}_{2}^{\rm s}} (\bm{J}_{3} , \bm{J}_{1} , \omega)} \, ,
\label{intrinsic_1/D}
\end{align}
which is straightforward to obtain given the basis decompositions of the susceptibility coefficients from equations~\eqref{definition_1/D} and~\eqref{simpler_A}, and the definition of the response matrix from equation~\eqref{Fourier_M}.
Equation~\eqref{collision_operator_2} then becomes
\begin{align}
& \, \mathscr{C} \big[ \oF \big] = \!\! \int_{0}^{+ \infty} \!\!\!\!\!\! \rd \tau' \, \!\! \int \!\! \rd \bm{J}_{1}' \rd \bm{J}_{2}' \!\! \int_{\mathcal{B}} \! \frac{\rd \omega}{2 \pi} \!\! \int_{\mathcal{B}'} \! \frac{\rd \omega'}{2 \pi} \, \re^{- \ri (\omega + \omega') \tau'} \, \frac{(2 \pi)^{2k - d}}{N}   \nonumber
\\
& \, \times \frac{\partial }{\partial \bm{J}_{1}^{\rm s}} \!\cdot\! \bigg\{ \sum_{\bm{m}_{1}^{\rm s}} \sum_{\bm{m}_{1}^{\rm s \prime} , \bm{m}_{2}^{\rm s \prime}} \frac{1}{\omega \!-\! \omega_{1}} \frac{1}{\omega' \!-\! \omega_{2}'} \bm{m}_{1}^{\rm s} \frac{1}{\mathcal{D}_{- \bm{m}_{1}^{\rm s} , \bm{m}_{2}^{\rm s \prime}} (\omega')}   \nonumber
\\
& \, \times \bigg[ \delta_{\bm{m}_{1}^{\rm s}}^{\bm{m}_{1}^{\rm s \prime}} \delta_{\rm D} (\bm{J}_{1} \!-\! \bm{J}_{1}') \!+\! (2 \pi)^{k} \frac{\bm{m}_{1}^{\rm s} \!\cdot\! \partial \oF / \partial \bm{J}_{1}^{\rm s}}{(\omega \!-\! \omega_{1}') \, \mathcal{D}_{\bm{m}_{1}^{\rm s} , \bm{m}_{1}^{\rm s \prime}} (\omega)} \bigg]   \nonumber
\\
& \, \times \bigg[ A_{\bm{m}_{1}^{\rm s \prime} , - \bm{m}_{2}^{\rm s \prime}} \bm{m}_{1}^{\rm s \prime} \!\cdot\! \frac{\partial \oF}{\partial \bm{J}_{1}^{\rm s \prime}} \oF (\bm{J}_{2}') + A_{\bm{m}_{2}^{\rm s \prime} , - \bm{m}_{1}^{\rm s \prime}} \bm{m}_{2}^{\rm s \prime} \!\cdot\! \frac{\partial \oF}{\partial \bm{J}_{2}^{\rm s \prime}} \oF(\bm{J}_{1}') \bigg] \bigg\} \, .
\label{collision_operator_3}
\end{align}

Next, the integration and sum with respect to ${\bm{J}_{1}}'$ and $\bm{m}_{1}^{\rm s \prime}$ are  performed. These only act on the two last lines of equation~\eqref{collision_operator_3}. As previously, one relies on the intrinsic definition of the dressed susceptibility coefficients from equation~\eqref{intrinsic_1/D}. Two different contributions have to be dealt with: the first one ${ \mathscr{C}_{1} \big[ \oF \big] }$ is associated witht the gradient ${ \bm{m}_{1}^{\rm s \prime} \!\cdot\! \partial \oF / \partial \bm{J}_{1}^{\rm s \prime} \oF (\bm{J}_{2}') }$, and the second one ${ \mathscr{C}_{2} \big[ \oF \big] }$ with the gradient ${ \bm{m}_{2}^{\rm s \prime} \!\cdot\! \partial \oF / \partial \bm{J}_{2}^{\rm s \prime} \oF (\bm{J}_{1}') }$. The first contribution ${ \mathscr{C}_{1} \big[ \oF \big] }$ takes the form
\begin{align}
\mathscr{C}_{1} \big[ \oF \big] & \, = \sum_{\bm{m}_{1}^{\rm s \prime}} \!\! \int \!\! \rd \bm{J}_{1}' \, \bigg[ \delta_{\bm{m}_{1}^{\rm s}}^{\bm{m}_{1}^{\rm s \prime}} \delta_{\rm D} (\bm{J}_{1} \!-\! \bm{J}_{1}') \!+\! (2 \pi)^{k} \frac{\bm{m}_{1}^{\rm s} \!\cdot\! \partial \oF / \partial \bm{J}_{1}^{\rm s}}{(\omega \!-\! \omega_{1}') \, \mathcal{D}_{\bm{m}_{1}^{\rm s} , \bm{m}_{1}^{\rm s \prime}} (\omega)} \bigg]  \nonumber
\\
&  \;\;\;\; \, \times A_{\bm{m}_{1}^{\rm s \prime} , - \bm{m}_{2}^{\rm s \prime}} \, \bm{m}_{1}^{\rm s \prime} \!\cdot\! \frac{\partial \oF}{\partial \bm{J}_{1}^{\rm s \prime}} \, \oF (\bm{J}_{2}')  \nonumber
\\
& \, = - \frac{1}{\mathcal{D}_{\bm{m}_{1}^{\rm s} , - \bm{m}_{2}^{\rm s \prime}} (\omega)} \, \bm{m}_{1}^{\rm s} \!\cdot\! \frac{\partial \oF}{\partial \bm{J}_{1}^{\rm s}} \, \oF (\bm{J}_{2}') \, .
\label{collision_last_1}
\end{align}
Similarly, the second contribution ${ \mathscr{C}_{2} \big[ \oF \big] }$ takes the form
\begin{align}
\mathscr{C}_{2} \big[ \oF \big] & \, = \sum_{\bm{m}_{1}^{\rm s \prime}} \!\! \int \!\! \rd \bm{J}_{1}' \, \bigg[ \delta_{\bm{m}_{1}^{\rm s}}^{\bm{m}_{1}^{\rm s \prime}} \delta_{\rm D} (\bm{J}_{1} \!-\! \bm{J}_{1}') \!+\! (2 \pi)^{k} \frac{\bm{m}_{1}^{\rm s} \!\cdot\! \partial \oF / \partial \bm{J}_{1}^{\rm s}}{(\omega \!-\! \omega_{1}') \, \mathcal{D}_{\bm{m}_{1}^{\rm s} , \bm{m}_{1}^{\rm s \prime}} (\omega)} \bigg]  \nonumber
\\
& \, \;\;\;\; \times A_{\bm{m}_{2}^{\rm s \prime} , - \bm{m}_{1}^{\rm s \prime}} \, \bm{m}_{2}^{\rm s \prime} \!\cdot\! \frac{\partial \oF}{\partial \bm{J}_{2}^{\rm s \prime}} \, \oF (\bm{J}_{1}')  \nonumber
\\
& \, = \bm{m}_{1}^{\rm s} \!\cdot\! \frac{\partial \oF}{\partial \bm{J}_{1}^{\rm s}} \, \bm{m}_{2}^{\rm s \prime} \!\cdot\! \frac{\partial \oF}{\partial \bm{J}_{2}^{\rm s \prime}} \, (2 \pi)^{k} \sum_{\bm{m}_{1}^{\rm s \prime}} \!\! \int \!\! \rd \bm{J}_{1}^{\prime} \, \frac{\oF (\bm{J}_{1}^{\prime}) \, A_{\bm{m}_{2}^{\rm s \prime} , - \bm{m}_{1}^{\rm s \prime}}}{(\omega \!-\! \omega_{1}') \mathcal{D}_{\bm{m}_{1}^{\rm s} , \bm{m}_{1}^{\rm s \prime}} (\omega)}   \nonumber
\\
& \, \;\;\;\; + A_{\bm{m}_{2}^{\rm s} , - \bm{m}_{1}^{\rm s}} \, \bm{m}_{2}^{\rm s \prime} \!\cdot\! \frac{\partial \oF}{\partial \bm{J}_{2}^{\rm s \prime}} \oF (\bm{J}_{1}) \, .
\label{collision_last_2}
\end{align}

Let us now rewrite equation~\eqref{collision_operator_3} while relying on the matrix method, that is by using the basis elements $\psi^{(p)}$. Within the basis, the bare and dressed susceptibility coefficients take the form of equation~\eqref{simpler_A} and allow for a rewrite of equation~\eqref{definition_1/D} as
\begin{equation}
 \, \frac{1}{\mathcal{D}_{\bm{m}_{1}^{\rm s} , \bm{m}_{2}^{\rm s}} (\bm{J}_{1} , \bm{J}_{2} , \omega)} = \opsi_{\bm{m}_{1}^{\rm s}}^{(\alpha)} (\bm{J}_{1}) \, \varepsilon_{\alpha \beta}^{-1} (\omega) \, \opsi_{\bm{m}_{2}^{\rm s}}^{(\beta) *} (\bm{J}_{2}) \, ,
\label{A_1/D_basis}
\end{equation}
where the sums over the greek indices are implied. Following more closely~\cite{Heyvaerts2010}, we introduced here the matrix ${ \bm{\varepsilon} (\omega) \!=\! \mathbf{I} \!-\! \widehat{\mathbf{M}} (\omega) }$, where the response matrix $\widehat{\mathbf{M}}$ is given by equation~\eqref{Fourier_M}. Finally, let us accordingly define the matrix ${ \mathbf{H} (\omega) }$ as
\begin{equation}
H_{\alpha \beta} (\omega) = (2 \pi)^{k} \sum_{\bm{m}^{\rm s}} \!\! \int \!\! \rd \bm{J} \, \frac{\oF (\bm{J})}{\omega \!-\! \bm{m}^{\rm s} \!\cdot\! \bm{\Omega}^{\rm s}} \, \opsi_{\bm{m}^{\rm s}}^{(\alpha) *} (\bm{J}) \, \opsi_{- \bm{m}^{\rm s}}^{(\beta) *} (\bm{J}) \, .
\label{definition_H_matrix}
\end{equation}
Combining the two contributions from equations~\eqref{collision_last_1} and~\eqref{collision_last_2}, and after some straightforward algebra, equation~\eqref{collision_operator_3} becomes
\begin{align}
\mathscr{C} \big[ \oF \big] = & \, - \!\! \int \!\! \rd \tau ' \!\! \!\!\ \int_{\mathcal{B}} \! \frac{\rd \omega}{2 \pi} \!\! \int_{\mathcal{B}'} \! \frac{\rd \omega'}{2 \pi} \, \re^{- \ri (\omega + \omega') \tau'} \frac{(2 \pi)^{k - d}}{N} \frac{\partial }{\partial \bm{J}_{1}^{\rm s}} \!\cdot\! \bigg[  \sum_{\bm{m}_{1}^{\rm s}} \frac{\bm{m}_{1}^{\rm s}}{\omega \!-\! \omega_{1}}  \nonumber
\\
& \, \times \bigg\{ \opsi^{(\alpha)}_{- \bm{m}_{1}^{\rm s}} (\bm{J}_{1}) \, \varepsilon_{\alpha \beta}^{-1} (\omega') \, H_{\beta \delta} (\omega') \, \varepsilon_{\gamma \delta}^{-1} (\omega) \, \opsi^{(\gamma)}_{\bm{m}_{1}^{\rm s}} (\bm{J}_{1}) \, \bm{m}_{1}^{\rm s} \!\cdot\! \frac{\partial \oF}{\partial \bm{J}_{1}^{\rm s}}  \nonumber
\\
& \, + \opsi^{(\alpha)}_{- \bm{m}_{1}^{\rm s}} (\bm{J}_{1}) \, \big[ \varepsilon_{\alpha \gamma}^{-1} (\omega') \!-\! \delta_{\alpha \gamma} \big] \, \opsi^{(\gamma) *}_{- \bm{m}_{1}^{\rm s}} (\bm{J}_{1}) \, \oF (\bm{J}_{1})  \nonumber
\\
& \, +\opsi^{(\alpha)}_{- \bm{m}_{1}^{\rm s}} (\bm{J}_{1}) \, \varepsilon_{\alpha \gamma}^{-1} (\omega') \, \varepsilon_{\delta \lambda}^{-1} (\omega)\, H_{\lambda \gamma} (\omega) \, \opsi^{(\delta)}_{\bm{m}_{1}^{\rm s}} (\bm{J}_{1}) \, \bm{m}_{1}^{\rm s} \!\cdot\! \frac{\partial \oF}{\partial \bm{J}_{1}^{\rm s}}   \nonumber
\\
& \, - \opsi^{(\alpha)}_{- \bm{m}_{1}^{\rm s}} (\bm{J}_{1}) \, \varepsilon_{\delta \lambda}^{-1} (\omega) \, H_{\lambda \alpha} (\omega) \, \opsi^{(\delta)}_{\bm{m}_{1}^{\rm s}} (\bm{J}_{1}) \, \bm{m}_{1}^{\rm s} \!\cdot\! \frac{\partial \oF}{\partial \bm{J}_{1}^{\rm s}} \bigg\} \bigg] \, .
\label{collision_operator_4}
\end{align}

Next the integrations with respect to ${\tau'}$ and ${\omega'}$ in equation~\eqref{collision_operator_4} should be performed, which is  technically demanding.  
This equation formally takes the form
 \begin{equation}
 \!\! \int_{0}^{+ \infty} \!\!\!\!\!\! \rd \tau ' \, \!\! \int_{\mathcal{B}'} \! \frac{\rd \omega'}{2 \pi} \, \re^{- \ri (\omega + \omega')\tau'} \, g (\omega , \omega') \, .
 \label{shape_integral_tau_omega}
 \end{equation}
The integration over ${\tau'}$ is straightforward provided that ${ \omega \!+\! \omega' }$ has a negative imaginary part. We therefore introduce ${p \!>\! 0}$ and perform the substitution ${ \omega \!+\! \omega' \!\to\! \omega \!+\! \omega' \!-\! \ri p }$, so that the integration may be computed as
\begin{equation}
\eqref{shape_integral_tau_omega} = \lim\limits_{p \to 0} \! \int_{\mathcal{B}'} \! \frac{\rd \omega'}{2 \pi} \, \frac{- \ri}{\omega \!+\! \omega' \!-\! \ri p} \, g (\omega , \omega') \, .
\label{shape_integral_tau_omega_2}
\end{equation}
As the system is assumed to be linearly stable, the poles of the function ${ \omega ' \!\mapsto\! g (\omega , \omega') }$ are all in the lower half complex plane and the Bromwich contour ${\mathcal{B}'}$ has to pass above all these singularities. The only pole in ${\omega'}$ which remains is then ${ \omega' \!=\! - \omega \!+\! \ri p }$ and is located in the upper half plane. We carry the integral over ${\omega'}$ using the residue theorem by closing the contour ${\mathcal{B}'}$ in the upper half complex plane -- this is possible because the integrands decreases sufficiently fast at infinity like ${ 1 / |\omega'|^{2} }$. One therefore gets
\begin{equation}
\eqref{shape_integral_tau_omega} = \lim\limits_{p \to 0} g (\omega , - \omega \!+\! \ri p) \, .
\label{shape_integral_tau_omega_3}
\end{equation}

We may now consider the integration with respect to $\omega$ in equation~\eqref{collision_operator_4}. First, we note that the fourth term of equation~\eqref{collision_operator_4} vanishes when integrated upon $\omega$. Indeed, by construction, the Bromwich contour $\mathcal{B}$ has to pass above all the singularities of the functions of ${ + \omega}$. This contour may then be closed in the upper half complex plane, and, because it surrounds no singularities, gives a vanishing result for this term. Equation~\eqref{collision_operator_4} may then be rewritten as
\begin{align}
\mathscr{C} \big[ \oF \big] = & \,  \lim\limits_{p \to 0} - \!\! \int_{\mathcal{B}} \! \frac{\rd \omega}{2 \pi} \frac{(2 \pi)^{k - d}}{N} \frac{\partial }{\partial \bm{J}_{1}^{\rm s}} \!\cdot\! \bigg[ \sum_{\bm{m}_{1}^{\rm s}} \frac{\bm{m}_{1}^{\rm s}}{\omega \!-\! \omega_{1}}  \nonumber
\\
& \, \times \bigg\{ \opsi^{(\alpha)}_{- \bm{m}_{1}^{\rm s}} (\bm{J}_{1}) \, \big[ \varepsilon_{\alpha \gamma}^{-1} (- \omega \!+\! \ri p) \!-\! \delta_{\alpha \gamma} \big] \, \opsi^{(\gamma) *}_{- \bm{m}_{1}^{\rm s}} (\bm{J}_{1}) \, \oF (\bm{J}_{1})  \nonumber
\\
& \, \;\;\;+ \opsi^{(\alpha)}_{- \bm{m}_{1}^{\rm s}} (\bm{J}_{1}) \, \varepsilon_{\alpha \beta}^{-1} (- \omega \!+\! \ri p) \, \varepsilon_{\gamma \delta}^{-1} (\omega) \, \opsi^{(\gamma)}_{\bm{m}_{1}^{\rm s}} (\bm{J}_{1})   \nonumber
\\
& \, \;\;\; \times \big[ H_{\beta \delta} (- \omega \!+\! \ri p) \!+\! H_{\delta \beta} (\omega) \big] \, \bm{m}_{1}^{\rm s} \!\cdot\! \frac{\partial \oF}{\partial \bm{J}_{1}^{\rm s}} \bigg\} \bigg] \, .
\label{collision_operator_5}
\end{align}
Let us now evaluate the term within brackets in the second term of equation~\eqref{collision_operator_5}. It reads
\begin{align}
\big[ H_{\beta \delta} (- \omega \!+\! \ri p) \!+\! H_{\delta \beta} (\omega) \big] = & \, (2 \pi)^{k} \! \sum_{\bm{m}_{2}^{\rm s}} \!\! \int \!\! \rd \bm{J}_{2} \, \opsi^{(\delta) *}_{\bm{m}_{2}^{\rm s}} (\bm{J}_{2}) \, \opsi^{(\beta) *}_{- \bm{m}_{2}^{\rm s}} (\bm{J}_{2}) \, \oF (\bm{J}_{2})   \nonumber
\\
& \, \times \bigg[ \frac{1}{\omega \!-\! \omega_{2}} \!-\! \frac{1}{\omega \!-\! (\omega_{2} \!+\! \ri p)} \bigg] \, ,
\label{evaluation_H}
\end{align}
using the notation ${ \omega_{2} \!=\! \bm{m}_{2}^{\rm s} \!\cdot\! \bm{\Omega}^{\rm s} (\bm{J}_{2}) }$. When considering the limit ${ p \!\to\! 0 }$, one should be careful with the fact ${ \omega \!=\! \omega_{2} }$ and ${ \omega \!=\! \omega_{2} \!+\! \ri p }$ are on opposite sides of the prescribed integration contour $\mathcal{B}$. Indeed, when lowering the integration contour to the real axis, the pole ${ \omega \!=\! \omega_{2} }$ remains below the contour, while the one in ${ \omega \!=\! \omega_{2} \!+\! \ri p }$ is above it. In this limit, the term in bracket in equation~\eqref{evaluation_H} becomes ${ \big[ 1/(\omega \!-\! \omega_{2} \!+\! \ri 0) \!-\! 1/(\omega \!-\! \omega_{2} \!-\! \ri 0)\big] }$. We may rely on Plemelj formula
\begin{equation}
\frac{1}{x \!\pm\! \ri 0^{+}} = \mathcal{P} \bigg( \frac{1}{x} \bigg) \mp \ri \pi \delta_{\rm D} (x) \, ,
\label{Plemelj_formula}
\end{equation}
where $\mathcal{P}$ stands for Cauchy principal value. Equation~\eqref{evaluation_H} can then be evaluated and reads
\begin{equation}
\eqref{evaluation_H} \!=\! - 2 \pi \ri  (2 \pi)^{k} \! \sum_{\bm{m}_{2}^{\rm s}} \!\! \int \!\! \rd \bm{J}_{2} \, \opsi^{(\delta *)}_{\bm{m}_{2}^{\rm s}} (\bm{J}_{2}) \, \opsi^{(\beta) *}_{- \bm{m}_{2}^{\rm s}} (\bm{J}_{2})  \oF (\bm{J}_{2}) \, \delta_{\rm D} (\omega \!-\! \omega_{2}) \, . \nonumber
\end{equation} 
When lowering the contour $\mathcal{B}$ to the real axis, one can also compute the integration with respect to $\omega$ of the first term in equation~\eqref{collision_operator_5}. Once again, the system being stable, the poles of ${ \varepsilon_{\alpha \gamma}^{-1} (- \omega \!+\! \ri p) }$ are all located in the upper half plane, and there remains only one pole on the real axis in ${ \omega \!=\! \omega_{1} }$. The contour $\mathcal{B}$ is closed in the lower half plane and only encloses this second pole. Accounting for the direction of integration, the residue theorem gives a factor ${ - 2 \ri \pi }$ and equation~\eqref{collision_operator_5} then becomes
\begin{align}
\mathscr{C} \big[ & \, \oF \big]  =  \ri \frac{(2 \pi)^{k - d}}{N} \frac{\partial }{\partial \bm{J}_{1}^{\rm s}} \!\cdot\! \bigg[ \sum_{\bm{m}_{1}^{\rm s}} \bm{m}_{1}^{\rm s} \, \opsi^{(\alpha)}_{- \bm{m}_{1}^{\rm s}} (\bm{J}_{1}) \, \opsi^{(\gamma) *}_{- \bm{m}_{1}^{\rm s}} (\bm{J}_{1}) \, \oF (\bm{J}_{1})   \nonumber
\\
& \,  \times \big[ \varepsilon_{\alpha \gamma}^{-1} (- \omega_{1} \!+\! \ri 0) \!-\! \delta_{\alpha \gamma} \big]  \nonumber
\\
& \, + (2 \pi)^{k} \! \sum_{\bm{m}_{1}^{\rm s} , \bm{m}_{2}^{\rm s}} \! \bm{m}_{1}^{\rm s} \!\! \int \!\! \rd \bm{J}_{2} \, \big[ \opsi^{(\alpha)}_{- \bm{m}_{1}^{\rm s}} (\bm{J}_{1}) \, \varepsilon_{\alpha \beta}^{-1} (- \omega_{2}) \, \opsi^{(\beta) *}_{- \bm{m}_{2}^{\rm s}} (\bm{J}_{2}) \big]  \nonumber
\\
& \, \times \big[ \opsi^{(\gamma)}_{\bm{m}_{1}^{\rm s}} (\bm{J}_{1}) \, \varepsilon_{\gamma \delta}^{-1} (\omega_{2}) \, \opsi^{(\delta) *}_{\bm{m}_{2}^{\rm s}} (\bm{J}_{2}) \big] \, \frac{\bm{m}_{1}^{\rm s} \!\cdot\! \partial \oF / \partial \bm{J}_{1}^{\rm s} \, \oF (\bm{J}_{2})}{\omega_{2} \!-\! \omega_{1} \!+\! \ri 0}  \bigg] \, ,
\label{collision_operator_6}
\end{align}
keeping track of the small positive imaginary part in the pole ${ 1/ (\omega_{2} \!-\! \omega_{1} \!+\! \ri 0) }$ associated with the fact that the contour $\mathcal{B}$ passed above the pole ${ \omega \!=\! \omega_{1} }$. Relying on the expression of the susceptibility coefficients from equation~\eqref{A_1/D_basis}, equation~\eqref{collision_operator_6} can immediately be rewritten as
\begin{align}
& \, \mathscr{C} \big[ \oF \big] = \ri \frac{(2 \pi)^{k - d}}{N} \frac{\partial }{\partial \bm{J}_{1}^{\rm s}} \!\cdot\! \bigg[  \nonumber
\\
& \, - \sum_{\bm{m}_{1}^{\rm s}} \bm{m}_{1}^{\rm s} \bigg(\! \frac{1}{\mathcal{D}_{\bm{m}_{1}^{\rm s} , \bm{m}_{1}^{\rm s}} (\bm{J}_{1} , \bm{J}_{1} , \omega_{1} \!+\! \ri 0)} \!+\! A_{\bm{m}_{1}^{\rm s} , \bm{m}_{1}^{\rm s}} (\bm{J}_{1} , \bm{J}_{1}) \!\bigg)  \, \oF (\bm{J}_{1})  \nonumber
\\
& \, + (2 \pi)^{k} \! \sum_{\bm{m}_{1}^{\rm s} , \bm{m}_{2}^{\rm s}} \!\! \bm{m}_{1}^{\rm s} \!\! \int \!\! \rd \bm{J}_{2} \, \frac{\bm{m}_{1}^{\rm s} \!\cdot\! \partial \oF / \partial \bm{J}_{1}^{\rm s} \, \oF (\bm{J}_{2})}{\omega_{2} \!-\! \omega_{1} \!+\! \ri 0}  \nonumber
\\
& \, \times \frac{1}{\mathcal{D}_{- \bm{m}_{1}^{\rm s} , - \bm{m}_{2}^{\rm s}} (\bm{J}_{1} , \bm{J}_{2} , - \omega_{2})} \frac{1}{\mathcal{D}_{\bm{m}_{1}^{\rm s} , \bm{m}_{2}^{\rm s}} (\bm{J}_{1} , \bm{J}_{2} , \omega_{2})} \bigg] \, ,
\label{collision_operator_7}
\end{align}
where we made the change ${ \bm{m}_{1}^{\rm s} \!\to\! - \bm{m}_{1}^{\rm s} }$ for the first term. We note that ${ A_{\bm{m}_{1}^{\rm s} , \bm{m}_{1}^{\rm s}} (\bm{J}_{1} , \bm{J}_{1}) }$ is real in equation~\eqref{A_1/D_basis}. 

Let us now rely on the fact that the collision term ${ \mathscr{C} \big[ \oF \big] }$ is also a real quantity. In equation~\eqref{collision_operator_7}, because of the prefactor ${``\ri"}$, we may restrict ourselves only to the imaginary part of the terms within brackets. The first term of equation~\eqref{collision_operator_7} requires us to study
\begin{align}
\text{Im} \bigg[ \frac{1}{\mathcal{D}_{\bm{m}_{1}^{\rm s} , \bm{m}_{1}^{\rm s}} (\bm{J}_{1} , \bm{J}_{1} , \omega_{1} \!+\! \ri 0)} & \,  \!\bigg] = \frac{1}{2 \ri} \opsi^{(\alpha)}_{\bm{m}_{1}^{\rm s}} (\bm{J}_{1}) \, \opsi^{(\beta) *}_{\bm{m}_{1}^{\rm s}} (\bm{J}_{1})  \nonumber
\\
& \!\!\!\!\!\!\! \times  \big[ \varepsilon_{\alpha \beta}^{-1} (\omega_{1} \!+\! \ri 0) \!-\! \varepsilon_{\beta \alpha}^{-1 *} (\omega_{1} \!+\! \ri 0) \big] \, . 
\label{term_1_collision_7}
\end{align}
In order to compute the term within brackets, we rely on the identity
\begin{equation}
\varepsilon^{-1} \!-\! (\varepsilon^{-1})^{\dag} = \varepsilon^{-1} (\varepsilon^{\dag} \!-\! \varepsilon) \, (\varepsilon^{\dag})^{-1} \, .
\label{identity_epsilon}
\end{equation}
The inner term within parenthesis in equation~\eqref{identity_epsilon} reads
\begin{align}
& \, \big[ \varepsilon^{\dag} \!-\! \varepsilon \big]_{\gamma \delta} (\omega_{1} \!+\! \ri 0) = - (2 \pi)^{k} \sum_{\bm{m}_{2}^{\rm s}} \!\! \int \!\! \rd \bm{J}_{2} \, \bm{m}_{2}^{\rm s} \!\cdot\! \frac{\partial \oF}{\partial \bm{J}_{2}^{\rm s}} \, \opsi^{(\gamma) *}_{\bm{m}_{2}^{\rm s}} (\bm{J}_{2}) \, \opsi^{(\delta)}_{\bm{m}_{2}^{\rm s}} (\bm{J}_{2})   \nonumber
\\
& \, \;\;\;\;\;\;\;\;\;\;\;\;\;\;\;\;\;\;\;\;\;\;\;\;\;\;\; \times \bigg[ \bigg(\! \frac{1}{\omega_{1} \!-\! \omega_{2} \!+\! \ri 0} \!\bigg)^{*} \!-\! \frac{1}{\omega_{1} \!-\! \omega_{2} \!+\! \ri 0} \bigg]  
\label{term_brackets_epsilon}
\\
& \, = - 2 \pi \ri (2 \pi)^{k} \sum_{\bm{m}_{2}^{\rm s}} \!\! \int \!\! \rd \bm{J}_{2} \, \delta_{\rm D} (\omega_{1} \!-\! \omega_{2}) \, \bm{m}_{2}^{\rm s} \!\cdot\! \frac{\partial \oF}{\partial \bm{J}_{2}^{\rm s}} \opsi^{(\gamma) *}_{\bm{m}_{2}^{\rm s}} (\bm{J}_{2}) \, \opsi^{(\delta)}_{\bm{m}_{2}^{\rm s}} (\bm{J}_{2}) \, ,  \nonumber
\end{align}
where Plemelj formula was used once again. Combining equations~\eqref{term_1_collision_7} and~\eqref{term_brackets_epsilon} yields
\begin{align}
\text{Im} \bigg[\! \frac{1}{\mathcal{D}_{\bm{m}_{1}^{\rm s} , \bm{m}_{1}^{\rm s}} (\bm{J}_{1} , \bm{J}_{1} , \omega_{1} \!+\! \ri 0)} \!\bigg] = & \, - \pi (2 \pi)^{k} \sum_{\bm{m}_{2}^{\rm s}} \!\! \int \!\! \rd \bm{J}_{2} \, \bm{m}_{2}^{\rm s} \!\cdot\! \frac{\partial \oF}{\partial \bm{J}_{2}^{\rm s}}  \nonumber
\\
& \, \times \frac{\delta_{\rm D} (\omega_{1} \!-\! \omega_{2})}{| \mathcal{D}_{\bm{m}_{1}^{\rm s} , \bm{m}_{2}^{\rm s}} (\bm{J}_{1} , \bm{J}_{2} , \omega_{1}) |^{2}} \, .
\label{term_1_final_7}
\end{align}
This contribution corresponds to the drift term in the Balescu-Lenard equation. 

To evaluate the second term in equation~\eqref{collision_operator_7}, we make use of the relation ${ 1 \!/\! \mathcal{D}_{- \bm{m}_{1}^{\rm s} , - \bm{m}_{2}^{\rm s}} \!(\bm{J}_{1} ,\! \bm{J}_{2} ,\! -\omega\!) \!=\! 1 \!/\! \mathcal{D}_{\bm{m}_{1}^{\rm s} , \bm{m}_{2}^{\rm s}}^{*} \!(\bm{J}_{1} ,\! \bm{J}_{2} ,\! \omega\!) }$, as demonstrated in note~[83] of~\cite{Chavanis2012}, while relying on Plemelj formula. This second term corresponds to the diffusion term in the Balescu-Lenard equation. All calculations are straightforward. Gathering these two contributions and keeping track of the signs of the various terms, we finally get the expression of the collision term ${ \mathscr{C} \big[ \oF \big] }$ as
\begin{align}
\mathscr{C} \big[ \oF \big] = & \, \frac{\pi (2 \pi)^{2 k -d}}{N} \frac{\partial }{\partial \bm{J}_{1}^{\rm s}} \!\cdot\! \bigg[ \sum_{\bm{m}_{1}^{\rm s} , \bm{m}_{2}^{\rm s}} \! \bm{m}_{1}^{\rm s} \frac{\delta_{\rm D} (\bm{m}_{1}^{\rm s} \!\cdot\! \bm{\Omega}_{1}^{\rm s} \!-\! \bm{m}_{2}^{\rm s} \!\cdot\! \bm{\Omega}_{2}^{\rm s})}{| \mathcal{D}_{\bm{m}_{1}^{\rm s} , \bm{m}_{2}^{\rm s}} (\bm{J}_{1} , \bm{J}_{2} , \bm{m}_{1}^{\rm s} \!\cdot\! \bm{\Omega}_{1}^{\rm s}) |^{2}}  \nonumber
\\
& \, \times \bigg(\! \bm{m}_{1}^{\rm s} \!\cdot\! \frac{\partial }{\partial \bm{J}_{1}^{\rm s}} \!-\! \bm{m}_{2}^{\rm s} \!\cdot\! \frac{\partial }{\partial \bm{J}_{2}^{\rm s}} \!\bigg) \, \oF (\bm{J}_{1}) \oF (\bm{J}_{2}) \bigg] \, .
\label{final_collision_term_derivation}
\end{align}
This collision term, together with equation~\eqref{rewrite_BBGKY_1_solving} finally yields the Balescu-Lenard equation~\eqref{BL_Kepler}.
It now only involves the divergence of a flux corresponding to a simple integration over action space, and a physically motivated resonant condition and amplification factor, as discussed in the main text.

\subsection{Multi-component Balescu-Lenard derivation}
\label{sec:multicase}

Let us explain how one can adapt the formalisms presented in the main text to the situation where the system is composed of multiple components. The different components are indexed by the letters ${``\ra"}$ and ${``\rb"}$. We assume that the component ${``\ra"}$ is made of $N_{\ra}$ particles of individual mass $\mu_{\ra}$, and the total mass of this component is $M_{\star}^{\ra}$. When accounting for multiple components and placing ourselves within the democratic heliocentric coordinates from equation~\eqref{democratic_coordinates}, the total Hamiltonian from equation~\eqref{Hamiltonian_democratic_simpler} becomes
\begin{align}
H = & \, \sum_{\ra} \sum_{i = 1}^{N_{\ra}} \frac{\mu_{\ra}}{2} ( \bm{v}_{i}^{\ra} )^{2} + \sum_{\ra} \mu_{\ra} M_{\bullet} \sum_{i = 1}^{N_{\ra}} U (|\bm{x}_{i}^{\ra}|)   \nonumber
\\
& \, + \sum_{\ra} \mu_{\ra}^{2} \sum_{i < j}^{N_{\ra}} U (|\bm{x}_{i}^{\ra} \!-\! \bm{x}_{j}^{\ra}|) + \sum_{{\ra} < {\rb}} \sum_{i = 1}^{N_{\ra}} \sum_{j = 1}^{N_{\rb}} \mu_{\ra} \mu_{\rb} U (|\bm{x}_{i}^{\ra} \!-\! \bm{x}_{i}^{\rb}|)   \nonumber
\\
& \, + \sum_{\ra} \mu_{\ra} M_{\star} \sum_{i = 1}^{N_{\ra}} \Phi_{\rr} (\bm{x}_{i}^{\ra}) \!+\! \frac{1}{2 M_{\bullet}} \bigg[ \sum_{\ra} \mu_{\ra} \sum_{i = 1}^{N_{\ra}} \bm{v}_{i}^{\ra} \bigg]^{2}
\label{total_Hamiltonian_multi}
\end{align}
where ${ \Gamma_{i}^{\ra} \!=\! (\bm{x}_{i}^{\ra} , \bm{v}_{i}^{\ra}) }$ stands for the position and velocity of the $i^{\rm th}$ particle of component ${``\ra"}$. The various terms appearing in equation~\eqref{total_Hamiltonian_multi} are respectively the kinetic energy of the particles, the Keplerian potential due to the central BH, the self-gravity among a given component, the interaction between particles of different components, the relativistic potential corrections $\Phi_{\rr}$, and finally the additional kinetic terms due to the change of coordinates from equation~\eqref{democratic_coordinates}. One should pay attention to the normalisation of the component $\Phi_{\rr}$. Indeed, we rewrite this potential as ${ \mu_{\ra} M_{\star} \Phi_{\rr} }$, where we introduce the total active mass of the system as ${ M_{\star} \!=\! \sum_{\ra} M_{\star}^{\ra} }$. This allows for a rewriting similar to equation~\eqref{Hamiltonian_democratic_simpler}. The dynamics of individual particles is then given by the Hamilton's equations associated with the Hamiltonian from equation~\eqref{total_Hamiltonian_multi}. We now introduce the system's total PDF ${ P_{\rm tot} (\Gamma_{1}^{\ra} , ... \Gamma_{N_{\ra}}^{\ra} , \Gamma_{1}^{\rb} , ... , \Gamma_{N_{\rb}}^{\rb},...) }$, which gives the probability of finding at time $t$, the particle $1$ of the component ${``\ra"}$ at position $\bm{x}_{1}^{\ra}$ with velocity $\bm{v}_{1}^{\ra}$, etc. As in equation~\eqref{normalisation_PN}, we normalise $P_{\rm tot}$ so that
\begin{equation}
 \int \!\! \rd \Gamma_{1}^{\ra} .. \rd \Gamma_{N_{\ra}}^{\ra} \rd \Gamma_{1}^{\rb} .. \rd \Gamma_{N_{\rb}}^{\rb} .. \, P_{\rm tot} (\Gamma_{1}^{\ra} , .. , \Gamma_{N_{\ra}}^{\ra} , \Gamma_{1}^{\rb} , .. , \Gamma_{N_{\rb}}^{\rb}, ..) \!=\! 1 \, .
\label{normalisation_Ptot}
\end{equation}
Following equation~\eqref{Liouville_equation}, the dynamics of $P_{\rm tot}$ is governed by Liouville's equation which becomes
\begin{equation}
\frac{\partial P_{\rm tot}}{\partial t} + \sum_{\ra} \sum_{i = 1}^{N_{\ra}} \bigg[ \dot{\bm{x}}_{i}^{\ra} \!\cdot\! \frac{\partial P_{\rm tot}}{\partial \bm{x}_{i}^{\ra}} + \dot{\bm{v}}_{i}^{\ra} \!\cdot\! \frac{\partial P_{\rm tot}}{\partial \bm{v}_{i}^{\ra}} \bigg] = 0 \, .
\label{Liouville_multi}
\end{equation}
We define the system's reduced PDFs $P_{n}^{\ra_{1}, ... , \ra_{n}}$  (see equation~\eqref{definition_Pn}) where one integrates $P_{\rm tot}$ over all particles, except $n$ particles belonging respectively to the components $\ra_{1}$, ..., $\ra_{n}$.
Our aim is now to write the two first equations of the associated BBGKY hierarchy. To get the evolution equation of $P_{1}^{\ra}$, one proceeds as in equation~\eqref{BBGKY_Pn}, by integrating equation~\eqref{Liouville_multi} over all particles except $\Gamma_{1}^{\ra}$. In order to clarify the upcoming calculations, we will from now on neglect any contributions associated with the last additional kinetic terms from equation~\eqref{total_Hamiltonian_multi}. Indeed, we justified in equation~\eqref{vanishing_quantity_BBGKY}, that, because of the ansatz from equations~\eqref{DL_DF} and~\eqref{DL_C}, once averaged over the fast Keplerian angle, these terms do not contribute the system's dynamics at the considered order of our kinetic developments. Relying on the symmetry of $P_{\rm tot}$ with respect to interchanges of particles of the same component, one gets
\begin{align}
& \, \frac{\partial P_{1}^{\ra}}{\partial t} \!+\! \bm{v}_{1}^{\ra} \!\cdot\! \frac{\partial P_{1}^{\ra}}{\partial \bm{x}_{1}^{\ra}} + \bigg[ M_{\bullet} \bm{\mathcal{F}}_{\! 1^{\ra} 0} \!+\! M_{\star} \bm{\mathcal{F}}_{\! 1^{\ra} \rr} \bigg] \!\cdot\! \frac{\partial P_{1}^{\ra}}{\partial \bm{v}_{1}^{\ra}}   \label{BBGKY_1_multi}
\\
& \, + (N \!-\! 1) \, \mu_{\ra} \!\! \int \!\! \rd \Gamma_{2}^{\ra} \, \bm{\mathcal{F}}_{\! 1^{\ra} 2^{\ra}} \!\cdot\! \frac{\partial P_{2}^{\ra \ra}}{\partial \bm{v}_{1}^{\ra}} \!+\! \!\sum_{\rb \neq \ra} N_{\rb} \, \mu_{\rb} \!\! \int \!\! \rd \Gamma_{2}^{\rb} \, \bm{\mathcal{F}}_{\! 1^{\ra} 2^{\rb}} \!\cdot\! \frac{\partial P_{2}^{\ra \rb}}{\partial \bm{v}_{1}^{\ra}} = 0 \, .  \nonumber
\end{align}
In equation~\eqref{BBGKY_1_multi}, we used the same notations as in equation~\eqref{BBGKY_Pn}, and introduced as $\bm{\mathcal{F}}_{\!1^{\ra} 0 }$ the force exerted by the BH on particle $1^{\ra}$, $\bm{\mathcal{F}}_{\! 1^{\ra} \rr}$ the force acting on particle $1^{\ra}$ associated with the relativistic corrections, and $\bm{\mathcal{F}}_{i j}$ the force between two particles.
To obtain the second equation of the hierarchy, one may proceed similarly and integrate equation~\eqref{Liouville_multi} for all particles, except $2$. Two different cases should be considered, depending on whether one is considering $P_{2}^{\ra \ra}$ or $P_{2}^{\ra \rb}$ (with ${ \ra \!\neq\! \rb }$). Let us first consider the diffusion equation satisfied by $P_{2}^{\ra \ra}$. Integrating equation~\eqref{Liouville_multi} with respect to all particles except $\Gamma_{1}^{\ra}$ and $\Gamma_{2}^{\ra}$, one gets
\begin{align}
& \, \frac{\partial P_{2}^{\ra \ra}}{\partial t} \!+\! \bm{v}_{1}^{\ra} \!\cdot\! \frac{\partial P_{2}^{\ra \ra}}{\partial \bm{x}_{1}^{\ra}} \!+\! \bm{v}_{2}^{\ra} \!\cdot\! \frac{\partial P_{2}^{\ra \ra}}{\partial \bm{x}_{2}^{\ra}} + \mu_{\ra} \bm{\mathcal{F}}_{\! 1^{\ra} 2^{\ra}} \!\cdot\! \frac{\partial P_{2}^{\ra \ra}}{\partial \bm{v}_{1}^{\ra}} \!+\! \mu_{\ra} \bm{\mathcal{F}}_{\! 2^{\ra} 1^{\ra}} \!\cdot\! \frac{\partial P_{2}^{\ra \ra}}{\partial \bm{v}_{2}^{\ra}}  \nonumber
\\
& \, + \bigg[ M_{\bullet} \bm{\mathcal{F}}_{\! 1^{\ra} 0} \!+\! M_{\star} \bm{\mathcal{F}}_{\! 1^{\ra} \rr} \bigg] \!\cdot\! \frac{\partial P_{2}^{\ra \ra}}{\partial \bm{v}_{1}^{\ra}} \!+\! \bigg[ M_{\bullet} \bm{\mathcal{F}}_{\! 2^{\ra} 0} \!+\! M_{\star} \bm{\mathcal{F}}_{\! 2^{\ra} \rr} \bigg] \!\cdot\! \frac{\partial P_{2}^{\ra \ra}}{\partial \bm{v}_{2}^{\ra}}   \nonumber
\\
& \, + (N_{\ra} \!-\! 2) \, \mu_{\ra} \!\! \int \!\! \rd \Gamma_{3}^{\ra} \, \bigg[ \bm{\mathcal{F}}_{\! 1^{\ra} 3^{\ra}} \!\cdot\! \frac{\partial P_{3}^{\ra \ra \ra}}{\partial \bm{v}_{1}^{\ra}} \!+\! \bm{\mathcal{F}}_{\! 2^{\ra} 3^{\ra}} \!\cdot\! \frac{\partial P_{3}^{\ra \ra \ra}}{\partial \bm{v}_{2}^{\ra}} \bigg]   \nonumber
\\
& \, + \sum_{\rb \neq \ra} N_{\rb} \, \mu_{\rb} \!\! \int \!\! \rd \Gamma_{3}^{\rb} \, \bigg[ \bm{\mathcal{F}}_{\! 1^{\ra} 3^{\rb}} \!\cdot\! \frac{\partial P_{3}^{\ra \ra \rb}}{\partial \bm{v}_{1}^{\ra}} \!+\! \bm{\mathcal{F}}_{\! 2^{\ra} 3^{\rb}} \!\cdot\! \frac{\partial P_{3}^{\ra \ra \rb}}{\partial \bm{v}_{2}^{\ra}} \bigg] = 0 \, .
\label{BBGKY_2_aa_multi}
\end{align}
Similarly, starting from equation~\eqref{Liouville_multi}, and integrating it with respect to $\Gamma_{1}^{\ra}$ and $\Gamma_{1}^{\rb}$ (for ${ \rb \!\neq\! \ra }$), one gets
\begin{align}
& \, \frac{\partial P_{2}^{\ra \rb}}{\partial t} \!+\! \bm{v}_{1}^{\ra} \!\cdot\! \frac{\partial P_{2}^{\ra \rb}}{\partial \bm{x}_{1}^{\ra}} \!+\! \bm{v}_{2}^{\rb} \!\cdot\! \frac{\partial P_{2}^{\ra \rb}}{\partial \bm{x}_{2}^{\rb}} \!+\! \mu_{\rb} \bm{\mathcal{F}}_{\! 1^{\ra} 2^{\rb}} \!\cdot\! \frac{\partial P_{2}^{\ra \rb}}{\partial \bm{v}_{1}^{\ra}} \!+\! \mu_{\ra} \bm{\mathcal{F}}_{\! 2^{\rb} 1^{\ra}} \!\cdot\! \frac{\partial P_{2}^{\ra \rb}}{\partial \bm{v}_{2}^{\rb}}   \nonumber
\\
& \, + \bigg[ M_{\bullet} \bm{\mathcal{F}}_{\!1^{\ra} 0} \!+\! M_{\star} \bm{\mathcal{F}}_{\! 1^{\ra} \rr} \bigg] \!\cdot\! \frac{\partial P_{2}^{\ra \rb}}{\partial \bm{v}_{1}^{\ra}} \!+\! \bigg[ M_{\bullet} \bm{\mathcal{F}}_{\! 2^{\rb} 0} \!+\! M_{\star} \bm{\mathcal{F}}_{\! 2^{\rb} \rr} \bigg] \!\cdot\! \frac{\partial P_{2}^{\ra \rb}}{\partial \bm{v}_{2}^{\rb}}  \nonumber 
\\
& \, + (N_{\ra} \!-\! 1) \, \mu_{\ra} \!\! \int \!\! \rd \Gamma_{3}^{\ra} \, \bigg[ \bm{\mathcal{F}}_{\! 1^{\ra} 3^{\ra}} \!\cdot\! \frac{\partial P_{3}^{\ra \rb \ra}}{\partial \bm{v}_{1}^{\ra}} \!+\! \bm{\mathcal{F}}_{\! 2^{\rb} 3^{\ra}} \!\cdot\! \frac{\partial P_{3}^{\ra \rb \ra}}{\partial \bm{v}_{2}^{\rb}} \bigg]  \nonumber
\\
& \, + (N_{\rb} \!-\! 1 ) \, \mu_{\rb} \!\! \int \!\! \rd \Gamma_{3}^{\rb} \, \bigg[ \bm{\mathcal{F}}_{\! 1^{\ra} 3^{\rb}} \!\cdot\! \frac{\partial P_{3}^{\ra \rb \rb}}{\partial \bm{v}_{1}^{\ra}} \!+\! \bm{\mathcal{F}}_{\! 2^{\rb} 3^{\rb}} \!\cdot\! \frac{\partial P_{3}^{\ra \rb \rb}}{\partial \bm{v}_{2}^{\rb}} \bigg]  \nonumber
\\
& \, + \sum_{\rc \neq \ra , \rb} N_{\rc} \, \mu_{\rc} \!\! \int \!\! \rd \Gamma_{3}^{\rc} \, \bigg[ \bm{\mathcal{F}}_{\! 1^{\ra} 3^{\rc}} \!\cdot\! \frac{\partial P_{3}^{\ra \rb \rc}}{\partial \bm{v}_{1}^{\ra}} \!+\! \bm{\mathcal{F}}_{\! 2^{\rb} 3^{\rc}} \!\cdot\! \frac{\partial P_{3}^{\ra \rb \rc}}{\partial \bm{v}_{2}^{\rb}} \bigg] = 0 \, .
\label{BBGKY_2_ab_multi}
\end{align}
As in equation~\eqref{definition_fn}, we now introduce the renormalised DFs $f_{1}^{\ra}$, $f_{2}^{\ra \rb}$, and $f_{3}^{\ra \rb \rc}$ as
\begin{align}
& \, f_{1}^{\ra} = \mu_{\ra} N_{\ra} P_{1}^{\ra}
\; ; \;
f_{2}^{\ra \ra} = \mu_{\ra}^{2} N_{\ra} (N_{\ra} \!-\! 1) P_{2}^{\ra \ra}
\; ; \;
f_{2}^{\ra \rb} = \mu_{\ra} \mu_{\rb} N_{\ra} N_{\rb} P_{2}^{\ra \rb}  \nonumber
\\
& \, f_{3}^{\ra \ra \ra} = \mu_{\ra}^{3} N_{\ra} (N_{\ra} \!-\! 1) (N_{\ra} \!-\! 2) P_{3}^{\ra \ra \ra}
\; ; \;
f_{3}^{\ra \ra \rb} = \mu_{\ra}^{2} \mu_{\rb} N_{\ra} (N_{\ra} \!-\! 1) N_{\rb} P_{3}^{\ra \ra \rb}  \nonumber
\\
& \, f_{3}^{\ra \rb \rc} = \mu_{\ra} \mu_{\rb} \mu_{\rc} N_{\ra} N_{\rb} N_{\rc} P_{3}^{\ra \rb \rc} \, ,
\label{definition_fn_multi}
\end{align}
where we assumed that ${``\ra"}$, ${``\rb"}$, and ${``\rc"}$ were associated with different components. With these new normalisations, equation~\eqref{BBGKY_1_multi} immediately becomes
\begin{align}
& \, \frac{\partial f_{1}^{\ra}}{\partial t} \!+\! \bm{v}_{1}^{\ra} \!\cdot\! \frac{\partial f_{1}^{\ra}}{\partial \bm{x}_{1}^{\ra}} \!+\! \bigg[ M_{\bullet} \bm{\mathcal{F}}_{\!1^{\ra} 0} \!+\! M_{\star} \bm{\mathcal{F}}_{\!1^{\ra} \rr} \bigg] \!\cdot\! \frac{\partial f_{1}^{\ra}}{\partial \bm{v}_{1}^{\ra}}  \nonumber
\\
& \, + \sum_{\rb} \!\! \int \!\! \rd \Gamma_{2}^{\rb} \, \bm{\mathcal{F}}_{\! 1^{\ra} 2^{\rb}} \!\cdot\! \frac{\partial f_{2}^{\ra \rb}}{\partial \bm{v}_{1}^{\ra}} = 0 \, ,
\label{BBGKY_1_f_multi}
\end{align}
where one should note that the sum over ${``\rb"}$ runs for all components, which allows for a generic writing.
Equations~\eqref{BBGKY_2_aa_multi} and~\eqref{BBGKY_2_ab_multi} can then be cast under the same generic form
\begin{align}
& \, \frac{\partial f_{2}^{\ra \rb}}{\partial t} \!+\! \bm{v}_{1}^{\ra} \!\cdot\! \frac{\partial f_{2}^{\ra \rb}}{\partial \bm{x}_{1}^{\ra}} \!+\! \bm{v}_{2}^{\rb} \!\cdot\! \frac{\partial f_{2}^{\ra \rb}}{\partial \bm{x}_{2}^{\rb}} \!+\! \mu_{\rb} \bm{\mathcal{F}}_{\!1^{\ra}2^{\rb}} \!\cdot\! \frac{\partial f_{2}^{\ra \rb}}{\partial \bm{v}_{1}^{\ra}} \!+\! \mu_{\ra} \bm{\mathcal{F}}_{\!2^{\rb} 1^{\ra}} \!\cdot\! \frac{\partial f_{2}^{\ra \rb}}{\partial \bm{v}_{2}^{\rb}}  \nonumber
\\
& \, + \bigg[ M_{\bullet} \bm{\mathcal{F}}_{\! 1^{\ra} 0} \!+\! M_{\star} \bm{\mathcal{F}}_{\! 1^{\ra} \rr} \bigg] \!\cdot\! \frac{\partial f_{2}^{\ra \rb}}{\partial \bm{v}_{1}^{\ra}} \!+\! \bigg[ M_{\bullet} \bm{\mathcal{F}}_{\! 2^{\rb}0} \!+\! M_{\star} \bm{\mathcal{F}}_{\! 2^{\rb} \rr} \bigg] \!\cdot\! \frac{\partial f_{2}^{\ra \rb}}{\partial \bm{v}_{2}^{\rb}}   \nonumber
\\
& \, + \sum_{\rc} \!\! \int \!\! \rd \Gamma_{3}^{\rc} \, \bigg[ \bm{\mathcal{F}}_{\! 1^{\ra} 3^{\rc}} \!\cdot\! \frac{\partial f_{3}^{\ra \rb \rc}}{\partial \bm{v}_{1}^{\ra}} \!+\! \bm{\mathcal{F}}_{\! 2^{\rb} 3^{\rc}} \!\cdot\! \frac{\partial f_{3}^{\ra \rb \rc}}{\partial \bm{v}_{2}^{\rb}} \bigg] = 0 \, .
\label{BBGKY_2_f_multi}
\end{align}
Let us insist on the fact that equation~\eqref{BBGKY_2_f_multi} holds for both the cases where ${``\ra"}$ and ${``\rb"}$ are equal or different, and the sum on ${``\rc"}$ runs for all components. As in equations~\eqref{definition_g2} and~\eqref{definition_g3}, one can now define the cluster representation of the DFs which, in this multi-component context, reads
\begin{equation}
f_{2}^{\ra \rb} (\Gamma_{1}^{\ra} , \Gamma_{2}^{\rb}) = f_{1}^{\ra} (\Gamma_{1}^{\ra}) \, f_{1}^{\rb} (\Gamma_{2}^{\rb}) \!+\! g_{2}^{\ra \rb} (\Gamma_{1}^{\ra} , \Gamma_{2}^{\rb}) \, ,
\label{definition_g2_multi}
\end{equation}
and
\begin{align}
& \, f_{3}^{\ra \rb \rc} (\Gamma_{1}^{\ra} , \Gamma_{2}^{\rb} , \Gamma_{3}^{\rc}) = f_{1}^{\ra} (\Gamma_{1}^{\ra}) \, f_{1}^{\rb} (\Gamma_{2}^{\rb}) \, f_{1}^{\rc} (\Gamma_{3}^{\rc})   \nonumber
\\
& \, + f_{1}^{\ra} (\Gamma_{1}^{\ra}) \, g_{2}^{\rb \rc} (\Gamma_{2}^{\rb} , \Gamma_{2}^{\rc}) \!+\! f_{1}^{\rb} (\Gamma_{2}^{\rb}) \, g_{2}^{\ra \rc} (\Gamma_{1}^{\ra} , \Gamma_{3}^{\rc}) \!+\! f_{1}^{\rc} (\Gamma_{3}^{\rc}) \, g_{2}^{\ra \rb} (\Gamma_{1}^{\ra} , \Gamma_{1}^{\rb})   \nonumber
\\
& \, + g_{3}^{\ra \rb \rc} (\Gamma_{1}^{\ra} , \Gamma_{2}^{\rb} , \Gamma_{3}^{\rc}) \, .
\label{definition_g3_multi}
\end{align}
Following equation~\eqref{normalisations_f1_g2_g3}, we assume that $g_{2}^{\ra \rb}$ scales like the inverse of the number of particles, while $g_{3}^{\ra \rb \rc}$ scales like the square of the inverse of the number of particles. Using the decompositions from equations~\eqref{definition_g2_multi} and~\eqref{definition_g3_multi}, and keeping only terms of order ${ 1/N_{\ra} }$ or larger (where ${``\ra"}$ runs over all the components), the first equation~\eqref{BBGKY_1_f_multi} of the BBGKY hierarchy becomes
\begin{align}
& \, \frac{\partial f_{1}^{\ra}}{\partial t} \!+\! \bm{v}_{1}^{\ra} \!\cdot\! \frac{\partial f_{1}^{\ra}}{\partial \bm{x}_{1}^{\ra}} \!+\! \bigg[ M_{\bullet} \bm{\mathcal{F}}_{\!1^{\ra} 0} \!+\! M_{\star} \bm{\mathcal{F}}_{\! 1^{\ra} \rr} \!+\! \sum_{\rb} \!\! \int \!\! \rd \Gamma_{2}^{\rb} \, \bm{\mathcal{F}}_{\!1^{\ra} 2^{\rb}} f_{1}^{\rb} (\Gamma_{2}^{\rb})\bigg] \!\cdot\! \frac{\partial f_{1}^{\ra}}{\partial \bm{v}_{1}^{\ra}}  \nonumber
\\
& \, + \sum_{\rb} \!\! \int \!\! \rd \Gamma_{2}^{\rb} \, \bm{\mathcal{F}}_{\!1^{\ra} 2^{\rb}} \!\cdot\! \frac{\partial g_{2}^{\ra \rb}}{\partial \bm{v}_{1}^{\ra}} = 0 \, .
\label{BBGKY_1_g_multi}
\end{align}
while the second equation~\eqref{BBGKY_2_f_multi} becomes
\begin{align}
& \, \frac{\partial g_{2}^{\ra \rb}}{\partial t} \!+\! \bm{v}_{1}^{\ra} \!\cdot\! \frac{\partial g_{2}^{\ra \rb}}{\partial \bm{x}_{1}^{\ra}} \!+\! \bm{v}_{2}^{\rb} \!\cdot\! \frac{\partial g_{2}^{\ra \rb}}{\partial \bm{x}_{2}^{\rb}}   \nonumber
\\
& \, + \mu_{\rb} \bm{\mathcal{F}}_{\!1^{\ra} 2^{\rb}} \!\cdot\! \frac{\partial f_{1}^{\ra}}{\partial \bm{v}_{1}^{\ra}} f_{1}^{\rb} (\Gamma_{2}^{\rb}) \!+\! \mu_{\ra} \bm{\mathcal{F}}_{\!2^{\rb} 1^{\ra}} \!\cdot\! \frac{\partial f_{1}^{\rb}}{\partial \bm{v}_{2}^{\rb}} f_{1}^{\ra} (\Gamma_{1}^{\ra}) \nonumber
\\
& \, + \bigg[ M_{\bullet} \bm{\mathcal{F}}_{\!1^{\ra} 0} \!+\! M_{\star} \bm{\mathcal{F}}_{\!1^{\ra} \rr} \bigg] \!\cdot\! \frac{\partial g_{2}^{\ra \rb}}{\partial \bm{v}_{1}^{\ra} } \!+\! \bigg[ M_{\bullet} \bm{\mathcal{F}}_{\!2^{\rb}0} \!+\! M_{\star} \bm{\mathcal{F}}_{\!2^{\rb} \rr} \bigg] \!\cdot\! \frac{\partial g_{2}^{\ra \rb}}{\partial \bm{v}_{2}^{\rb}}  \nonumber
\\
& \, + \bigg[ \sum_{\rc} \!\! \int \!\! \rd \Gamma_{3}^{\rc} \, \bm{\mathcal{F}}_{\!1^{\ra} 3^{\rc}} f_{1}^{\rc} (\Gamma_{3}^{\rc}) \bigg] \!\cdot\! \frac{\partial g_{2}^{\ra \rb}}{\partial \bm{v}_{1}^{\ra}} \!+\! \bigg[ \sum_{\rc} \!\! \int \!\! \rd \Gamma_{3}^{\rc} \, \bm{\mathcal{F}}_{\! 2^{\rb} 3^{\rc}} f_{1}^{\rc} (\Gamma_{3}^{\rc}) \bigg] \!\cdot\! \frac{\partial g_{2}^{\ra \rb}}{\partial \bm{v}_{2}^{\rb}}   \nonumber
\\
& \, + \bigg[ \sum_{\rc} \!\! \int \!\! \rd \Gamma_{3}^{\rc} \, \bm{\mathcal{F}}_{\!1^{\ra} 3^{\rc}} g_{2}^{\rb \rc} (\Gamma_{2}^{\rb} , \Gamma_{3}^{\rc}) \bigg] \!\cdot\! \frac{\partial f_{1}^{\ra}}{\partial \bm{v}_{1}^{\ra}}  \nonumber
\\
& \,  + \bigg[ \sum_{\rc} \!\! \int \!\! \rd \Gamma_{3}^{\rc} \, \bm{\mathcal{F}}_{\!2^{\rb} 3^{\rc}} g_{2}^{\ra \rc} (\Gamma_{1}^{\ra} , \Gamma_{3}^{\rc}) \bigg] \!\cdot\! \frac{\partial f_{1}^{\rb}}{\partial \bm{v}_{2}^{\rb}} = 0 \, .
\label{BBGKY_2_g_multi}
\end{align}
Much like for equation~\eqref{definition_F_C}, let us introduce the system's ${1-}$body DF $F^{\ra}$ and ${2-}$body autocorrelation $\mathcal{C}^{\ra \rb}$ as
\begin{equation}
F^{\ra} = \frac{f_{1}^{\ra}}{M_{\star}} \;\;\; ; \;\;\; \mathcal{C}^{\ra \rb} = \frac{g_{2}^{\ra \rb}}{M_{\star}^{2}} \, ,
\label{definition_F_C_multi}
\end{equation}
noting the slightly different normalisations of $\mathcal{C}^{\ra \rb}$, so as to ensure a symmetric rescaling with respect to ${``\ra"}$ and ${``\rb"}$. We also follow equations~\eqref{rescaling_U} and~\eqref{rescale_Phi_a} to rescale the interaction potential as well as the relativistic corrections with the mass of the BH. Given these various renormalisations, equation~\eqref{BBGKY_1_g_multi} becomes
\begin{align}
& \, \frac{\partial F^{\ra}}{\partial t} \!+\! \bm{v}_{1}^{\ra} \!\cdot\! \frac{\partial F^{\ra}}{\partial \bm{x}_{1}^{\ra}} \!+\! \bm{\mathcal{F}}_{\!1^{\ra}0} \!\cdot\! \frac{\partial F^{\ra}}{\partial \bm{v}_{1}^{\ra}} \!+\! \varepsilon \bigg[ \sum_{\rb} \!\! \int \!\! \rd \Gamma_{2}^{\rb} \, \bm{\mathcal{F}}_{\!1^{\ra}2^{\rb}} F^{\rb} (\Gamma_{2}^{\rb}) \bigg] \!\cdot\! \frac{\partial F^{\ra}}{\partial \bm{v}_{1}^{\ra}} \nonumber
\\
& \, \!+\! \varepsilon \bm{\mathcal{F}}_{\!1^{\ra} \rr} \!\cdot\! \frac{\partial F^{\ra}}{\partial \bm{v}_{1}^{\ra}}  \!+\! \varepsilon \sum_{\rb} \!\! \int \!\! \rd \Gamma_{2}^{\rb} \, \bm{\mathcal{F}}_{\!1^{\ra} 2^{\rb}} \!\cdot\! \frac{\partial \mathcal{C}^{\ra \rb}}{\partial \bm{v}_{1}^{\ra}} = 0 \, ,
\label{BBGKY_1_rescaled_multi}
\end{align}
where the small parameter ${ \varepsilon \!=\! M_{\star} / M_{\bullet} }$ was introduced. Similarly, equation~\eqref{BBGKY_2_g_multi} becomes
\begin{align}
& \, \frac{\partial \mathcal{C}^{\ra \rb}}{\partial t} \!+\! \bm{v}_{1}^{\ra} \!\cdot\! \frac{\partial \mathcal{C}^{\ra \rb}}{\partial \bm{x}_{1}^{\ra}} \!+\! \bm{v}_{2}^{\rb} \!\cdot\! \frac{\partial \mathcal{C}^{\ra \rb}}{\partial \bm{x}_{2}^{\rb}} \!+\! \bm{\mathcal{F}}_{\!1^{\ra}0} \!\cdot\! \frac{\partial \mathcal{C}^{\ra \rb}}{\partial \bm{v}_{1}^{\ra}} \!+\! \bm{\mathcal{F}}_{\!2^{\rb}0} \!\cdot\! \frac{\partial \mathcal{C}^{\ra \rb}}{\partial \bm{v}_{2}^{\rb}}  \nonumber
\\
& \, + \varepsilon \bm{\mathcal{F}}_{\!1^{\ra} \rr} \!\cdot\! \frac{\partial \mathcal{C}^{\ra \rb}}{\partial \bm{v}_{1}^{\ra}} \!+\! \varepsilon \bm{\mathcal{F}}_{\! 2^{\rb} \rr} \!\cdot\! \frac{\partial \mathcal{C}^{\ra \rb}}{\partial \bm{v}_{2}^{\rb}} 
  \nonumber
\\
 & \, + \varepsilon \eta_{\rb} \bm{\mathcal{F}}_{\! 1^{\ra} 2^{\rb}} \!\cdot\! \frac{\partial F^{\ra}}{\partial \bm{v}_{1}^{\ra}} F^{\rb}(\Gamma_{2}^{\rb}) \!+\! \varepsilon \eta_{\ra} \bm{\mathcal{F}}_{\! 2^{\rb} 1^{\ra}} \!\cdot\! \frac{\partial F^{\rb}}{\partial \bm{v}_{2}^{\rb}} F^{\ra} (\Gamma_{1}^{\ra}) \nonumber
\\
& \, + \varepsilon \bigg[\! \sum_{\rc} \!\! \int \!\!\! \rd \Gamma_{3}^{\rc} \bm{\mathcal{F}}_{\! 1^{\ra} 3^{\rc}} F^{\rc} (\Gamma_{3}^{\rc}) \bigg] \!\cdot\! \frac{\partial \mathcal{C}^{\ra \rb}}{\partial \bm{v}_{1}^{\ra}} \!+\! \varepsilon\bigg[\! \sum_{\rc} \!\! \int \!\!\! \rd \Gamma_{3}^{\rc}  \bm{\mathcal{F}}_{\!2^{\rb} 3^{\rc}} F^{\rc} (\Gamma_{3}^{\rc}) \bigg] \!\cdot\! \frac{\partial \mathcal{C}^{\ra \rb}}{\partial \bm{v}_{2}^{\rb}}  \nonumber
\\
& \, + \varepsilon \bigg[ \sum_{\rc} \!\! \int \!\! \rd \Gamma_{3}^{\rc} \, \bm{\mathcal{F}}_{\!1^{\ra} 3^{\rc}} \mathcal{C}^{\rb \rc} (\Gamma_{2} , \Gamma_{3}^{\rc}) \bigg] \!\cdot\! \frac{\partial F^{\ra}}{\partial \bm{v}_{1}^{\ra}}   \nonumber
\\
& \, + \varepsilon \bigg[ \sum_{\rc} \!\! \int \!\! \rd \Gamma_{3}^{\rc} \, \bm{\mathcal{F}}_{\! 2^{\rb} 3^{\rc}} \mathcal{C}^{\ra \rc} (\Gamma_{1}^{\ra} , \Gamma_{3}^{\rc}) \bigg] \!\cdot\! \frac{\partial F^{\rb}}{\partial \bm{v}_{2}^{\rb}} = 0 \, ,
\label{BBGKY_2_rescaled_multi}
\end{align}
where we introduced the small parameter ${ \eta_{\ra} \!=\! \mu_{\ra} / M_{\star} }$ of order ${ 1/ N_{\ra} }$.
Equations~\eqref{BBGKY_1_rescaled_multi} and~\eqref{BBGKY_2_rescaled_multi} are the direct analogs of equations~\eqref{BBGKY_1_rescaled} and~\eqref{BBGKY_2_rescaled}, when one considers a system with multiple components.

As was done in section~\ref{sec:AA_coordinates}, one may now rewrite the two previous evolution equations within the appropriate angle-action coordinates for the BH-induced Keplerian motion. We perform a degenerate angle-average as defined in equation~\eqref{definition_degenerate_angle_average}, and assume that $F^{\ra}$ and $\mathcal{C}^{\ra \rb}$ satisfy the ansatz from equations~\eqref{DL_DF} and~\eqref{DL_C}. It is then straightforward to rewrite equation~\eqref{BBGKY_1_rescaled_multi} as
\begin{equation}
\frac{\partial \oFa}{\partial \tau} \!+\! \big[ \oFa , \oP \!+\! \oP_{\rr} \big] \!+\! \sum_{\rb} \!\! \int \!\! \rd \bR_{2} \, \big[ \oCab (\bR_{1} , \bR_{2}) , \oU_{12} \big]_{(1)} = 0 \, ,
\label{BBGKY_1_multi_final}
\end{equation}
where we used the rescaled time ${ \tau \!=\! (2 \pi)^{d -k} \varepsilon t }$ from equation~\eqref{definition_tau}, with ${ \varepsilon \!=\! M_{\star} / M_{\bullet} }$. Following equation~\eqref{definition_averaged_Phi}, we also introduced the total averaged self-consistent potential $\oP$ as
\begin{equation}
\oP = \sum_{\ra} \oPa \, ,
\label{total_selfconsistent_potential}
\end{equation}
where the averaged potential $\oPa$ is given by
\begin{equation}
\oPa (\bR_{1}) = \!\! \int \!\! \rd \bR_{2} \, \oFa (\bR_{2}) \, \oU_{12} (\bR_{1} , \bR_{2}) \, ,
\label{definition_Phi_self_multi}
\end{equation}
In equation~\eqref{definition_Phi_self_multi}, the averaged interaction potential $\oU_{12}$ introduced in equation~\eqref{definition_Ubar12} was used. One can similarly rewrite equation~\eqref{BBGKY_2_rescaled_multi} as
\begin{align}
& \, \frac{\partial \oCab}{\partial \tau} \!+\! \big[ \oCab (\bR_{1} , \bR_{2}), \oP (\bR_{1}) \!+\! \oP_{\rr} (\bR_{1}) \big]_{(1)}   \nonumber
\\
& \, + \big[ \oCab (\bR_{1} , \bR_{2}) , \oP (\bR_{2}) \!+\! \oP_{\rr} (\bR_{2}) \big]_{(2)}  \nonumber
\\
& \, + \sum_{\rc} \!\! \int \!\! \rd \bR_{3} \, \oCbc (\bR_{2} , \bR_{3}) \big[ \oFa (\bR_{1}) , \oU_{13} \big]_{(1)}   \nonumber
\\
& \, + \sum_{\rc} \!\! \int \!\! \rd \bR_{3} \, \oCac (\bR_{1} , \bR_{3}) \big[ \oFb (\bR_{2}) , \oU_{23} \big]_{(2)}   \nonumber
\\
& \, + \frac{1}{(2 \pi)^{d - k}} \bigg\{ \eta_{\rb} \big[ \oFa (\bR_{1}) \oFb (\bR_{2}) , \oU_{12} \big]_{(1)}   \nonumber
\\
& \, + \eta_{\ra} \big[ \oFa (\bR_{1}) \oFb (\bR_{2}) , \oU_{21} \big]_{(2)} \bigg\} = 0 \, .
\label{BBGKY_2_multi_final}
\end{align}
With the two coupled evolution equations~\eqref{BBGKY_1_multi_final} and~\eqref{BBGKY_2_multi_final} while keeping track of the different mass prefactors, one can follow the path presented in the previous subsection to derive   equation~\eqref{LB_multi}, the appropriate closed kinetic equation for $\oFa$.

\section{From Fokker-Planck to Langevin}
\label{sec:FP_to_Langevin}

Following~\cite{Risken1996}, let us briefly recall how one may obtain the stochastic Langevin equation describing the individual dynamics of a test particle starting from a Fokker-Planck equation describing the diffusion of the system's DF as a whole. We start from the generic writing of the degenerate Balescu-Lenard equation from equation~\eqref{rewrite_BL_Kepler}, written as an anisotropic self-consistent Fokker-Planck equation. It reads
\begin{equation}
\frac{\partial \oF}{\partial \tau} = \frac{\partial }{\partial \bm{J}^{\rm s}} \!\cdot\! \bigg[ \bm{A} (\bm{J} , \tau) \, \oF (\bm{J} , \tau) \!+\! \bm{D} (\bm{J} , \tau) \!\cdot\! \frac{\partial \oF}{\partial \bm{J}^{\rm s}} \bigg] \, ,
\label{anistropic_FP}
\end{equation}
where, following the notations from equation~\eqref{drift_diff_Kepler},  the drift vector ${ \bm{A} (\bm{J} , \tau) }$ and diffusion tensor ${ \bm{D} (\bm{J} , \tau) }$ are introduced as
\begin{equation}
\bm{A} (\bm{J} , \tau) \!=\! \! \sum_{\bm{m}^{\rm s}} \! \bm{m}^{\rm s} A_{\bm{m}^{\rm s}} (\bm{J} , \tau) \; ; \; \bm{D} (\bm{J} , \tau) \!=\! \! \sum_{\bm{m}^{\rm s}} \! \bm{m}^{\rm s} \!\otimes\! \bm{m}^{\rm s} \, D_{\bm{m}^{\rm s}} (\bm{J} , \tau) \, .
\label{drift_vector_diffusion_tensor}
\end{equation}
One should keep in mind that in equation~\eqref{rewrite_BL_Kepler} the drift and diffusion coefficients also secularly depend on the system's DF $\oF$, but this was not written out to simplify the notations. Following the notations of equation~(4.94a) in~\cite{Risken1996}, we may immediately rewrite equation~\eqref{anistropic_FP} as
\begin{equation}
\frac{\partial \oF}{\partial \tau} \!=\! \frac{\partial }{\partial \bm{J}^{\rm s}} \!\cdot\! \bigg[ - \bm{D}^{(1)} \!(\bm{J} , \tau) \, \oF (\bm{J} , \tau) \!+\! \frac{\partial }{\partial \bm{J}^{\rm s}} \!\cdot\! \bigg[ \bm{D}^{(2)} \!(\bm{J} , \tau) \, \oF (\bm{J} , \tau) \bigg] \, \bigg] \, ,
\label{anisotropic_FP_Risken}
\end{equation}
where the first- and second-order diffusion coefficients read
\begin{equation}
\bm{D}^{(1)} (\bm{J} , \tau) \!=\! - \bm{A} (\bm{J} , \tau) \!+\! \frac{\partial }{\partial \bm{J}^{\rm s}} \!\cdot\! \bm{D} (\bm{J} , \tau) \; ; \; \bm{D}^{(2)} (\bm{J} , \tau) \!=\! \bm{D} (\bm{J} , \tau) \, . 
\label{link_A_D}
\end{equation}
One should pay attention to the fact that the diffusion of stars takes place in the full action domain $\bm{J}$, while only gradients with respect to the slow actions $\bm{J}^{\rm s}$ are present in equation~\eqref{anisotropic_FP_Risken}. Of course, by enlarging the diffusion coefficients $\bm{D}^{(1)}$ and $\bm{D}^{(2)}$ with zero coefficients for all the adiabatically conserved fast actions $\bm{J}^{\rm f}$, it is straightforward to rewrite equation~\eqref{anisotropic_FP_Risken} as a diffusion equation in ${\bm{J}-}$space involving derivatives with respect to $\bm{J}$.

Let us now focus on the individual dynamics of a given test particle. We denote as ${ \bm{\mathcal{J}} (\tau) }$ its position in action space at time $\tau$. This test particle then undergoes a stochastic diffusion consistent with the averaged diffusion captured by the Fokker-Planck equation~\eqref{anisotropic_FP_Risken}, namely a Langevin equation reading
\begin{equation}
\frac{\rd \bm{\mathcal{J}}}{\rd \tau} = \bm{h} (\bm{\mathcal{J}} , \tau) + \bm{g} (\bm{\mathcal{J}} , \tau) \!\cdot\! \bm{\Gamma} (\tau) \, ,
\label{Langevin_equation}
\end{equation}
where we introduced the Langevin vector and tensor $\bm{h}$ and $\bm{g}$, as well as the stochastic Langevin forces ${ \bm{\Gamma} (\tau) }$, whose statistics satisfy
\begin{equation}
\big< \bm{\Gamma} (\tau) \big> = 0 \;\;\; ; \;\;\; \big< \bm{\Gamma} (\tau) \!\otimes\! \bm{\Gamma} (\tau') \big> = 2 \, \bm{I} \, \delta_{\rm D} (\tau \!-\! \tau') \, ,
\label{Langevin_forces}
\end{equation}
with $\bm{I}$ the identity matrix.
Following equation~(3.124) of~\cite{Risken1996}, we may now express the Langevin coefficients from equation~\eqref{Langevin_equation} as a function of the coefficients appearing in the Fokker-Planck equation~\eqref{anisotropic_FP_Risken}. The second-order diffusion tensor $\bm{D}^{(2)}$ being definite positive, we introduce as ${ \sqrt{\bm{D}}^{(2)} }$ one of its square root, so that one has the component relations
\begin{equation}
\bm{h}_{i} = \bm{D}^{(1)}_{i} \!-\! \sum_{j,k} \big(\! \sqrt{\bm{D}}^{(2)} \big)_{kj} \frac{\partial \big(\! \sqrt{\bm{D}}^{(2)} \big)_{ij}}{\partial x_{k}} \;\;\; ; \;\;\; \bm{g}_{ij} = \big(\! \sqrt{\bm{D}}^{(2)} \big)_{ij} \, .
\label{g_h_relation}
\end{equation}

Equation~\eqref{g_h_relation} therefore allows us to fully specify the properties of the diffusion of an individual particle as captured by the Langevin equation~\eqref{Langevin_equation}. Of course, self-consistency requires that the diffusion coefficients $\bm{D}^{(1)}$ and $\bm{D}^{(2)}$, and therefore the Langevin coefficients $\bm{h}$ and $\bm{g}$ should be updated as the system's DF $\oF$ changes on secular timescales,
as mentioned in the main text.

\end{document}